\documentclass[twocolumn]{aastex6}
\usepackage{multirow}
\usepackage{amsbsy}
\usepackage{wasysym}
\usepackage{newfloat}

\DeclareFloatingEnvironment[name={Supplementary Figure}]{suppfigure}
\begin{document}

\title{The Gould's Belt Distances Survey (GOBELINS)  I. \\
Trigonometric parallax distances and depth of the Ophiuchus complex}

\author{Gisela N.\ Ortiz-Le\'on\altaffilmark{1}, 
Laurent Loinard\altaffilmark{1,2}, 
Marina A.\ Kounkel\altaffilmark{3},
Sergio A.\ Dzib\altaffilmark{2},
Amy J.\ Mioduszewski\altaffilmark{4},
Luis F.\ Rodr\'{\i}guez\altaffilmark{1,5},
Rosa M.\ Torres\altaffilmark{6},
Rosa A.\ Gonz\'alez-L\'opezlira\altaffilmark{1,13,14},
Gerardo Pech\altaffilmark{1,7},
Juana L.\ Rivera\altaffilmark{1},
Lee Hartmann\altaffilmark{3},
Andrew F.\ Boden\altaffilmark{8},
Neal J.\ Evans II\altaffilmark{9},
Cesar Brice\~no\altaffilmark{10},
John J.\ Tobin\altaffilmark{11},
Phillip A.~B.\ Galli\altaffilmark{12,15},
and Donald Gudehus\altaffilmark{16}
}

\email{g.ortiz@crya.unam.mx}

\altaffiltext{1}{Instituto de Radioastronom\'ia y Astrof\'isica, 
Universidad Nacional Aut\'onoma de Mexico,
Morelia 58089, Mexico}
\altaffiltext{2}{Max Planck Institut f\"ur Radioastronomie, Auf dem H\"ugel 69, 
D-53121 Bonn, Germany}
\altaffiltext{3}{Department of Astronomy, University of Michigan, 500 Church Street, 
Ann Arbor, MI 48105,  USA}
\altaffiltext{4}{National Radio Astronomy Observatory, Domenici Science Operations 
Center, 1003 Lopezville Road, Socorro, NM 87801, USA}
\altaffiltext{5}{King Abdulaziz University, P.O. Box 80203, Jeddah 21589, Saudi Arabia}
\altaffiltext{6}{Centro Universitario de Tonal\'a, Universidad de Guadalajara,
Avenida Nuevo Perif\'erico No. 555, Ejido San Jos\'e
Tatepozco, C.P. 48525, Tonal\'a, Jalisco, M\'exico. }
\altaffiltext{7}{ The Academia Sinica Institute of Astronomy and Astrophysics,
AS/NTU. No.1, Sec. 4, Roosevelt Rd, Taipei 10617,
Taiwan, R.O.C.}
\altaffiltext{8}{Division of Physics, Math and Astronomy, California Institute of Technology, 
1200 East California Boulevard, Pasadena, CA 91125, USA} 
\altaffiltext{9}{Department of Astronomy, The University of Texas at Austin, 
2515 Speedway, Stop C1400, Austin, TX 78712-1205, USA}
\altaffiltext{10}{Cerro Tololo Interamerican Observatory, Casilla 603, La Serena, Chile}
\altaffiltext{11}{Leiden Observatory, PO Box 9513, NL-2300 RA, Leiden, The Netherlands}
\altaffiltext{12}{Instituto de Astronomia, Geof\'isica e Ci\^encias Atmosf\'ericas, Universidade de S\~ao Paulo,
Rua do Mat\~ao 1226, Cidade Universit\'aria, S\~ao Paulo, Brazil }
\altaffiltext{13}{Helmhotz-Institute f\"ur Strahlen-und Kernphysik (HISKP), Universit\"at Bonn, Nussallee 14-16, D-53115 Bonn, Germany}
\altaffiltext{14}{Argelander-Institut f\"ur Astronomie, Auf dem H\"ugel 71, D-53121, Bonn, Germany}
\altaffiltext{15}{Univ. Grenoble Alpes, IPAG, 38000, Grenoble, France}
\altaffiltext{16}{Department of Physics \& Astronomy, Georgia State University, Atlanta, GA 30303, USA}

\begin{abstract}

We present the first results of the \emph{Gould's Belt Distances Survey (GOBELINS)}, a project aimed at measuring the proper motion and trigonometric parallax of a large sample of young stars in nearby regions using multi-epoch Very Long Baseline Array (VLBA) radio observations. Enough VLBA detections have now been obtained for 16 stellar systems in Ophiuchus to derive their parallax and proper motion. This leads to distance determinations for individual stars with an accuracy of 0.3 to a few percent. In addition, the orbits of 6 multiple systems were modelled by combining absolute positions with VLBA (and in some cases, near infrared) angular separations. Twelve stellar systems are located in the dark cloud Lynds~1688; the individual distances for this sample are highly consistent with one another, and yield a mean parallax for Lynds~1688 of $\varpi=7.28\pm0.06$ mas, corresponding to a distance $d=137.3\pm1.2$~pc. This represents an accuracy better than 1\%. Three systems for which astrometric elements could be measured are located in the eastern streamer (Lynds 1689) and yield an estimate of $\varpi=6.79\pm0.16$~mas, corresponding to a distance $d=147.3\pm3.4$~pc. This suggests that the eastern streamer is located about 10 pc farther than the core, but this conclusion needs to be confirmed by observations (currently being collected) of additional sources in the eastern streamer. From the measured proper motions, we estimate the one-dimensional velocity dispersion in Lynds 1688 to be 2.8$\pm$1.8 and 3.0$\pm$2.0~${\rm km~s}^{-1}$, in R.A. and DEC., respectively; these are larger than, but still consistent within $1\sigma$, with those found in other studies.

 \end{abstract}


\section{Introduction}\label{sec:intro}

\subsection{The Gould's Belt}

The Gould's Belt (see \citealt{Poppel_1997} for a comprehensive review) is a local Galactic structure containing much of the dense interstellar matter and many of the young stars
within a few hundred parsecs of the Sun. It was originally identified by John Herschel (circa 1847) and Benjamin Gould (in the 1870s), who noticed that most of the brightest stars were neither randomly distributed on the sky, nor associated with the Galactic plane, but instead concentrated along a great circle tilted by about 18$^{\circ}$ from the Galactic equator. Modern studies \citep[e.g.,][]{Perrot_2003} have shown that the Gould's Belt is a broad elliptical ring of young stars and interstellar matter with semimajor and semiminor axes of 375 pc and 235 pc, respectively. The center of the structure is located at about 105 pc from the Sun, in the direction of the Galactic anti-center. There is ample evidence that the Gould's Belt is expanding and has a dynamical age of order 30 Myr -- \cite{Perrot_2003} find 26.4 $\pm$ 0.4 Myr. The oldest stars associated with the Gould's Belt are also about 30 Myr old \citep[e.g.,][]{Stothers_1974}, but T Tauri stars (age 10$^6$--10$^7$ yr), as well as protostars ($\lesssim$~10$^5$ yr) and pre-stellar cores, are also present, showing that star-formation is still on-going. 

The Gould's Belt contains several million Solar masses of interstellar material and includes all the nearby sites of active star-formation (Orion, Ophiuchus, Perseus, etc.). These have been the benchmarks against which theories of star-formation have been tested. Indeed, numerous ``Gould's Belt surveys" targeting these regions have been carried out over the years --for instance, the James Clerk Maxwell Telescope Legacy Survey of Nearby Star-forming Regions in the Gould Belt \citep{Ward-Thompson_2007}, the Spitzer Gould Belt \citep{Dunham_2015} and c2d \citep{Evans_2009} Legacy Surveys, and the Herschel Gould's Belt Survey  \citep{Andre_2010}. 
To take full advantage of this wealth of high quality information, it is fundamental to have accurate distance measurements to each of the regions in the Gould's Belt. In addition, these regions are a few hundred parsecs away and typically a few tens of parsecs across --and therefore presumably also a few tens of parsecs deep. As a consequence, using a single mean distance (however accurately measured) for all young stellar objects (YSOs) in a given region will result in typical distance errors in excess of 10\% for the individual YSOs. A case in point is that of the Taurus star-forming region, which is located at a mean distance of about 145 pc, but is about 30 pc deep \citep{Loinard_2007,Torres_2007,Torres_2009,Torres_2012}. Using the mean distance to Taurus to calculate luminosities for YSOs located on the near side of the complex (at 130 pc) results in an error of 25\%. Thus, it is not sufficient to have an accurate mean distance for each region. Rather, it is highly desirable to have accurate distances to a substantial sample of individual objects within each region. Such detailed information makes it possible, in addition, to reconstruct the internal 3-dimensional (3D) structure of the clouds. 

Recently, \cite{Bouy_2015} used 
stars from the Hipparcos catalogue to determine the 3D distribution of the spatial density of OB stars within 500 pc from the Sun. They found no evidence for a ring-like structure and claimed that the Gould's Belt is the result of a 2D projection effect. They also propose that the apparent rotation and expansion of the belt is due to relative motions associated with galactic dynamics, but this needs to be investigated with accurate measurements of the dynamical state of the Belt. 

\subsection{VLBI distance determinations}

Understanding the processes of star formation requires accurate observational constraints. The observational signatures predicted by star formation models have to be compared to actual observations, but a direct comparison can only be performed when the stellar properties, such as source size, luminosity and mass, are well determined. Frequently the distances to star-forming regions are poorly constrained because they are obscured by molecular gas and dust. In such cases, inaccurate distances are often the main source of error on intrinsic parameter determinations. 

Numerous indirect methods can be used to estimate the distance to young stars \cite[e.g.,][]{de_Grijs_2011}, but they typically result in systematic uncertainties in excess of 20\%. Only trigonometric parallaxes can provide unbiased distance measurements, but they are notoriously challenging to obtain. For instance, the trigonometric parallax of a star at 200 pc is 5 milli-arcseconds (mas), so an astrometric accuracy of 50 micro-arcseconds ($\mu$as) on the parallax would be required to measure that distance to 1\% accuracy. This is more than one order of magnitude better than the astrometry delivered by the Hipparcos satellite \citep{Perryman_1997}. Indeed, Hipparcos did not significantly improve our knowledge of the distance to star-forming regions in the Gould's Belt \citep[e.g.,][]{Bertout_1999}. Also, the Hipparcos result on the distance to the Pleiades cluster, which is commonly used  for testing theoretical stellar models, disagrees with all distance determinations obtained through other methods \citep{Melis_2014,David_2016}.  The upcoming {\it Gaia} astrometric mission \citep{de_Bruijne_2012} will likely reach an accuracy of a few tens of $\mu$as, sufficient for percent accuracy determinations of distances in the Gould's Belt. However, 
since it operates at optical wavelengths, {\it Gaia} will be limited to stars that have low extinction.
This will be an issue in star-forming regions like Orion, Ophiuchus, or Serpens, where values of $A_{V}$ larger than 10 are common \citep{Ridge_2006,Cambresy_1999}.

For accurate astrometry, an alternative to optical-wavelength space missions is provided by Very Long Baseline Interferometry \citep[VLBI; e.g.,][]{Thompson_2007,Reid_2014}. VLBI observations at centimeter wavelengths typically reach an angular resolution of order 1 mas. When VLBI observations are phase-referenced to a bright nearby source, the angular offset between the target and the reference source can be measured to an accuracy of $\sim$ 20 to 300 $\mu$as, depending on the signal-to-noise ratio of the detection, the declination of the source, and the distance between the target and the reference source \citep{Pradel_2006}. The reference sources are usually distant quasars that are very nearly fixed on the celestial sphere. Thus, the measured offset between the reference source and the target can be transformed into accurate coordinates for the target. When several such observations collected over one year or more are combined, the parallax and proper motion of the target can be measured with high accuracy. Also, the astrometry quality of both VLBI and {\it Gaia} observations will be tested by considering objects that both instruments can detect.

Two technical points are worth mentioning here. The first is that a systematic error on the target coordinates will obviously occur if the reference quasar position is not well known. The positional errors of reference calibrators used in VLBI observations are typically between 0.5 and 10 mas, so this is the level of accuracy that can be expected on absolute coordinates derived from VLBI data. However, this additive error will equally affect all observations of a given target (as long as the same calibrator was used), and hence have no measurable effect on the parallax and proper motion measurements obtained from multi-epoch observations. The second, potentially more serious issue, is that, {because of  emerging jet components}, the photocenter of the quasars may shift with time
when accuracies of a few $\mu$as on positions and a few $\mu$as yr$^{-1}$ on proper motions are reached \citep[e.g.,][]{Reid_2004}. Because our typical positional errors are $100-300~\mu{\rm as}$, this problem will not be relevant for the data presented here, and can be mitigated by including several reference sources in the observations and monitoring their relative positions as a function of time \citep[e.g.,][]{Reid_2014}. 

VLBI astrometry can only be applied to a specific class of targets if they are detectable in VLBI observations \citep[e.g.,][]{Thompson_2007,Reid_2014}. 
This requires that the potential targets not only be radio sources, but also have an average brightness temperature in excess of $\sim$ 10$^6$  K within the synthesized beam (i.e.\ be non-thermal sources), 
as VLBI arrays do not have sufficient sensitivity to detect weaker emission.\footnote[1]{VLBI arrays are equivalent to telescopes thousands of kilometers in diameter in terms of angular resolution, but emphatically {\bf not} in terms of collecting area.}  
A summary of the mechanisms that produce non-thermal radio emission in YSOs is given in  Appendix \ref{sec:mechanisms}. VLBI observations of non-thermal continuum emission from young stars have been used to measure very accurate trigonometric parallaxes to individual YSOs and star-forming regions \citep{Loinard_2005,Loinard_2007,Loinard_2008,Torres_2007,Torres_2009,Torres_2012,Dzib_2010,Dzib_2011,Menten_2007,Dzib_2016}. These observations focused on YSOs that were previously known to be non-thermal radio emitters. Building upon these successes, we have initiated a large project (the {\bf GOuld's BELt dIstaNces Survey, hereafter GOBELINS}\footnote[2]{A french word that refers to the tapestries from the Gobelin factory in Paris, France.}) aimed at measuring the trigonometric parallax and proper motions of a large sample of magnetically active young stars in the Gould's Belt (specifically in Taurus, Ophiuchus, Orion, Perseus, and Serpens) using VLBI observations. 

\subsection{GOBELINS}\label{sec:gould}

GOBELINS was approved by the  Telescope Allocation Committee of the {\em National Radio Astronomy Observatory} in the Spring of 2010. It followed a two-stage strategy. During the first phase, large maps of each of the regions of interest were obtained, using conventional interferometry observations, with the Karl G.\ Jansky Very Large Array (VLA; we called this first phase of the project the {\bf Gould's Belt Very Large Array Survey}). These maps (published by \citealt{Dzib_2013,Dzib_2015,Kounkel_2014,Ortiz_15}, and \citealt{Pech_2016}) enabled us to identify radio-bright young stellar objects in each region and attempt a first separation between thermal and non-thermal sources. For instance, in Ophiuchus, \cite{Dzib_2013} identified 56 radio sources associated with young stellar objects and proposed that for $\gtrsim$ 50\% of them, the emission is of non-thermal origin. The second stage consists in multi-epoch VLBI observations of the selected targets with the Very Long Baseline Array \citep[VLBA;][]{Napier_1994}, to measure the astrometric elements (trigonometric parallax and proper motion) of each target. In this paper, we report on the first VLBI observations of the sources in the Ophiuchus region.

The results from GOBELINS  will be used first and foremost to pinpoint the location of the regions of star-formation within the Gould's Belt, as well as their internal three-dimensional structure. In addition, since the proper motion of each target will be measured simultaneously with its trigonometric parallax, the transverse component of the velocity vector will be obtained. In many cases, the radial velocity will be available from the literature or could be measured with dedicated optical or near-infrared (NIR) spectroscopy. Thus, GOBELINS will also provide the complete velocity vector for many targets. This will enable us to examine both the internal dynamics of each region and the large-scale relative motions of the different clouds in the Gould's Belt \citep[see][for a preliminary example]{Rivera_2015}. In particular, these measurements will help to characterize the overall dynamics of the Gould's Belt and will be relevant to the understanding of its very origin. 

GOBELINS will also provide radio images of a large sample of young stellar objects at milli-arcsecond resolution. This is unparalleled at any other wavelength, and will enable us to characterize the population of young, very tight, binary and multiple systems (see \citealt{Torres_2012} and \citealt{Dzib_2010} for examples of young multiple systems characterized by VLBI observations), as well as the magnetic structures around young stars (R.M.\ Torres et al., in preparation). Finally, these results will enable us to study the physical processes underlying the radio emission. For instance, the Gould's Belt Very Large Array Survey data \citep{Dzib_2013,Dzib_2015, Kounkel_2014,Ortiz_15, Pech_2016} have shown that the radio emission from YSO's is reasonably correlated with their X-ray luminosity, following the so-called G\"udel-Benz relation \citep{Guedel_1993,Benz_1994}. The VLBI observations will enable us to unambiguously separate the thermal and non-thermal components and re-examine this relation in more detail. It will also allow us to examine the prevalence of non-thermal radio emission in young stars as a function of their age and mass, providing clues on the magnetic evolution of young stellar objects.

\subsection{The Ophiuchus region}\label{sec:intro_oph}


\begin{figure*}[!ht]
\begin{center}
 \includegraphics[width=0.85\textwidth,angle=0]{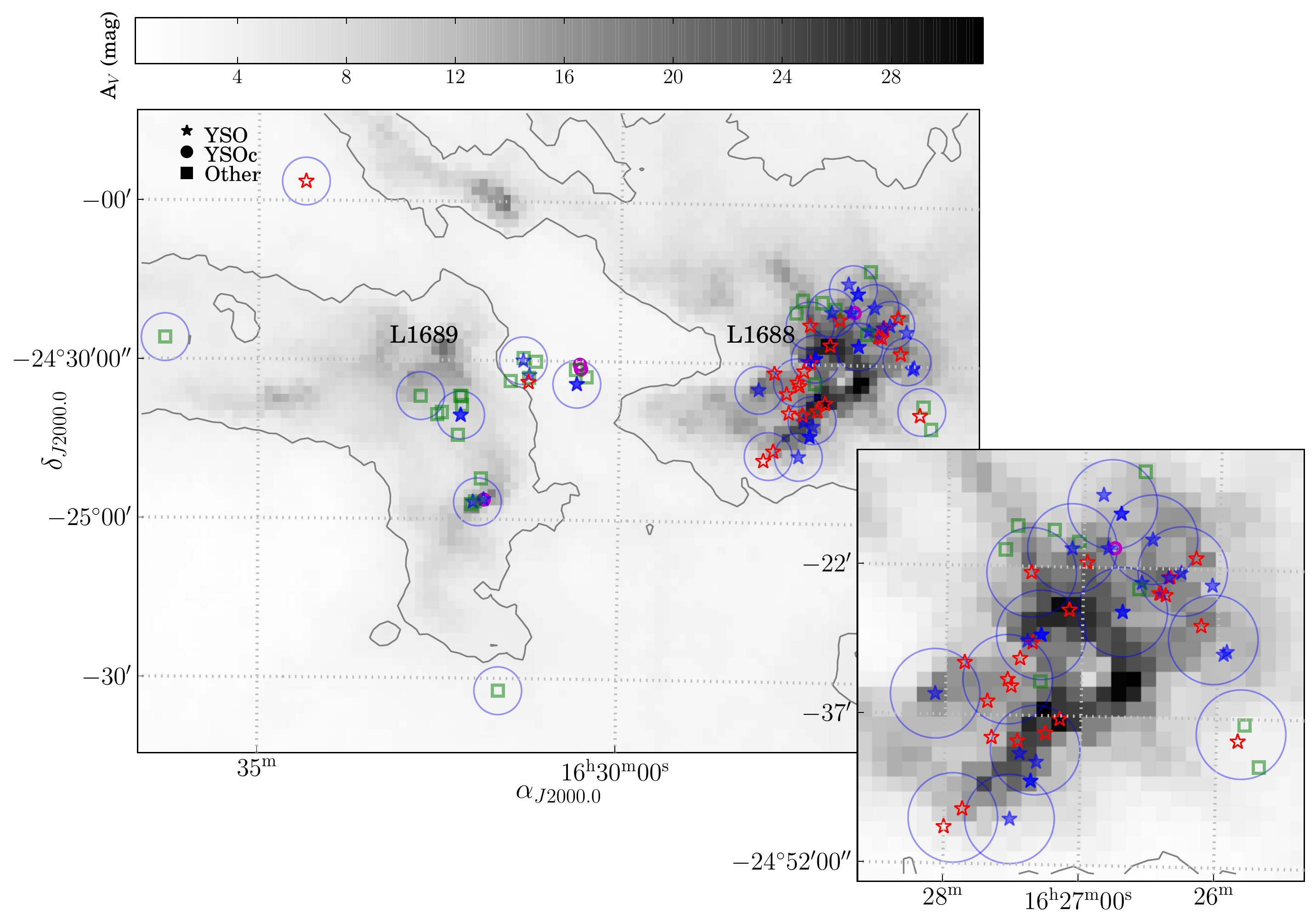}
\end{center}
 \caption{Spatial distribution of sources discussed in this work. Detected YSOs are shown as blue solid stars, YSO candidates as small magenta open circles, and other sources as green open squares. The YSOs not detected in our observations are shown as red open stars. The large blue circles indicate the position and size of representative
VLBA fields used to observe our targets. The grey scale represents the extinction map obtained as part of the COMPLETE project \citep{Ridge_2006}, based on 2MASS data \citep{Skrutskie_2006}.  The grey contour indicates an $A_{\rm V}$ of 4. The inset shows an enlargement of the Lynds 1688 area.}
\label{fig:mapOph}
\end{figure*}

As mentioned earlier, in the present paper we will focus on the GOBELINS observations of the Ophiuchus region.  Ophiuchus is one of the best-studied regions of star-formation (see \citealt{Wilking_2008} for a recent review). It consists of a centrally condensed core associated with the dark cloud Lynds 1688 (where $A_{V}$ = 50 to 100 magnitudes; \citealt{Wilking_2008}) and several filamentary clouds (collectively known as the ``streamers") extending toward the east (Lynds 1689 is a particularly prominent dark cloud associated with the eastern streamer) and the north-east (see Figure 1 in this paper and Figure 1 in \citealt{Dzib_2013}). 

The distance to Ophiuchus has been discussed in some detail by \cite{Wilking_2008}, \cite{Lombardi_2008}, \cite{Loinard_2008}, and \cite{Mamajek_2008}. The canonical value of 160 pc \citep{Bertiau_1958,Whittet_1974,Chini_1981} remained in use until very recently. Evidence for a somewhat shorter distance (120--145 pc) started to emerge from optical photometric and astrometric studies of the nearby Upper Scorpius subgroup \citep{de_Geus_1989,de_Zeeuw_1999}. The implications for Ophiuchus itself, however, were limited by the unclear relation between Upper Scorpius and Ophiuchus (see \citealt{Wilking_2008} for a discussion of this topic). More recently, \cite{Mamajek_2008} used the trigonometric parallaxes of the stars illuminating seven reflection nebulae within 5$^\circ$ of the Ophiuchus core to derive an estimate of 135 $\pm$ 8 pc. Both \cite{Knude_1998}, and \cite{Lombardi_2008} combined Hipparcos parallaxes and extinction measurements to conclude that Ophiuchus is at a distance of about 120 pc. \cite{Lombardi_2008}, in particular, report a mean distance of 120 $\pm$ 6 pc for the entire region, with some evidence that the streamers might be $\sim$ 10 pc closer than the core. This would be consistent with the distance of 96 $\pm$ 9 pc derived by \cite{Le_Bouquin_2014}; (see also \citealt{Schaefer_2008}) for the pre-main sequence binary Haro 1-14c, located in the north-eastern streamer.

It is important to note that none of the measurements mentioned so far involve direct trigonometric parallaxes to Ophiuchus cluster members. This is, of course, because the stars in that cluster are too deeply embedded to be detectable with Hipparcos or ground-based optical telescopes. Indeed, to date, there are only two published trigonometric parallaxes for Ophiuchus and both were obtained through VLBI observations. The first measurement was reported by \cite{Imai_2007} and targeted water masers associated with the Class~0 protostar IRAS~16293--2422, located in the northern part of Lynds 1689. They derive a distance of 178$^{+18}_{-34}$ pc, significantly larger than the 120--140 pc estimates that seem to emerge from the recent studies, described above, of the Ophiuchus core. It is not clear if this discrepancy stems from issues with one or more of the distance measurements, or if it is indicative that the eastern streamer is significantly more distant than the core. The second parallax measurement was reported by \cite{Loinard_2008}, and focused on two young stars (DoAr 21 and S1) located toward the Ophiuchus core. They obtain 120 $\pm$ 5 pc for the mean distance to these two stars, and adopt this value as the best estimate of the distance to the Ophiuchus core. 

In summary, there is a growing consensus that the Ophiuchus core is at 120--140 pc, but reducing the actual uncertainty on the distance has proven difficult. In addition, there are some conflicting results regarding the orientation of the streamers relative to the core. This unsatisfactory state of affairs largely results from the scarcity of direct parallax measurements to Ophiuchus members. 

\bigskip

In this paper we present new VLBA observations taken over a period of 4 years as  part of GOBELINS, and report on the detection of 26 young stellar systems in the Ophiuchus region (corresponding to 34 individual young stars, as some of the systems are multiple). The target sample and observing strategy are described in Section \ref{sec:obs}, the detections are described in Section \ref{sec:detections}, and the properties of the detected radio emission are analyzed in  Appendix  \ref{sec:results}. 
Section \ref{sec:astrometry} focuses on a subset of this sample and presents the astrometry of 16 stellar systems. We first analyze single objects in Section \ref{sec:singles}, and then stars in multiple systems in Section \ref{sec:multi} (other detected sources that are not known young stars are analyzed in Appendix \ref{sec:bs}). Finally, we provide a new improved distance to the core of Ophiuchus, and a description of the cloud depth in Section \ref{sec:distance2oph}.

\section{Observations, correlation and data reduction}\label{sec:obs}

The observations were obtained with the National Radio Astronomy Observatory's VLBA at $\nu=$ 5 and 8 GHz. We report on a total of 86 projects (code BL175), observed between 2012 March and 2016 April, and scheduled either dynamically or on a fixed date basis. Observations were usually obtained within 3 weeks of the equinoxes (March 21 and September 22) in each year; this corresponds to the maximum elongation of the parallax ellipse. The data were recorded in dual polarization mode with 256 MHz of bandwidth in each polarization, covered by 8 separate 32-MHz intermediate frequency (IF) channels. Projects observed during the first $\sim$1.5 years of our program were taken at 8 GHz (Table \ref{tab:observations}).  We switched to 5 GHz after the upgrade of the C-band receivers of the VLBA, which resulted in an increase of the bandwidth and sensitivity at that frequency.

A brief parenthesis about pointing positions and fields of view is in order here, as these concepts can be somewhat ambiguous for VLBI instruments. Observing with VLBI arrays involves two steps: (i) the actual observations when the antennas are all pointed toward a given direction ($\alpha_0$, $\delta_0$) and the data are recorded, and (ii) the correlation step (often carried out days or even weeks after the observations) when the data from the individual antennas are combined to form visibilities (see \citealt{Thompson_2007} for details). The field of view relevant for the observation step corresponds to the primary beam ($\Omega_{\text{PB}}$) of the individual telescopes. For the 25-meter dishes conforming the VLBA, the primary beam has a diameter of order 10$'$ and 6$'$, at 5 and 8 GHz, respectively. During correlation, however, the useful field of view is limited by coherence losses, due to beamwidth and time smearing, to a small {\em patch} typically only a few arcseconds in diameter. The center coordinates of a patch are specified during the correlation step, and can be chosen anywhere within the primary beam. In particular, they do not need to coincide with the position ($\alpha_0$, $\delta_0$) where the telescopes were pointing, as long as they are within $\Omega_{\text{PB}}$ of that position. By running multiple correlations on the same data, one can reconstruct an arbitrary number of patches, each at different locations within the primary beam. These different locations are usually called {\em phase centers}. The VLBA correlation is now performed by a DifX digital correlator \citep{Deller_2011}, that can simultaneously reconstruct multiple patches in a single pass through the data. A given VLBA observation is then defined by specifying (i) a pointing center ($\alpha_0$, $\delta_0$) where all antennas will point during the observations, and (ii) multiple phase centers at coordinates ($\alpha_{0,i}$, $\delta_{0,i}$) where correlations will be performed. In this mode, the correlator produces independent files containing the different phase centers. The first  file contains the first (primary) phase center listed for each pointing center. Often, but not always, the primary phase center in a given observation corresponds to the pointing center itself.

Accounting for the foregoing discussion of positions and fields of view, our observations were set up as follows. From the {\em Gould's Belt Very Large Array Survey} observations of Ophiuchus reported by \citet[see Section \ref{sec:gould}]{Dzib_2013},  a sample of YSOs with potentially non-thermal radio emission (our primary target list) was compiled. Here, we call YSOs  those sources that have been associated with young stars in infrared and X-ray surveys, and young stellar object candidates (YSOc)  those sources not classified as young stars by these surveys, but that show evidence of coronal magnetic activity in the radio (for instance flux variability).
 All of the YSOs in our sample have been accommodated in 44 different pointing positions of the VLBA (Table \ref{tab:observations}); representative 
fields are distributed across the region as shown in Figure \ref{fig:mapOph}. In some instances, a few primary targets could be observed simultaneously (as different phase centers) in the same observation. Within each of the 44 observed primary beams, we then included additional phase centers at the position of {\bf all} the sources reported by \cite{Dzib_2013} within the primary beam, independently of whether those sources were classified as YSOs, candidate YSOs, or extragalactic; and independently of whether the radio emission was anticipated to be thermal or non-thermal. In total, 118 sources toward the Ophiuchus region have been observed during our program, of which 50 are known YSOs.

The observations were organized in observing sessions, each with a different code, during which one or two pointing positions were observed (Table \ref{tab:observations}).  The observing sessions consisted of cycles alternating between the target(s) and the main phase calibrator J1627$-$2427: $\{$target --- J1627$-$2427$\}$ for single-target sessions, and $\{$target 1 --- J1627$-$2427 --- target 2 --- J1627$-$2427$\}$ for those sessions where two targets were observed simultaneously. The target to calibrator angular separations were in the range from 0.1 degrees for sources in Lynds 1688 to 1.2 degrees for targets in the streamers. The on-source time was $\sim$110 s for each target and  $\sim$50 s for the calibrator in every cycle. The total on-source time during each observing session was $\sim 1.6$ hours in projects that observed at 8 GHz, and $\sim$1 hour at 5 GHz. Scans on the secondary calibrators,  J1625$-$2417, J1625$-$2527, and J1633$-$2557, were also taken every $\sim$50 minutes during the observations. Unfortunately, one of the secondary calibrators, J1625$-$2417, was too weak to be detected in any of our observations at both 5 and 8 GHz. Finally, geodetic blocks were also included in each project, usually observed before and after the regular session.

The data reduction was done using AIPS \citep{Greisen_2003} and following standard procedures for phase referencing VLBA observations. Initial calibration was performed as follows. Scans having elevations below $10$ degrees were flagged.  The delays introduced by the ionospheric content were removed, and corrections to the Earth Orientation Parameters used by the correlator were then applied. Corrections for the rotation of the RCP and LCP feeds, as well as for voltage offsets in the samplers,  were also applied. Amplitude calibration was done with the $T_{\rm sys}$ method, using the provided gain curves and system temperatures to derive the System Equivalent Flux Density (SEFD) of each antenna. Instrumental single-band delays were then determined and removed using fringes detected on a single scan on the calibrator J1625$-$2527 or J1627$-$2427. Global fringe  fitting  was run on the  main phase calibrator in order to find residual phase rates. This was done in two steps. First we used the task FRING without giving a specific source model, applied the solutions derived, split and imaged the phase calibrator data.  Then we ran FRING again on the data set with all the calibration applied except global fringe fitting, and using as a source model the self-calibrated image of the phase calibrator. Finally, the phase calibrator was phase-referenced to itself, and the secondary calibrators, as well as the program sources, were phase-referenced to the phase calibrator. The rms errors in source positions achieved with this initial calibration were as good as 0.01-0.02 mas for the strongest sources (a few mJy in flux density), and of the order of 0.1-0.3 mas for sub-mJy sources. However, these errors misrepresent the true errors because they do not incorporate systematic errors, which are dominated by unmodelled tropospheric zenith delays,  ionospheric content delays, and atmospheric fluctuations above the VLBA antennas  \citep{Pradel_2006}.

Two calibration strategies can be adopted in order to deal with these systematic errors. One method consists in removing the tropospheric and clock errors using the all-sky calibrator blocks \citep{Reid_2004}. These blocks consisted of observations of many calibrators over a wide range of elevations taken with 512 MHz total bandwidth covered by 16 IFs. The multi-band delay, i.e., the phase slope with frequency, was derived for each scan and antenna, and used to model the clock and zenith-path delay errors using the AIPS task DELZN. The corrections were then exported and applied to the phase referencing data set before  global fringe-fitting. The second method uses the scans on the secondary calibrators to determine the phase gradient across the sky. The data of the secondary calibrators are split and self-calibrated after initial and DELZN corrections are applied. The position offsets of the secondary calibrators  from their respective phase centers are determined and removed, and residual phases are determined for all calibrators with the task CALIB. Finally, the AIPS task ATMCA is used to determine the phase gradients across the sky and then to correct the phase of all sources. We found that the corrections incorporated with DELZN decreased the rms error positions by a factor of up to $\sim 2$ when applied to sources at more than 1 degree from the main calibrator. On the other hand, the non-detection of the secondary calibrator J1625$-$2417 prevented us from applying the corrections from ATMCA in most projects. We attempted to derive these corrections using the only detected calibrators  J1627$-$2427 and J1625$-$2527, but this was limited to targets that are in line (within an angle of $45^{\rm o}$) with the two calibrators, and no significant improvement in the rms position error or image quality was achieved. Consequently, for the epochs taken during the Fall of 2015 and Spring of 2016, we replaced the secondary calibrator J1625$-$2417 with  J1633$-$2557. This calibrator is well detected at both 5 and 8 GHz, and enabled us to apply the ATMCA corrections in the most recent projects. After application of ATMCA, the rms error of the position decreased,  in some cases, to a quarter of its original value. 

For observations where several phase centers are observed within a given primary beam, the calibration strategy described above was applied to the primary phase center data. The other phase center data were calibrated by simply copying the final calibration (CL) tables, after appropriate editing with the AIPS task TABED to account for different source ID numbering.
 
Finally, we imaged the calibrated visibilities using a pixel size in the range of $50-100~\mu$as and pure natural weighting (ROBUST = $+$5 in AIPS). We constructed maps as large as $\sim 1.2''$ to search for our sources.  Typical angular resolutions are 4 mas  $\times$ 2 mas ($\sim 0.4$ AU at the distance of Ophiuchus) and 3 mas $\times$ 0.9 mas ($\sim 0.3$ AU) at 5 and 8 GHz, respectively. The best noise level was achieved in the images at 5 GHz, and was of order of $25~\mu\rm{Jy~ beam}^{-1}$. The fluxes of sources observed in data with multiple field centers were corrected for primary beam attenuation. In doing this, we assumed that the primary beam response of the VLBA 25 meter antennas is similar to that for the VLA 25 meter antennas. The new AIPS task CLVLB, which incorporates antenna beam parameters for the VLBA, could be used for this purpose but its performance is still under testing.

\section{VLBA detections}\label{sec:detections}

In Table \ref{tab:yso} we list the YSOs detected in the Ophiuchus region.
Columns (1)  and (2) give the VLA position of the sources, and their names, respectively. We report sources with flux densities above a 6$\sigma$ detection threshold, if they are detected in only one epoch. On the other hand, for sources with more than one detection, a threshold (in individual epochs) of $5\sigma$ was used. We give the minimum and maximum total flux densities measured at both frequencies in columns (3) to (6), but we note that some sources were not observed at 8 GHz. In epochs where sources were observed but not detected, we give an upper flux density limit of 3$\sigma$. Six objects are resolved into multiple components; for those, we report the flux densities for each component separately. Brightness temperature (see Appendix \ref{sec:results} for details) is given in column (7). 
The evolutionary status of the detected YSOs is indicated in column (8),
and the number of detections and observed epochs in column (9). Notice that the number of observations carried out toward each source differs considerably between sources. This is partly because our program was running in dynamic scheduling during the first 1.5 years, and partly because observations on all the 50 targeted YSOs were not fully completed in each equinox even during the fixed-date observations. 
The spatial distribution of our VLBA-detected sources is shown in Figure \ref{fig:mapOph}. The majority of the YSOs belong to the core, with only 5 YSOs distributed across the  eastern streamer cloud Lynds 1689. The other detected sources are more evenly distributed across the core and Lynds 1689.

\section{Astrometry}\label{sec:astrometry}

For all the objects detected with the VLBA, we have measured the source positions at each epoch by fitting two-dimensional Gaussians to the images, using the task JMFIT in AIPS.  The resulting values are listed in Table \ref{tab:positions}. Having identified single, double, and multiple sources in our images, we used different approaches for the determination of the source astrometric parameters. We will first describe the approach followed for sources that appear to be single stars, or sources that show evidence of multiplicity but for which we do not have enough data to perform a more complex analysis.

\subsection{Single stars}\label{sec:singles}

Source positions were modeled to derive the trigonometric parallax ($\varpi$), proper motion ($\mu_\alpha$, $\mu_\delta$), and mean position ($\alpha_0$, $\delta_0$). The fits were performed
by minimizing the  associated $\chi^2$ 
(e.g., \citealt{Loinard_2007}) 
and solving for the five astrometric elements simultaneously. 
For the errors in the positions at each epoch, we used the statistical errors provided by JMFIT, which roughly represent the expected theoretical precision of VLBA astrometry. However, systematic errors may significantly contribute to the astrometric accuracy \citep{Pradel_2006} and affect the derived astrometric parameters.

We estimate the systematic errors in two different ways. First, we use the empirical relation found by \cite{Pradel_2006}, according to which the VLBA astrometric accuracy scales linearly with the angular separation between the source and the phase calibrator. In the core, sources are separated from the phase calibrator, J1627$-$2427, by up to 0.4 degrees. Given this angular separation, and a typical declination of $\sim -25^{\rm o}$, the expected VLBA rms astrometric errors $\sqrt{(\Delta\alpha\cos\delta)^2 + (\Delta\delta)^2}$ are  $\sim 210~\mu$as. For this calculation, we have assumed that the rms errors for source coordinates, VLBA station coordinates, Earth Orientation parameters (EOPs), and wet tropospheric zenith path delays all contribute together (Tables 3 and 4 and Equation 2 in \citealt{Pradel_2006}).

The systematic errors were also estimated by quadratically adding an error to the statistical errors given by JMFIT until a reduced $\chi^2$  of 1 was achieved in the astrometric fits. These systematic errors are in general several times larger  than those predicted by the empirical relation. We note, however, that in their simulations, \cite{Pradel_2006} assumed a full track on the source, while in our observations the source is tracked during less than 4 hours, resulting in a poorer $(u-v)$ coverage. 
We used the latter approach (based on a measured reduced $\chi^2$) to deal with systematic errors. As stated above, these errors were added quadratically to the statistical errors of each individual epoch and used in the last iteration of the fits.

In the following sub-sections, we will comment separately on a few of the critical sources  describing the additional data that were taken from the VLBA archive, when available, and detailing the quality of the fits. In Table \ref{tab:parallaxes}, we provide the resulting astrometric parameters and distances for the complete sample. The corresponding measured source positions and best fits are shown in Figure \ref{fig:all}.

\subsubsection{DROXO 71}

The model that assumes a uniform and linear proper motion produces a good fit to the data. However, we discarded the last detection because it degrades the quality of the fit. We ignore the source of error that may be introduced in this particular epoch, but we found that the ``expected'' position from the fit to the first 7 epochs is on a sidelobe.  

\begin{figure*}[!bht]
 \hspace{2cm}
 {\includegraphics[width=0.27\textwidth,angle=0]{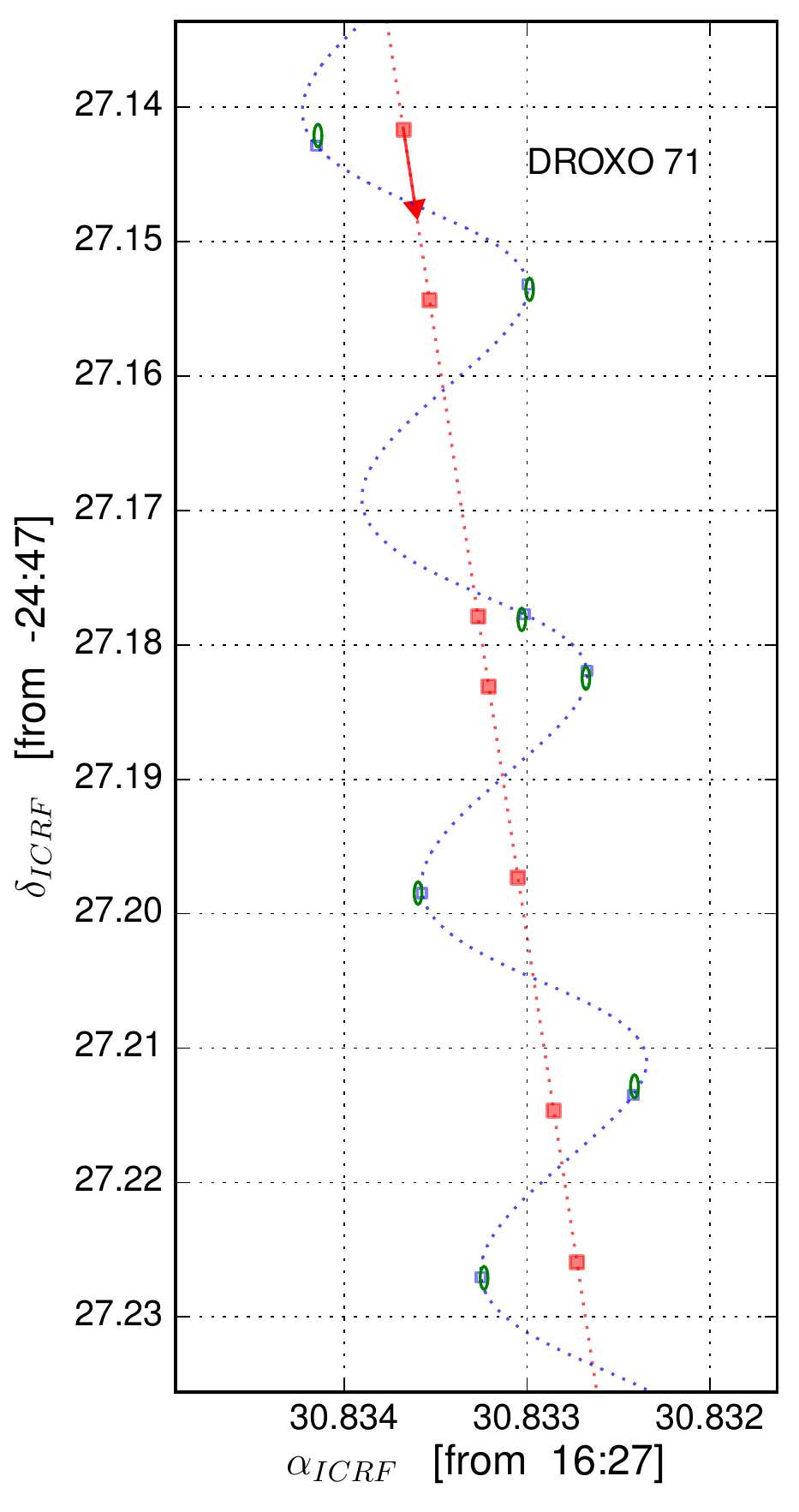}}
 {\includegraphics[width=0.48\textwidth,angle=0]{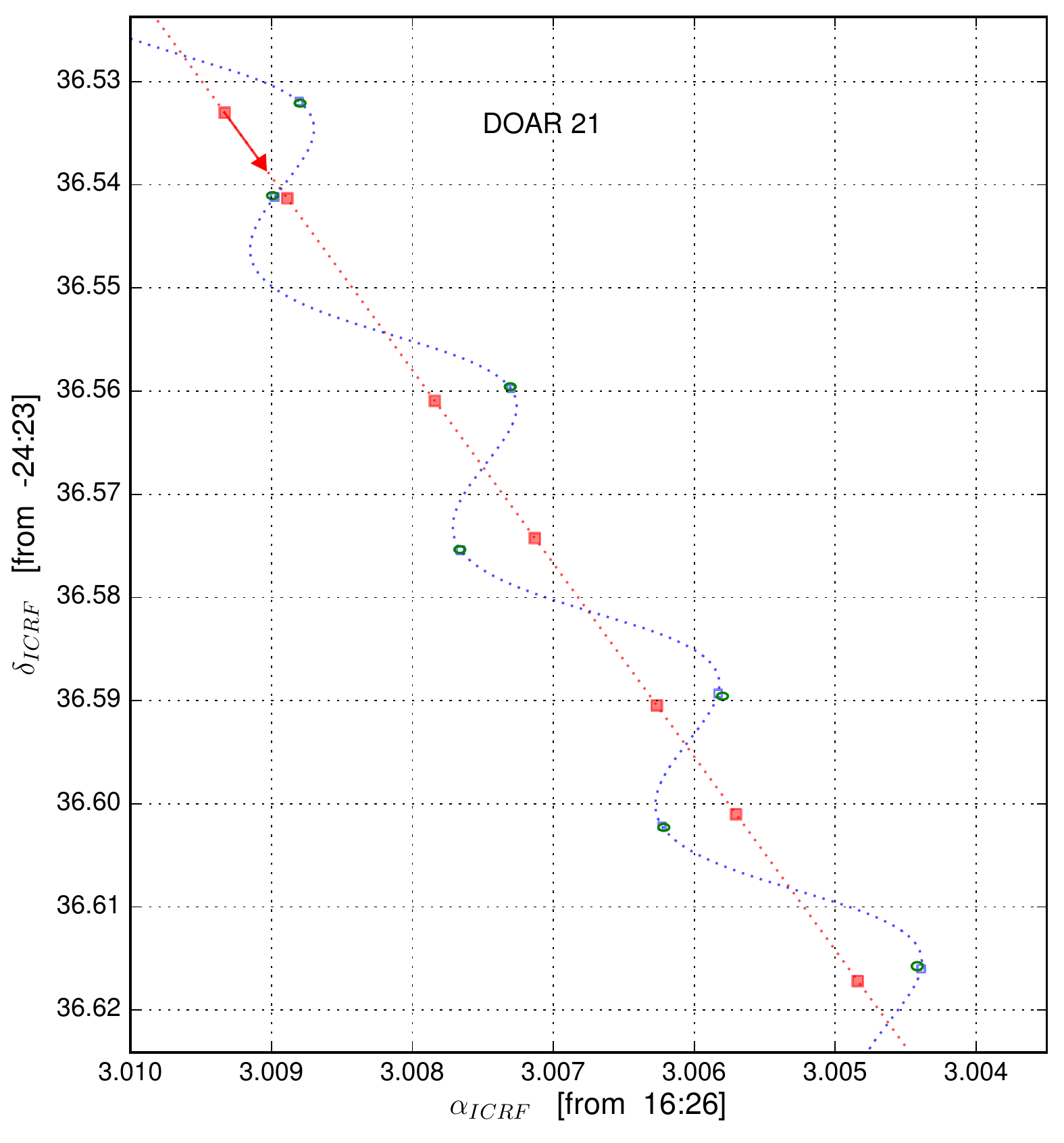}} \\
 
 \hspace{1.0cm}
 {\includegraphics[width=0.26\textwidth,angle=0]{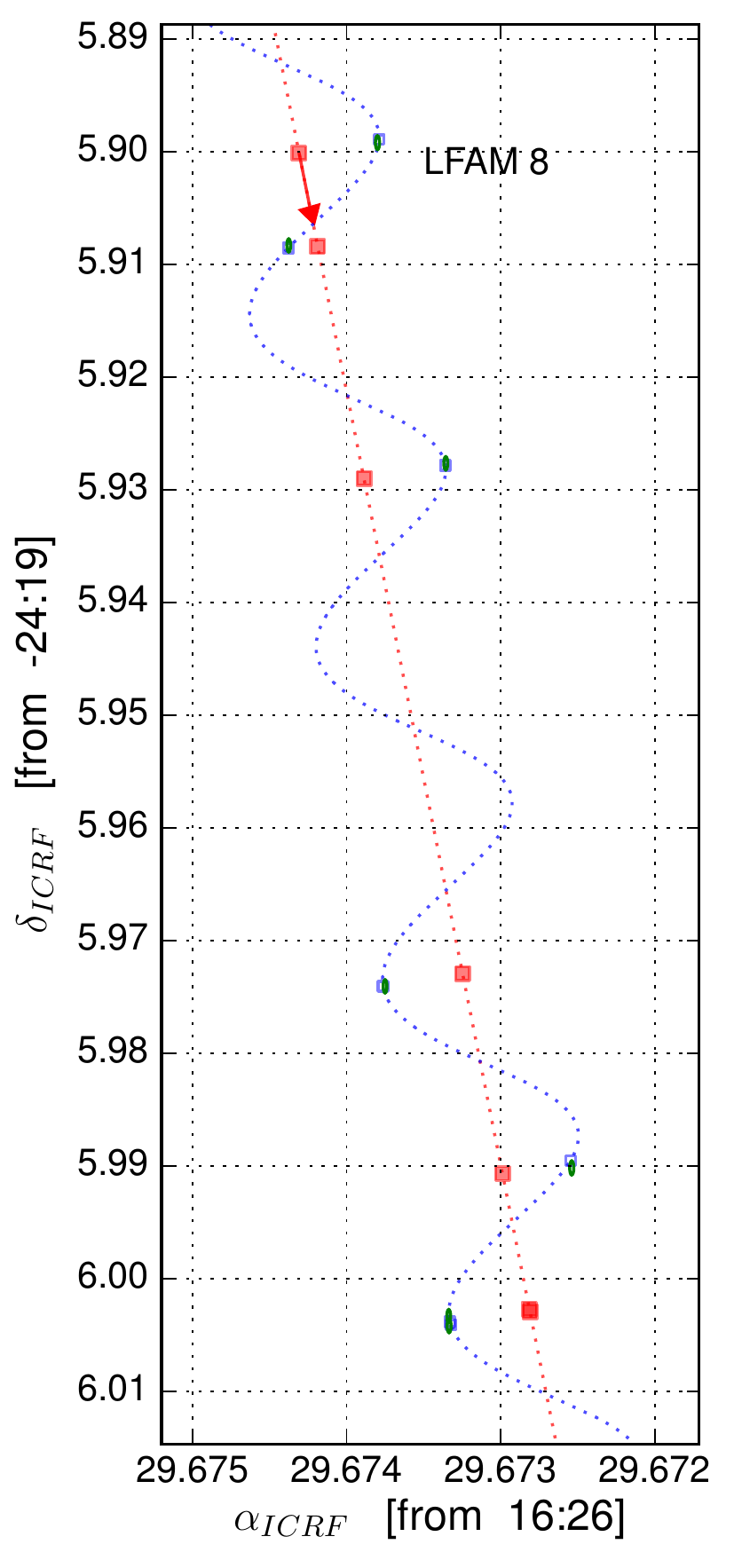}}
  {\includegraphics[width=0.307\textwidth,angle=0]{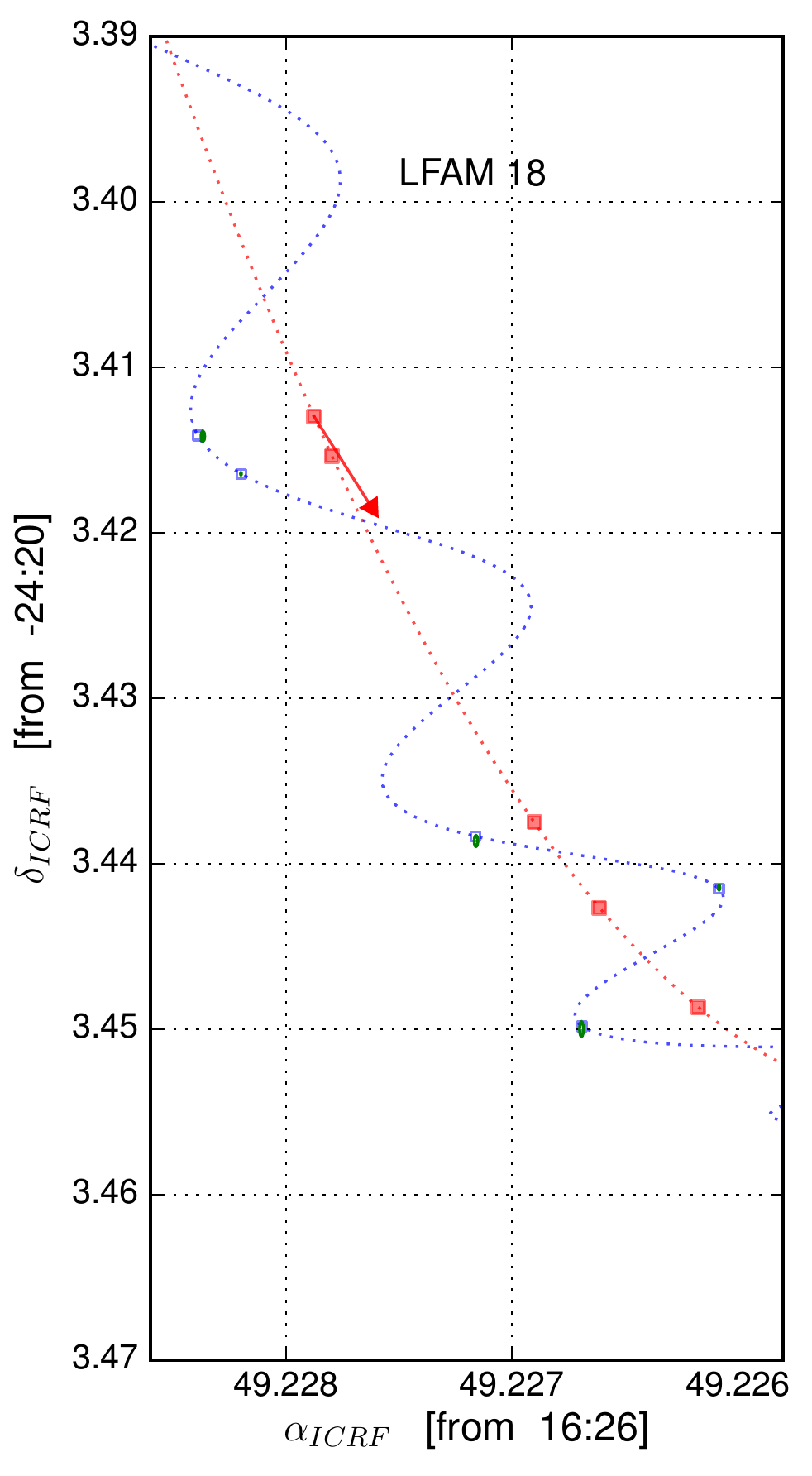}} 
   {\includegraphics[width=0.294\textwidth,angle=0]{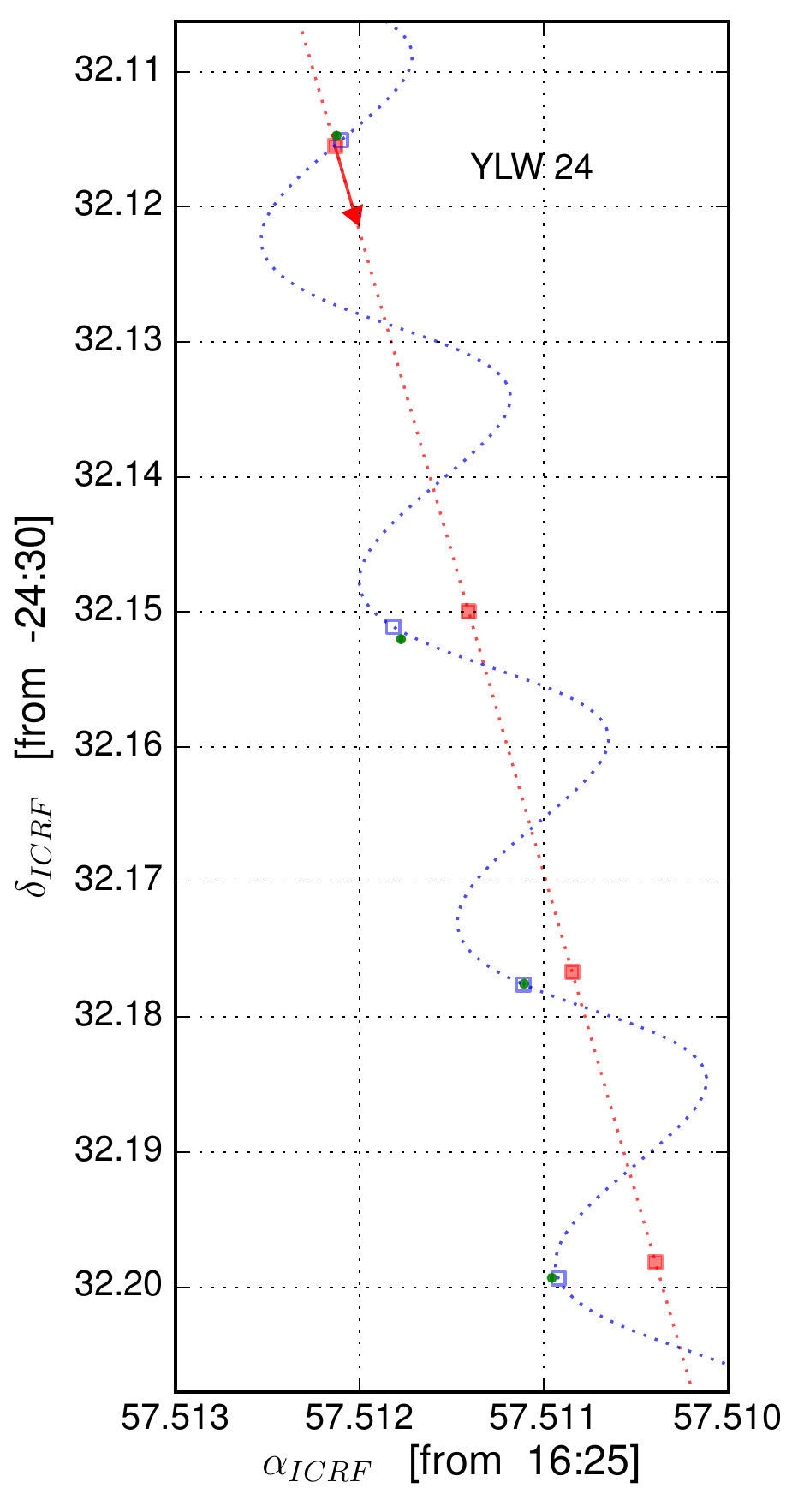}} 
\caption{Observed positions and best fit for single sources. Measured positions are shown as green ellipses, the size of which represents the magnitude of the errors. The expected positions from the fit are shown as blue open squares. The blue dotted line is the full model, and the red line is the model with the parallax signature removed. The red squares indicate the position of the source expected from the model without parallax. 
When the systematic errors on source positions can be estimated from the fits, these are included in the error bars shown in the plots. The arrow shows the direction of position change with time.
}
\label{fig:all}
\end{figure*}

\setcounter{figure}{0}
\renewcommand{\thefigure}{2}
\begin{figure*}[!htb]
\hspace{1cm}
  {\includegraphics[width=0.275\textwidth,angle=0]{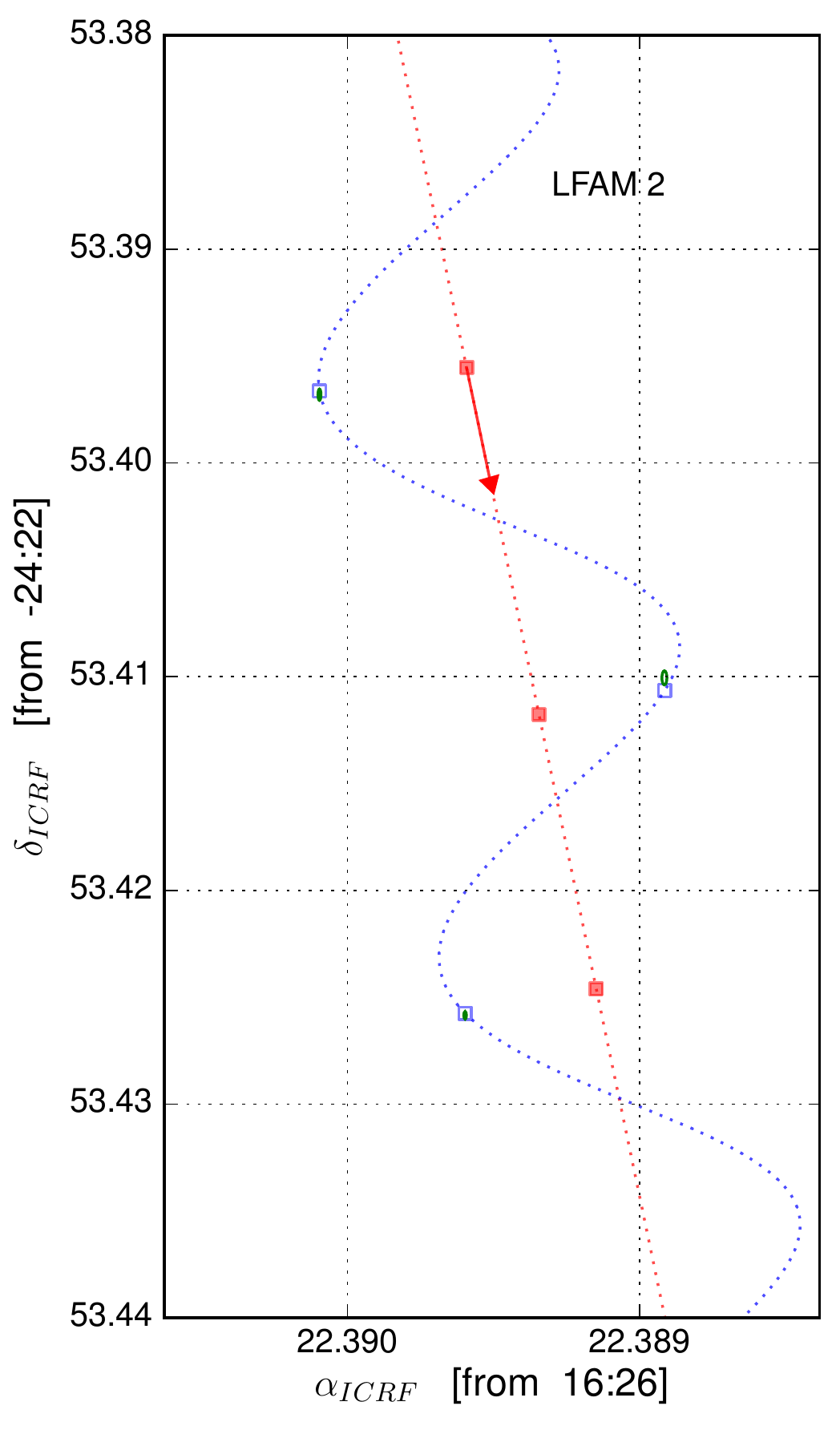}} 
 {\includegraphics[width=0.32\textwidth,angle=0]{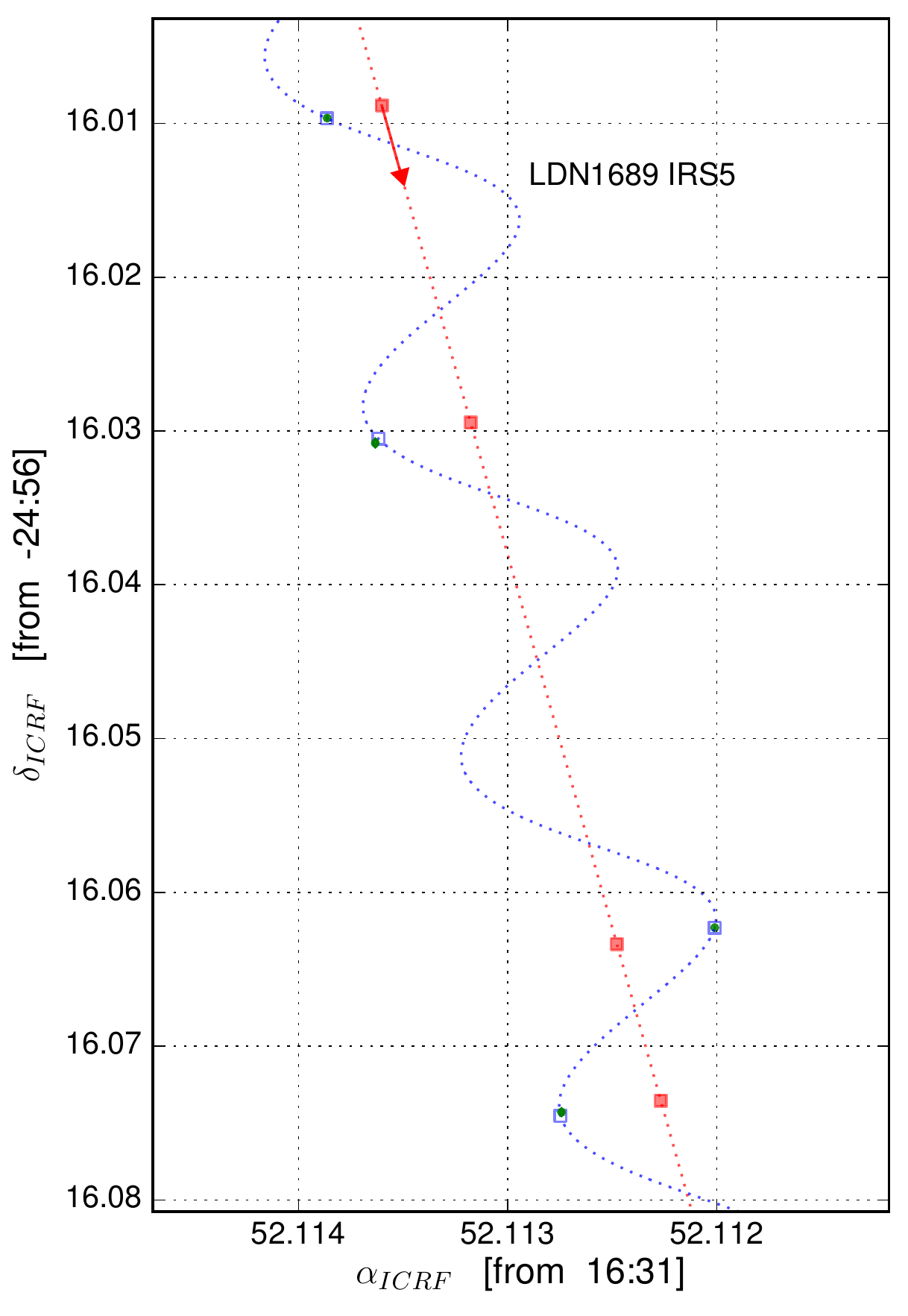}}
 {\includegraphics[width=0.285\textwidth,angle=0]{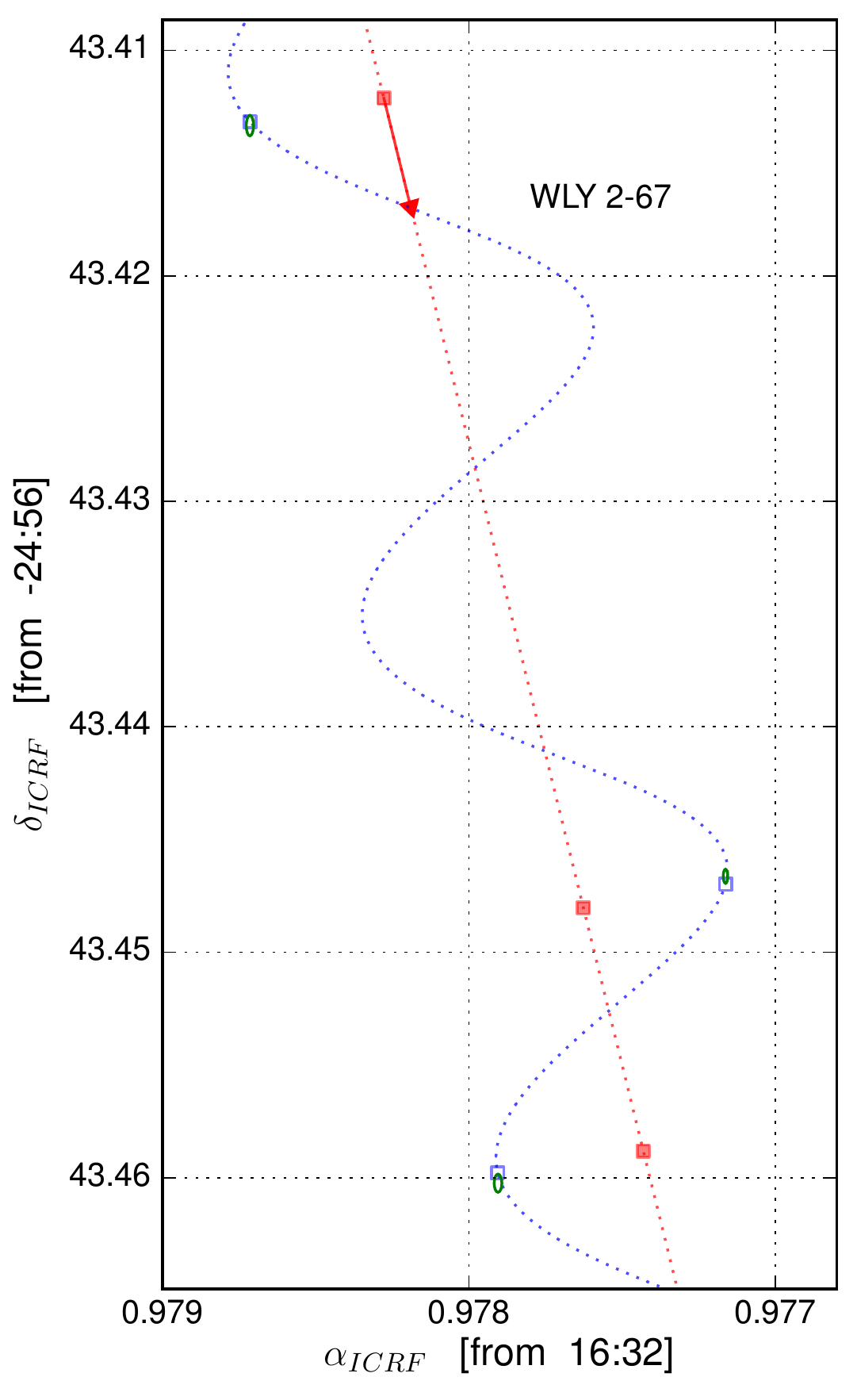}}
\caption{ \it Continued. }
\label{fig:all2}
\end{figure*}

\setcounter{figure}{2}
\renewcommand{\thefigure}{\arabic{figure}}

\subsubsection{DoAr 21}\label{sec:doar21}
DoAr 21 has been observed and detected at 7 epochs.  Additionally to these observations, the source was observed with the VLBA prior to 2012 in projects BL128 (9 epochs from 2005 September to 2006 December) and BT093 (8 epochs from 2007 July to 2007 September). Seven epochs from project BL128 were analyzed and published in \cite{Loinard_2008}.  Because the observing and calibration strategy adopted for GOBELINS is not exactly the same as in \cite{Loinard_2008}, we perform a consistency check as follows. We downloaded, calibrated and revised these prior observations, by applying the same calibration procedures as to the data from our own project. We note, however, that a large fraction of the observations taken between 2006 December and 2007 September experienced very poor weather conditions (Table \ref{tab:doar21}). Such conditions affected the image quality and the rms error position was relatively high, in comparison with that measured in observations collected under optimal weather. In the present analysis, we do not include the epochs when data were highly affected, and only 5 additional epochs to the seven reported by \cite{Loinard_2008} were included. Hence, a total of 12 epochs from 2005 September to 2007 September are considered in our analysis. 

Unlike the GOBELINS observations, projects BL128 and BT093 did not include geodetic-like scans, and therefore no DELZN correction was applied to those projects. On the other hand, the secondary calibrators J1633$-$2557 and J1634$-$2058 were detected and used only in projects  BT093CD and BT093CE, for which an improvement in the image quality and the rms error position was found when using the multi-calibrator strategy. 

Projects BL128 and BT093 observed J1625$-$2527 (located 1 degree south of DoAr 21) as the main phase calibrator, while GOBELINS observed J1627$-$2427, which is only 0.2 degrees away from the science target. Before combining data from the three projects, it was necessary to correct the target source positions, so that all positions were measured relative to the main calibrator J1627$-$2427. The mean position of  J1625$-$2527 measured in observations from 2005 to 2007 is $\alpha_{\rm J2000.0}$= 16$^{\rm h}$25$^{\rm m}$46.891640$^{\rm s}$; $\delta_{\rm J2000.0}$=$-$25$^{\rm o}$27$'$38.32684$''$.  In the images corresponding to GOBELINS, we found that J1625$-$2527 shows a position offset relative to the phase center, as a result of a separation between the source and the 
main phase calibrator of 1.07 degrees. The mean position of J1625$-$2527 relative to J1627$-$2427 in those data is $\alpha_{\rm J2000.0}$= 16$^{\rm h}$25$^{\rm m}$46.891617$^{\rm s}$; $\delta_{\rm J2000.0}$=$-$25$^{\rm o}$27$'$38.32808$''$. Thus, an offset $\Delta \alpha = -2.3\times10^ {-5}$ s and $\Delta \delta = -0.0012''$ had to be applied to the positions of DoAr 21 measured from projects BL128 and BT093.

We fit the data from these 12 epochs from BL128 and BT093, 
and obtain  a distance $d = 123.4^{+16.3}_{-12.9}$ pc, fully consistent with the results reported by \cite{Loinard_2008} and consistent within 1 sigma with the new determination based solely on GOBELINS data. It is noteworthy, however, that the errors reported by \cite{Loinard_2008} on the parallax obtained from the data corresponding to BL128 alone are significantly better than those we obtained when combining BL128 and BT093. We argue that 
\cite{Loinard_2008} underestimated their systematic errors, which
resulted in artificially small quoted errors.

When the values derived from GOBELINS data and those obtained from the older BL128+BT093 data are weighted-averaged, the resulting astrometric elements are nearly identical to those derived from GOBELINS data alone. This is expected, of course, as the accuracy of these more recent observations greatly surpass those of BL128 or BT093. In the rest of the paper, we therefore will use the results based solely on GOBELINS.


\subsubsection{LFAM 18}

LFAM 18 has been detected in 5 of 9 observations. The model assuming a linear and uniform proper motion produces a poor fit to the data. We then consider a model with an accelerated and uniform proper motion, and found that it produces a better fit to the data. Indeed, \cite{Cheetham_2015} state that the source has evidence of multiplicity, and this may explain why the detected source follows an accelerated rather than a linear motion. The second component of the system is, however, not detected in our VLBA observations. We report the astrometric parameters from the latter fit, and caution that the errors may be underestimated. This is because the method we use to estimate the systematic errors that is described above can only be applied when more than 5 detections are available. 

\bigskip
\bigskip

\subsubsection{YLW 24, LFAM 2, LDN 1689 IRS5 and WLY 2-67}

These four sources 
have been detected just in 3 to 5 epochs and the errors in astrometric parameters may be underestimated. Their corresponding fits give parallaxes that are consistent with the rest of the sources (see also Section \ref{sec:distance2oph}). For these 4 sources, we assume a uniform and linear proper motion, since they do not show any evidence of multiplicity.

\subsection{Multiple stars}\label{sec:multi}

Our VLBA observations (combined with past astrometric observations) have detected a total of 8 multiple systems (cf.\ Table \ref{tab:yso}). In 6 of them, the individual components  have been detected at sufficient epochs such that we can model their orbital motion, in addition to parallax and proper motion.  Two different fits were performed as follows. In the first  ``Full model'', we use all available absolute VLBA positions of individual components (including data from epochs where a single component is detected), together with relative positions, to solve for the orbital elements, center of mass at first  epoch of the GOBELINS observations where the primary is detected ($\alpha_{\rm{CM},0}$, $\delta_{\rm{CM},0}$), parallax ($\varpi$), and  proper motion ($\mu_\alpha$, $\mu_\delta$) of the system. The orbital free parameters in this model are period ($P$), time of periastron passage ($T$), eccentricity ($e$), angle of line of nodes ($\Omega$), inclination ($i$), angle from node to periastron ($\omega$), semimajor axis of primary ($a_1$), and mass ratio $m_2/m_1$.
Combining the mass ratio with Kepler's third law (that contains the sum of masses), and because we know the distance to the system from the parallax solution, the masses of each component are also inferred. The orbital solutions are derived by minimizing  $\chi^2$, which is computed for a grid of initial guesses of model parameters.  
The errors in the parameters were calculated by taking the average, weighted by $\chi^2$, of the output uncertainties over values across the entire grid.

\begin{figure*}[!t]
\begin{center}
 \includegraphics[width=0.34\textwidth,angle=0]{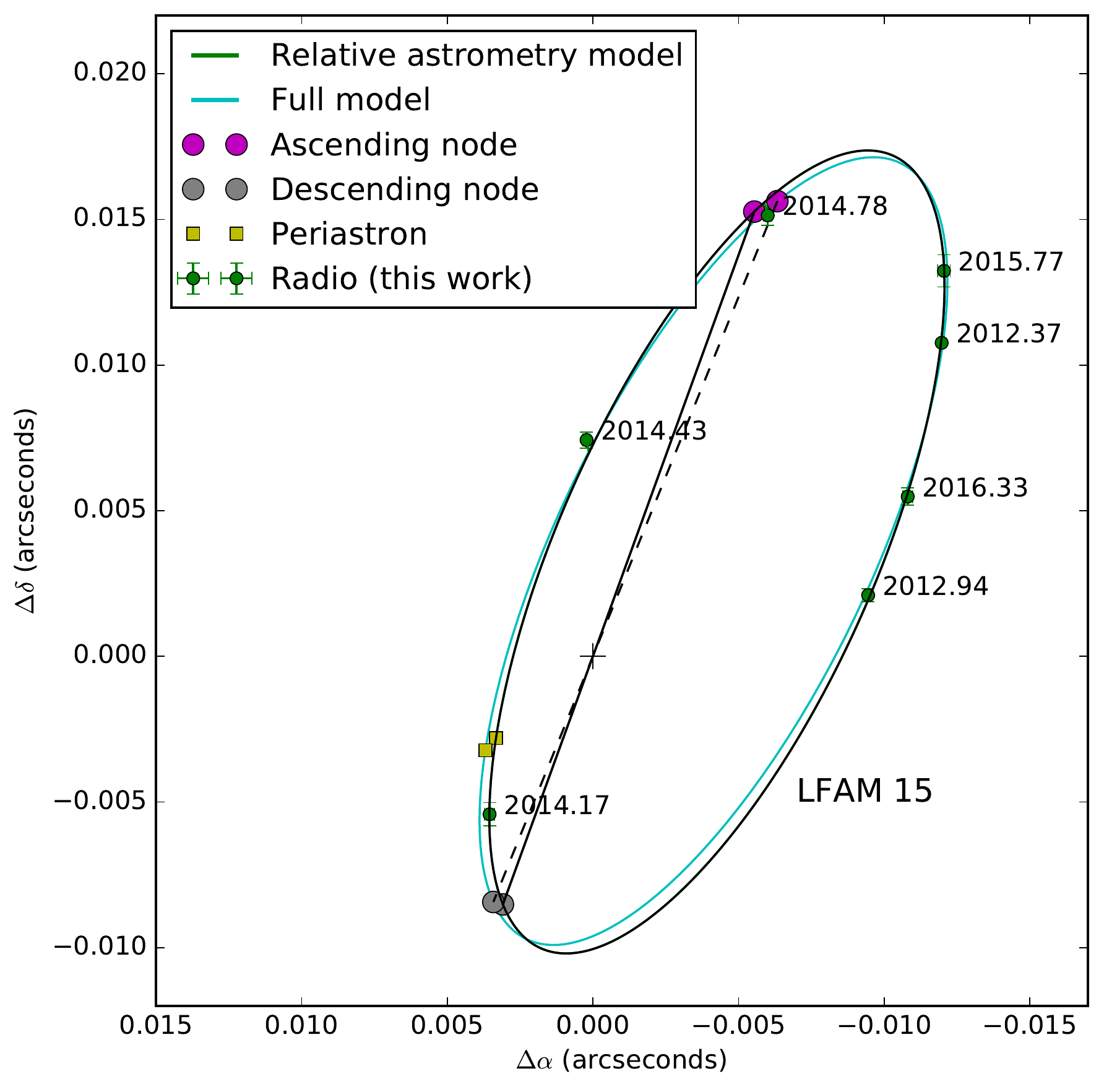} 
\includegraphics[width=0.4\textwidth,angle=0]{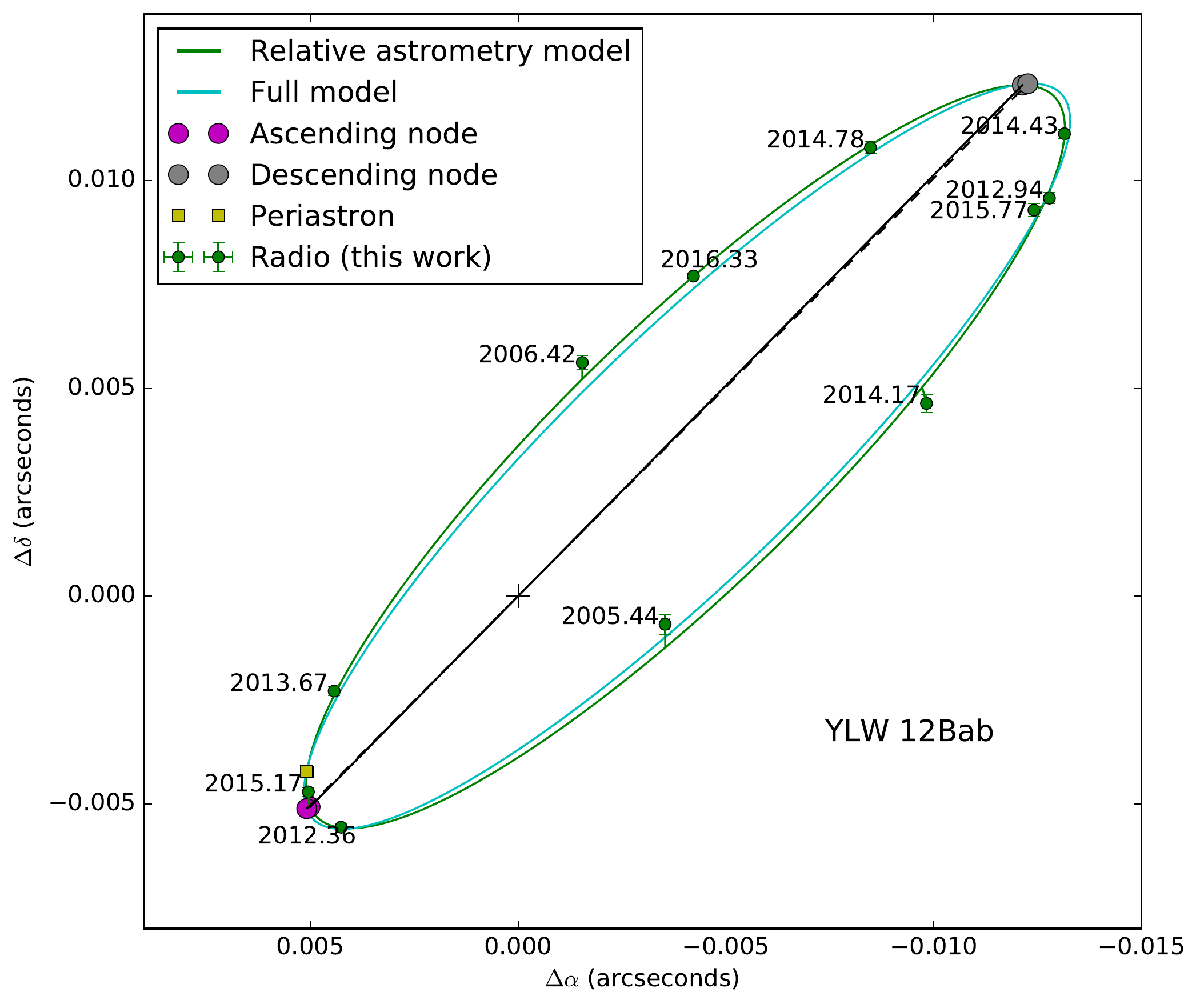} \\
 \includegraphics[width=0.35\textwidth,angle=0]{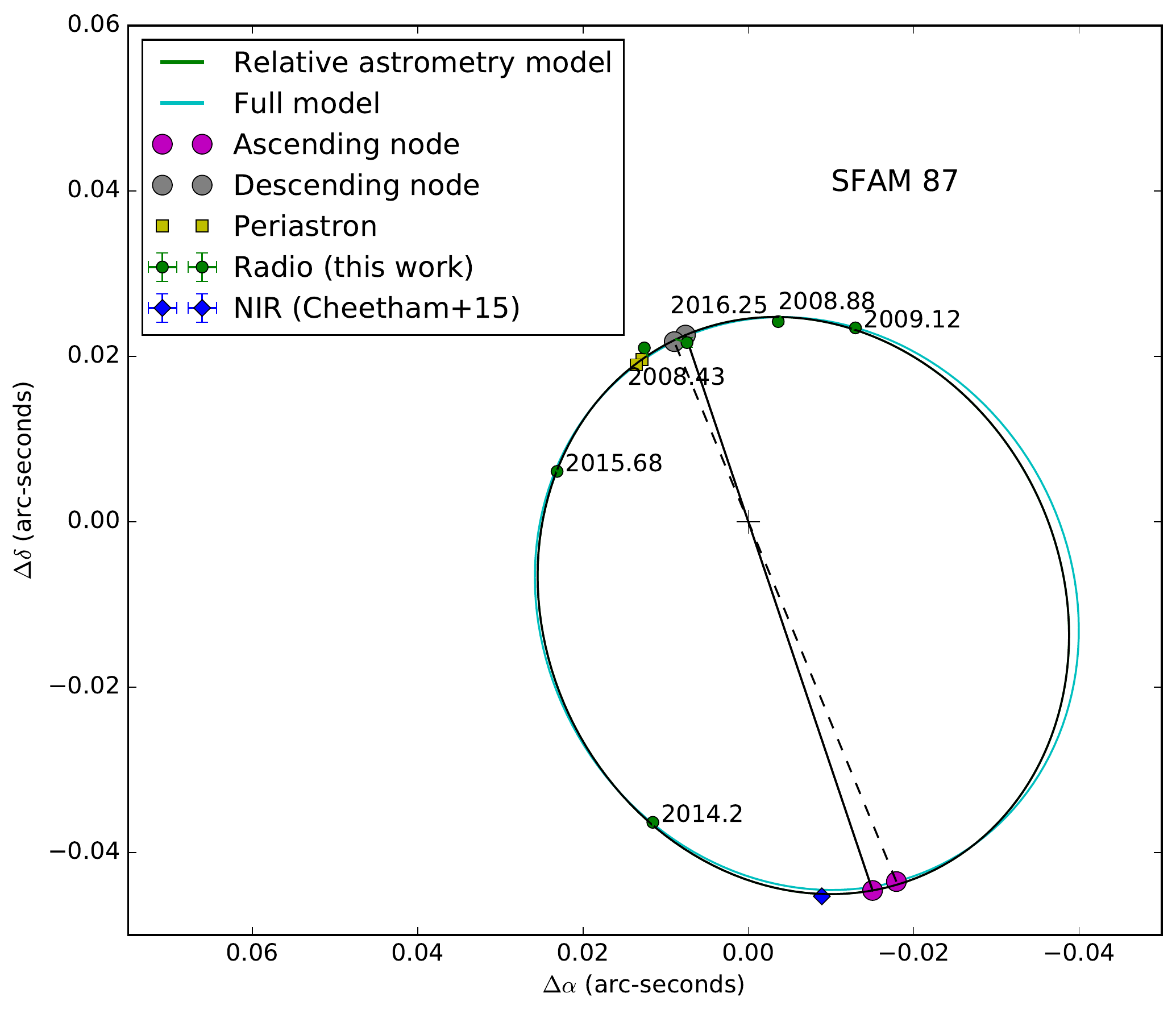} 
 \includegraphics[width=0.51\textwidth,angle=0]{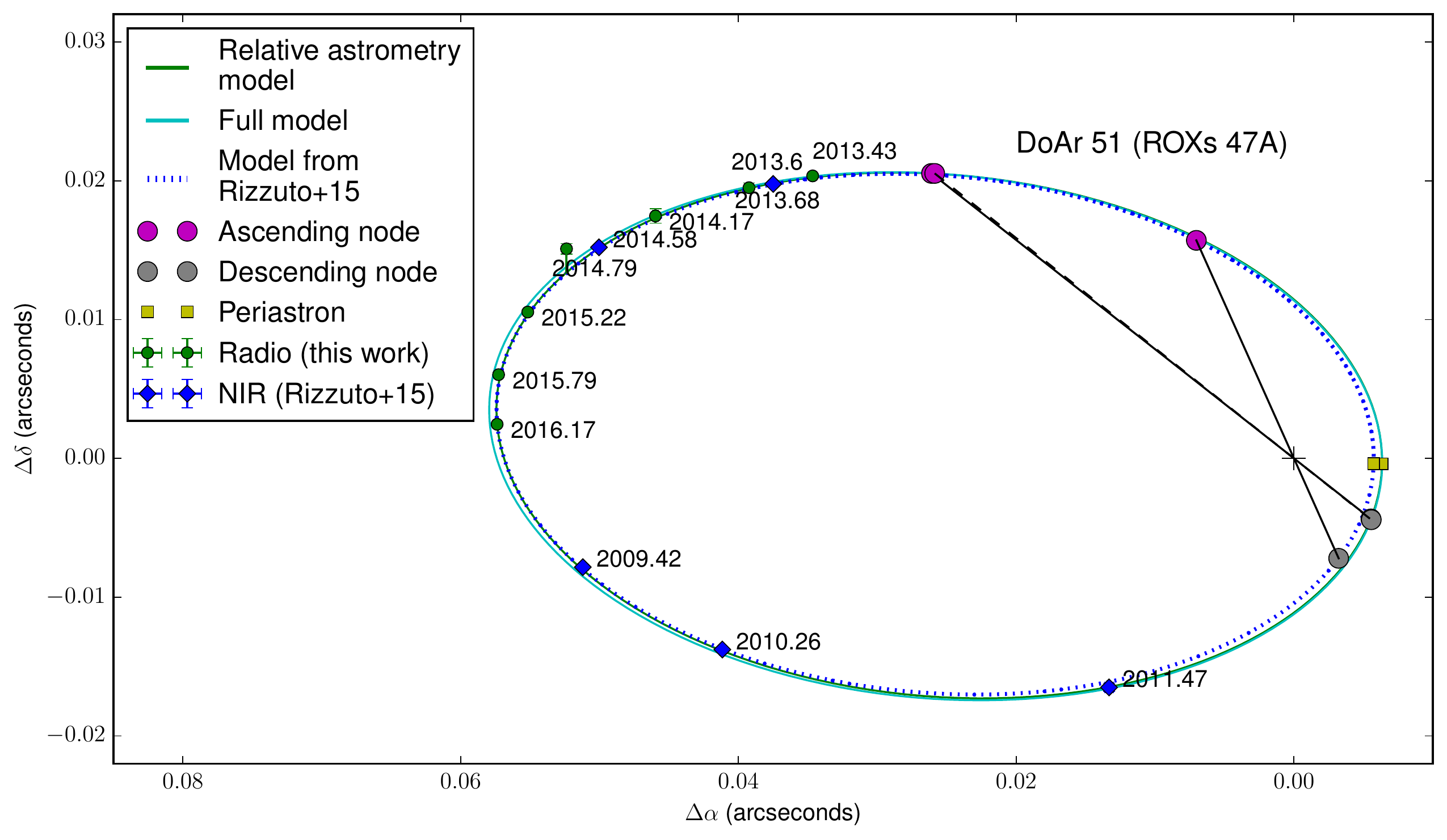}  \\
  \includegraphics[width=0.4\textwidth,angle=0]{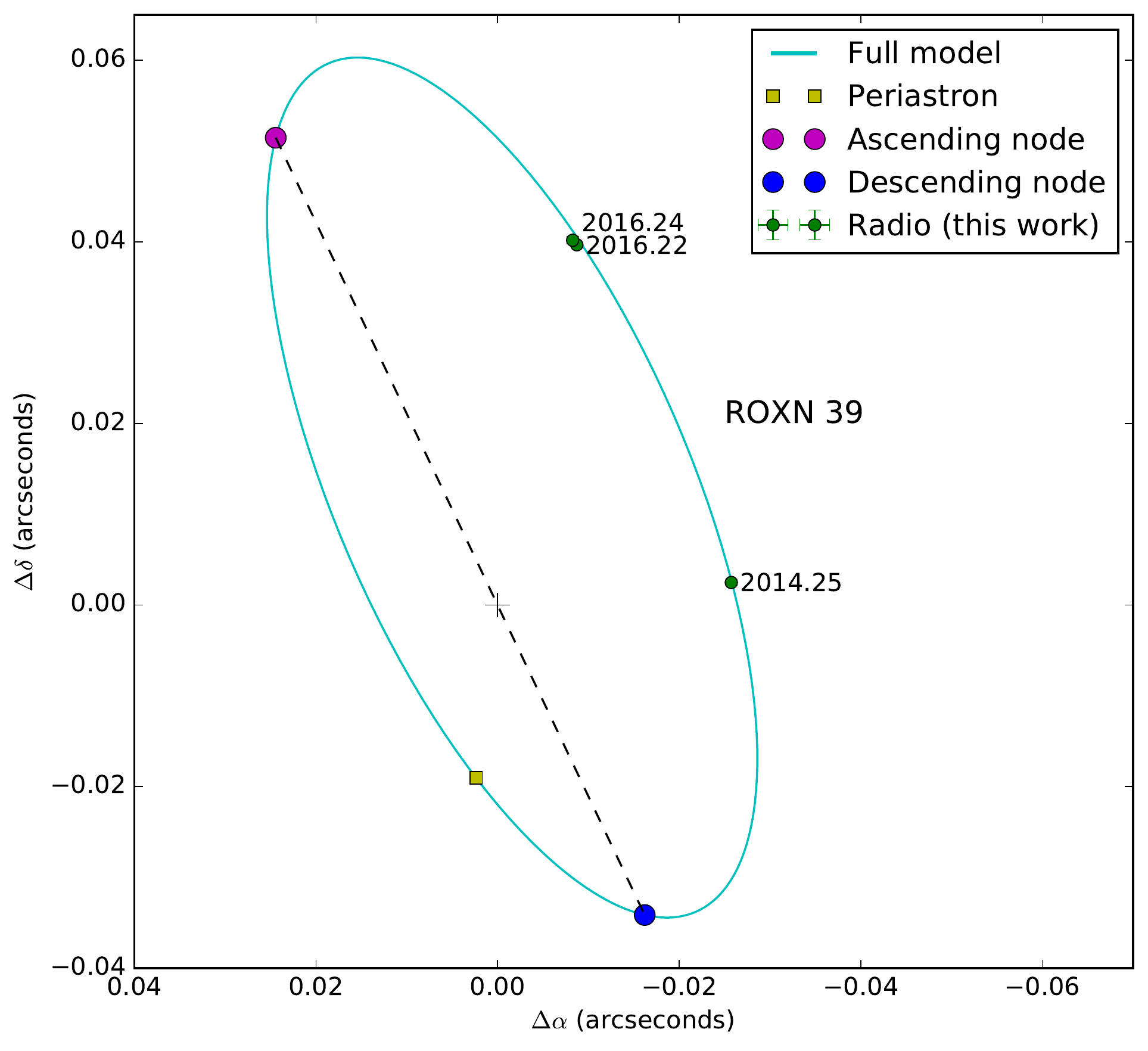}
\caption{Relative positions of the components of the binary systems. The green points mark the detections with the VLBA, while blue points indicate the detections in the NIR by \cite{Cheetham_2015} and  \cite{Rizzuto_2016}. Green and cyan solid lines correspond to the ``Relative astrometry" and ``Full'' model orbital fits, respectively. The blue dotted line in the fourth panel is the fit by \cite{Rizzuto_2016} to NIR only data. The black  solid and dashed lines trace the line of nodes of the  ``Relative astrometry" and ``Full'' model, respectively. The cross marks the position of   the primary source. }
\label{fig:orbits}
\end{center}
\end{figure*}

From the VLBA  images,  we compute, for each system, the angular separation and position angle of the secondary relative to the primary star. In addition,  we have compiled from the literature separations measured with near-infrared (NIR) observations.  These data are shown in Figure \ref{fig:orbits} for each source separately. In the second ``Relative astrometry model''  we only use the separation and position angle of secondary relative to primary, measured at epochs when both components are detected simultaneously. We use the orbital elements determined from the  ``Full model'' as initial parameters for the \cite{Gudehus_2001} code, the {\it Binary Star Combined Solution Package,\footnote[3]{This package is available at http://www.astro.gsu.edu/{$\sim$}gudehus/binary.html.}} to solve for  $P$, $T$, $e$, $\Omega$, $i$, $\omega$, and $a$. The total mass of the system,  $M_T$, is then obtained from Kepler's law, but we are not able to determine individual masses in this case. For this fit, the uncertainties in the orbital elements are computed from the scatter on model parameters. 
 
The results from both fits are given in Table \ref{tab:orbits}, and the resulting orbits are shown in Figure \ref{fig:orbits}. We find that both fits are consistent with each other within the errors. 
We will comment on each system separately in the following paragraphs.

\subsubsection{LFAM 15}

\begin{figure*}[!tb]
\begin{center}
 \includegraphics[width=0.8\textwidth,angle=0]{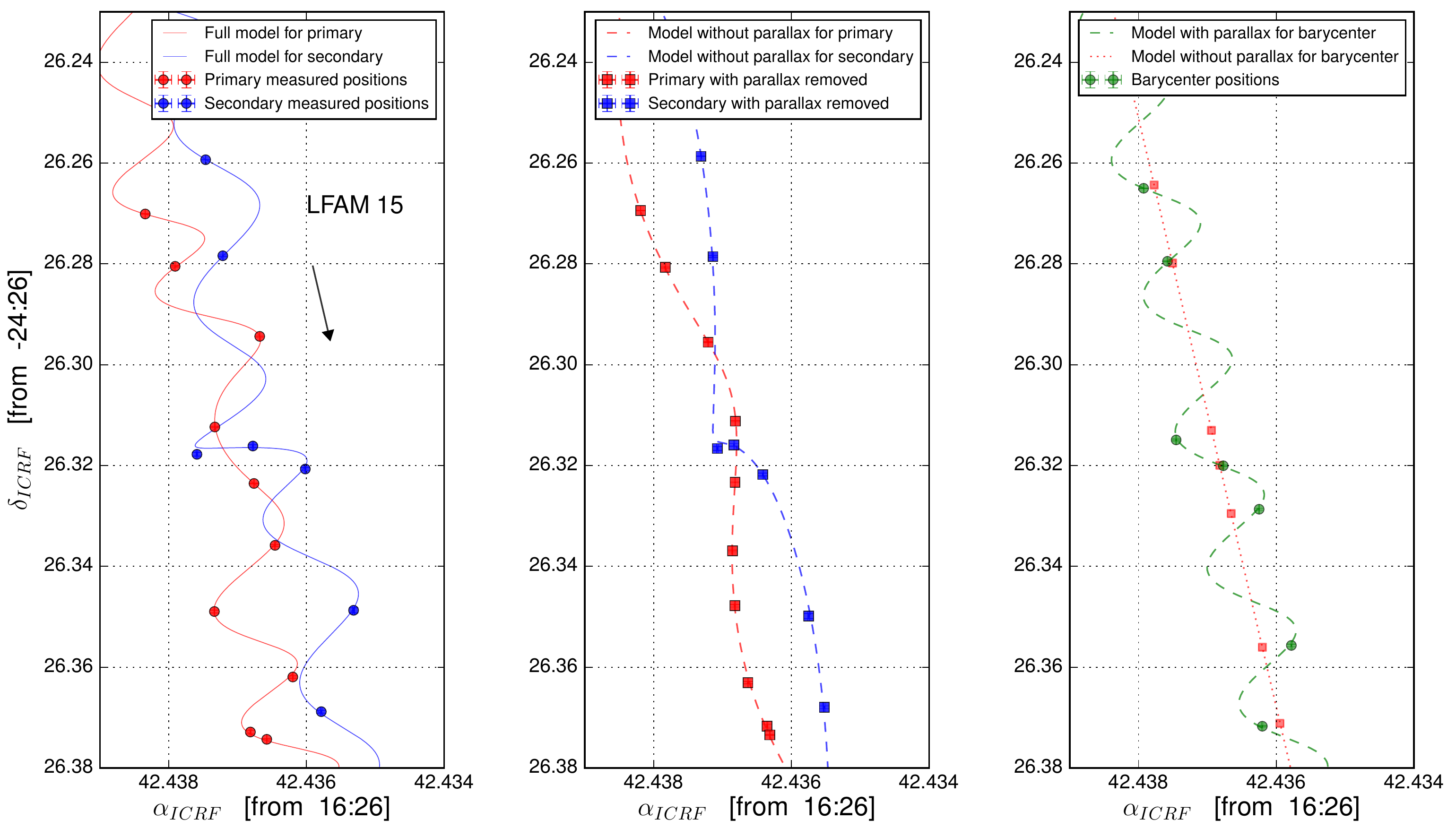}
\caption{Observed positions and best fit for LFAM 15.  {\it Left:} Measured positions of each component are shown as red and blue circles. The solid lines show the fit corresponding to the ``Full model'' described in the text. {\it Middle:} The squares mark the measured positions with the parallax signature removed, while the dashed lines are the fits from the ``Full model'',  also without parallax. {\it Right:} Green dots mark the position of the center of mass derived using the solutions from the orbital model for the mass ratio.  The green dashed line is the model for the motion of the center of mass of the system, while the red line is this same model with the parallax signature removed.  The red squares indicate the position of the center of mass expected from the model without parallax. The arrow shows the direction of position change with time. Positional errors, as delivered by JMFIT, are smaller than the size of the symbols. } 
\label{fig:lfam15}
\end{center}
\end{figure*}

The source LFAM 15 has been found to be double in 7 out of our 10 observed epochs. LFAM 15 was also observed at 4 epochs as part of project BL128  (Table \ref{tab:doar21}), between 2005 June and 2006 March. We have calibrated these additional data and detected the source in three epochs, albeit always as a single component. In Figure \ref{fig:lfam15}, we show the measured positions for each component and the resulting  best ``Full model'' fit that, as mentioned before, consists of the sum  of orbital motion around the center of mass,  proper motion, and parallax of the system.

\subsubsection{YLW 12B}

\begin{figure*}[!t]
\begin{center}
{\includegraphics[width=0.85\textwidth,angle=0]{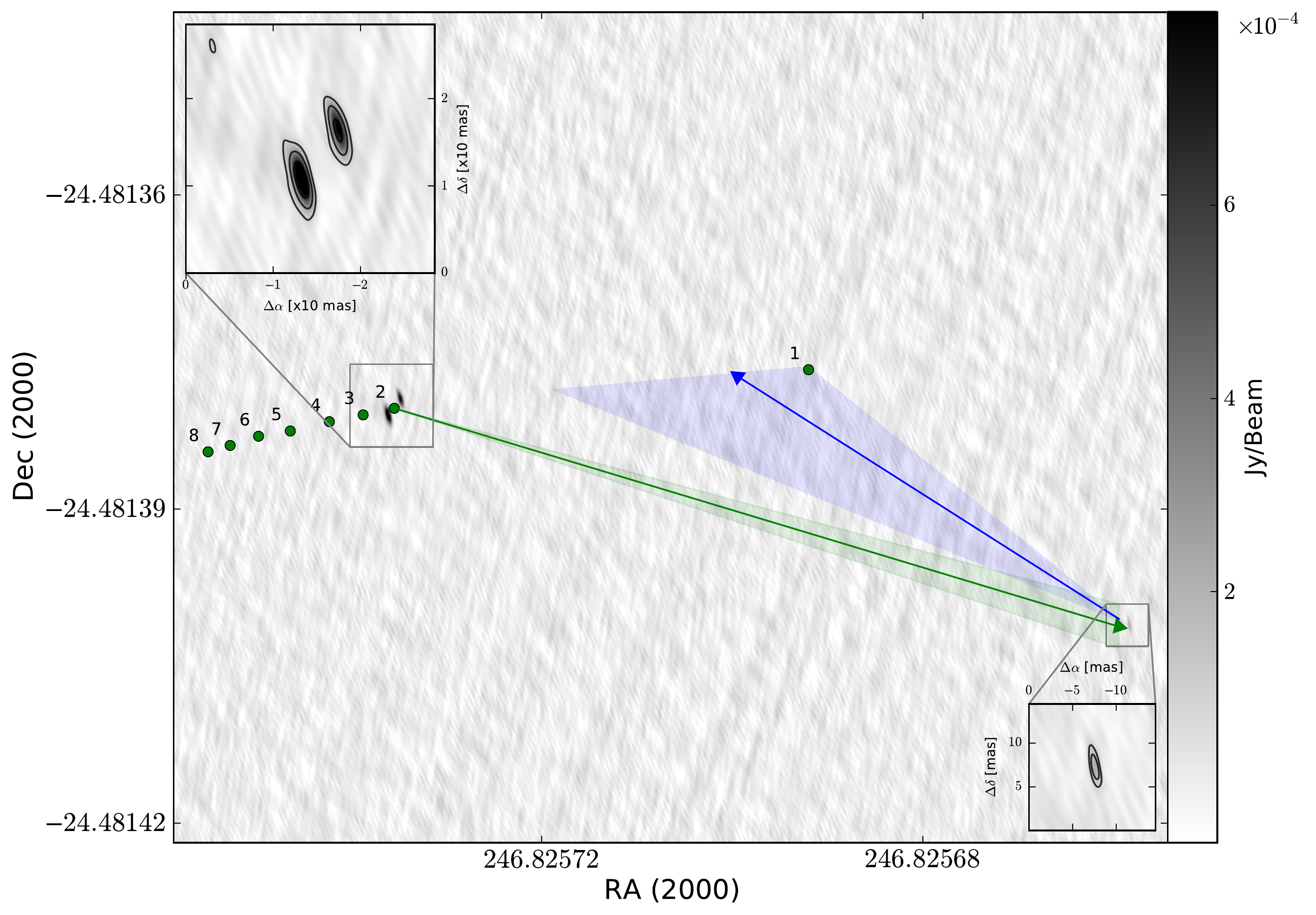}}
\caption{VLBA image of the multiple system YLW 12B, from data obtained on 2012, May 11 (first epoch observed by GOBELINS). Insets show zooms on the tight binary YLW~12Bab, top left, and the southwestern companion YLW~12Bc, bottom right. Contour levels are 3, 9 and 21 (in top left inset), and  3 and 5 (in bottom right inset) times 32$\mu\rm{Jy~beam}^{-1}$, the rms noise in the image. The green dots mark the position of YLW~12Bab relative to YLW~12Bc at the epochs when the three sources in the system are detected. These are 8 epochs, which correspond to the Julian dates listed in Table \ref{tab:positions} for the third component in the system, as follows: 1= 2453529.77467, 2=2456058.83738, 3=2456269.26236, etc. Notice that, at epoch with JD = 2453818.98287, the third component was detected but  the compact binary was not.
The arrows with their error cones show the acceleration vectors at the first epoch observed by GOBELINS. }
\label{fig:ylw12b_system}
\end{center}
\end{figure*}

YLW~12B is found to be a hierarchical triple source formed by a close binary (separation of 4--17 milli-arcseconds, equivalent to 0.6--2.4 AU), detected in 9 epochs, and a third component, located a few hundred milli-arcseconds to the southwest of the binary and detected in 7 epochs (Figure \ref{fig:ylw12b_system}). Hereafter, we call the close binary  YLW~12Bab and the third component, YLW~12Bc. Six archival VLBA observations, obtained between 2005 and 2006 as part of project BL128, have also been found for this source; they have been calibrated and analyzed (Table \ref{tab:doar21}). One of these older epochs was discarded due to poor weather conditions during the observations. In these archival observations, both components of YLW~12Bab  were detected on 2005 June 8 and 2006 June 1, while YLW~12Bc was detected on 2005 June 8 and 2006 March 24.

\begin{figure*}[!tb]
\begin{center}
 \includegraphics[width=0.9\textwidth,angle=0]{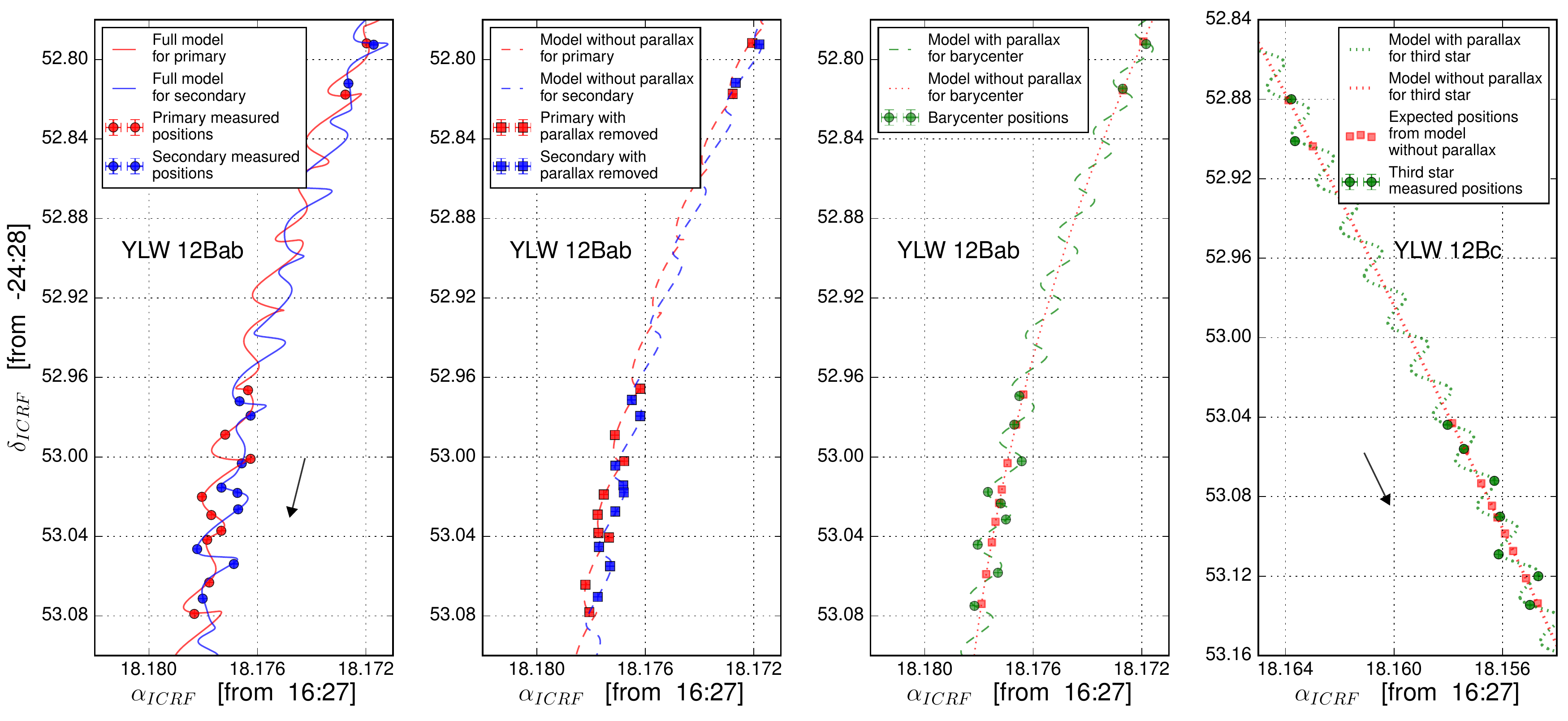}
\caption{First to third panels are like Figure \ref{fig:lfam15}, but for YLW 12Bab. Fourth panel shows the observed positions and best independent fit for YLW 12Bc, with parallax fixed to the value of YLW 12Bab.}
\label{fig:ylw12b}
\end{center}
\end{figure*}

Given the distance to the source, the angular separation between YLW~12Bab and YLW~12Bc of $\sim 140$ and $\sim 320$ mas in 2005 and 2016, respectively, corresponds to 20 to 45 AU. This suggests that the sources  form a bound multiple system, and later in this section we provide a stronger evidence that supports this conclusion.
To fit the positions  of the YLW~12Bab components, and to take the effect of the third companion into account, we add  two more free parameters to the ``Full model''. These parameters are the acceleration of the center of mass of YLW~12Bab in each direction, a$_\alpha$ and a$_\delta$, which we consider to be uniform. As in the case of LFAM 15, three plots were constructed to visualize the best fit solution. In the first panel of Figure \ref{fig:ylw12b}, we show the observed positions of the compact binary and the best fit, while in the second panel, we show this fit and source positions with the effect of parallax removed.   Using the solution for the mass ratio from the ``Full model'',  we now compute the positions of the center of mass  of YLW~12Bab, and plot them along with the parallax plus proper motion model in the third panel of Figure \ref{fig:ylw12b}.  It is clear that the compact binary follows a curved motion as a result of the gravitational force exerted by the third companion. Indeed, we find that the acceleration is statistically different from zero at $> 8\sigma$ in both directions.

Let us now discuss the third star of the system. The positions of YLW~12Bab relative to YLW~12Bc, as well as the acceleration vector of YLW 12Bab, are shown in Figure \ref{fig:ylw12b_system}. We see that the acceleration vector of YLW~12Bab points toward YLW~12Bc, 
as would be expected of a gravitationally bound system. This plot also shows that our assumption of a uniform accelerated motion is a reasonable approximation, because our observations cover only a small fraction of the orbit expected for the wider system.   We attempted to fit the orbit of this source around the center of mass of the whole system with a simultaneous distance and proper motion fit for the three stars. However, we were not able to constrain most of the orbital parameters because the VLBA detections of the third companion are still insufficient. We only find  solutions for the following parameters:  $\Omega\sim 85^{\rm o}$, $\varpi = 7.190 \pm 0.088$~mas, $\mu_{{\rm CM}, \alpha} \cos\delta = -4.55\pm 0.03$~mas~yr$^{-1}$, $\mu_{{\rm CM}, \delta} = -24.27\pm0.06$~mas~yr$^{-1}$; limits on the mass, $M_3 >3~{\rm M}_\odot$, period, $P\sim300-400$~yr, and that inclination is consistent with the compact binary. Even though we do not have enough data for modeling the orbit of YLW~12Bc, we can still constrain its proper motion and acceleration using its absolute positions measured with the VLBA. In order to do so, we fit separately the third star using the astrometric code for single sources to solve solely for proper motion and acceleration terms, while fixing the parallax at the value derived for YLW~12Bab. We show this last fit in the fourth panel of Figure \ref{fig:ylw12b}, and  give the solution of the astrometric parameters in Table \ref{tab:parallaxes}.

As we mentioned above, the trajectory of YLW~12Bab is somewhat curved. That of YLW~12Bc is, on the other hand, more linear. This results in a smaller measured acceleration for YLW~12Bc ($\sim$ 0.2 mas yr$^{-2}$) than for YLW~12Bab (0.64 mas yr$^{-2}$), and suggests that YLW~12Bc is the most massive member of the system. Indeed, we find that $M_3\!\!>\!\!3~{\rm M}_\odot$, while $M_1 = 1.26~{\rm M}_\odot$, and $M_2 =1.40~{\rm M}_\odot$. This is the reason why we plot the position of YLW~12Bab relative to YLW~12Bc in Figure \ref{fig:ylw12b_system}, rather than the converse.  We also see that, as expected, the acceleration of YLW 12Bc points, within the errors, toward YLW 12Bab (see Figure \ref{fig:ylw12b_system}). 

\subsubsection{SFAM 87}

\begin{figure*}[!tb]
\begin{center}
 \includegraphics[width=0.8\textwidth,angle=0]{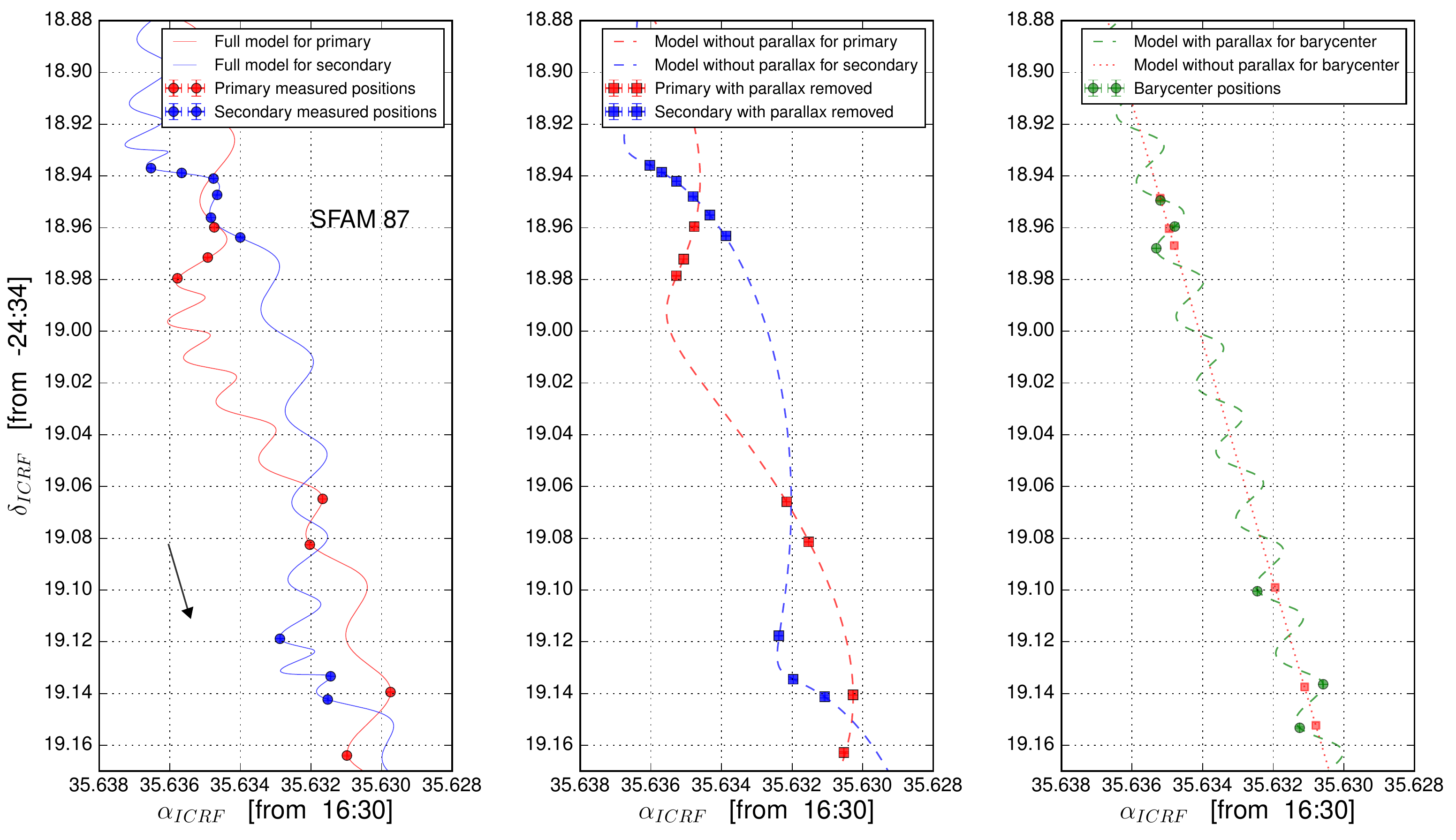}
\caption{Same as Figure \ref{fig:lfam15}, but for SFAM 87.}
\label{fig:sfam87}
\end{center}
\end{figure*}

SFAM 87 (also called ROX 39) has been detected in 4 epochs of GOBELINS. This source was also observed in 6 epochs from 2008 March to 2009 May as part of project BT097, but these data have not been published yet. We calibrated these additional epochs (Table \ref{tab:doar21}) and combined them with our more recent data for the astrometric fits. The source is resolved into two components that are simultaneously detected in 6 out of the 10 total epochs (3 in BT097 and 3 in GOBELINS). We note that \cite{Cheetham_2015} identified a companion to SFAM 87 in 2013, March, using  NIR aperture masking. This source appears to be the counterpart of the secondary source detected in our VLBA images (Figure \ref{fig:orbits}). We used all available VLBA and NIR data to fit jointly  orbital and proper motion,  as well as parallax. The resulting best fit is shown in  Figure \ref{fig:sfam87} in a similar fashion to LFAM~15.

\subsubsection{DoAr 51}

\begin{figure*}[!tb]
\begin{center}
 \includegraphics[width=0.8\textwidth,angle=0]{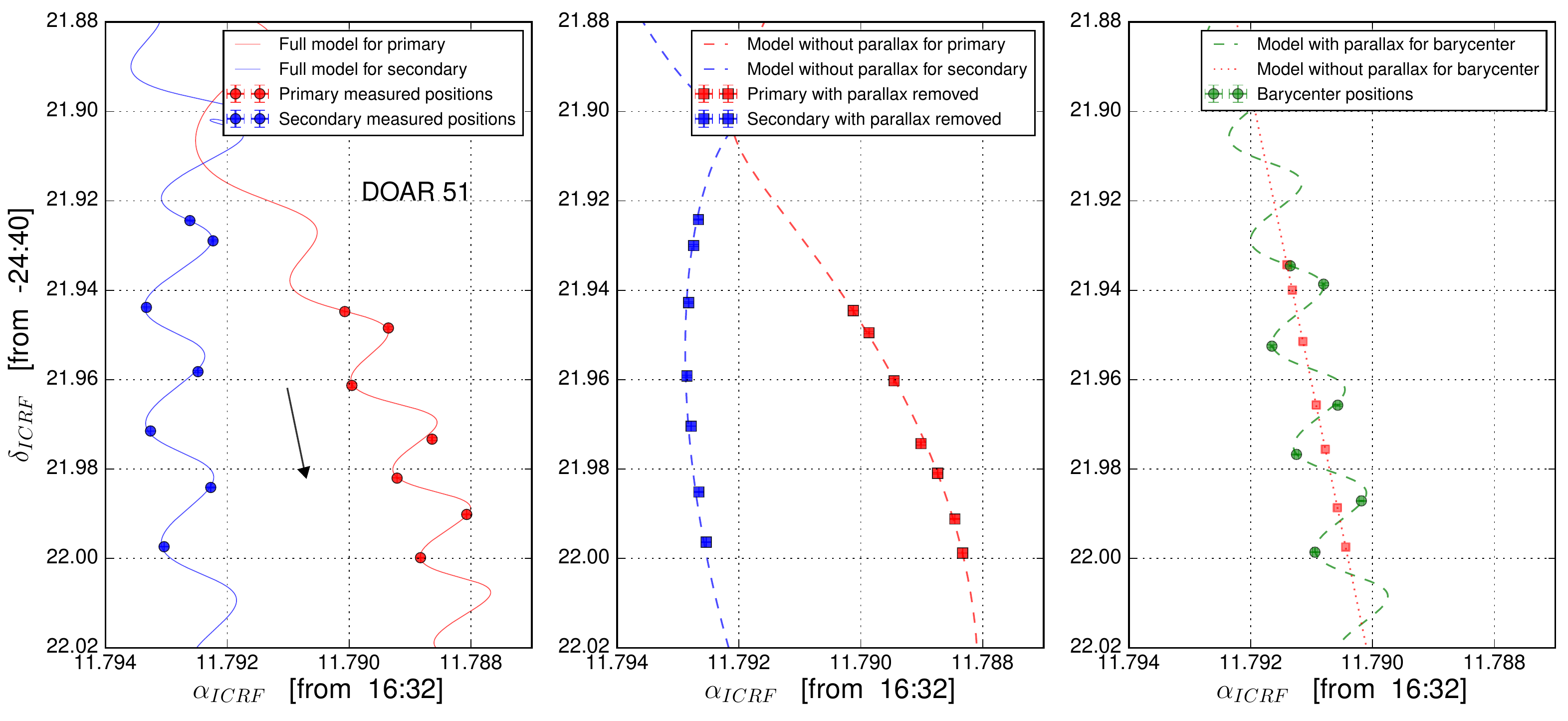}
\caption{Same as Figure \ref{fig:lfam15}, but for DOAR 51.}
\label{fig:doar51}
\end{center}
\end{figure*}

DoAr 51 (also called ROXs 47A) is located in the Lynds 1689 eastern streamer, 1.2$^{\rm o}$ east of the Ophiuchus core.  The source has been detected in 7 epochs, and was found to be double in all of them.  DoAr 51  was identified as a hierarchical triple system composed of a tight binary with separation of about 40 mas and a third component about 0.8 arcsec away by \cite{Barsony_2003}. \cite{Cheetham_2015} confirmed the tight binary using NIR aperture masking observations. They resolved the source into two components, with an angular separation of $51.5\pm0.20$ mas in June 2009 and $43.38\pm0.18$ mas in April 2010. This is to be compared with a separation of 40 $\pm$ 30 mas reported by \cite{Barsony_2003} in May 2002.  More recently, \cite{Rizzuto_2016} used the two positions measured by \cite{Cheetham_2015} and three new detections in the NIR to model the orbit of the system. These 5 detections are shown in Figure \ref{fig:orbits}, together with the  positional offsets between the components of the system as seen in our VLBA images. It is clear that our radio sources are the counterparts of the NIR sources, as they lie along the orbit derived by \cite{Rizzuto_2016}.  We model all data available from VLBA and NIR observations to better constrain the orbit of the system. The resulting best fit using the ``Full model'' is shown in Figure \ref{fig:doar51}, and the corresponding orbit in Figure \ref{fig:orbits}.  For comparison, we also show in this latter figure the orbit derived by \cite{Rizzuto_2016}.

\subsubsection{ROXN 39}

\begin{figure*}[!ht]
\begin{center}
\includegraphics[width=0.8\textwidth,angle=0]{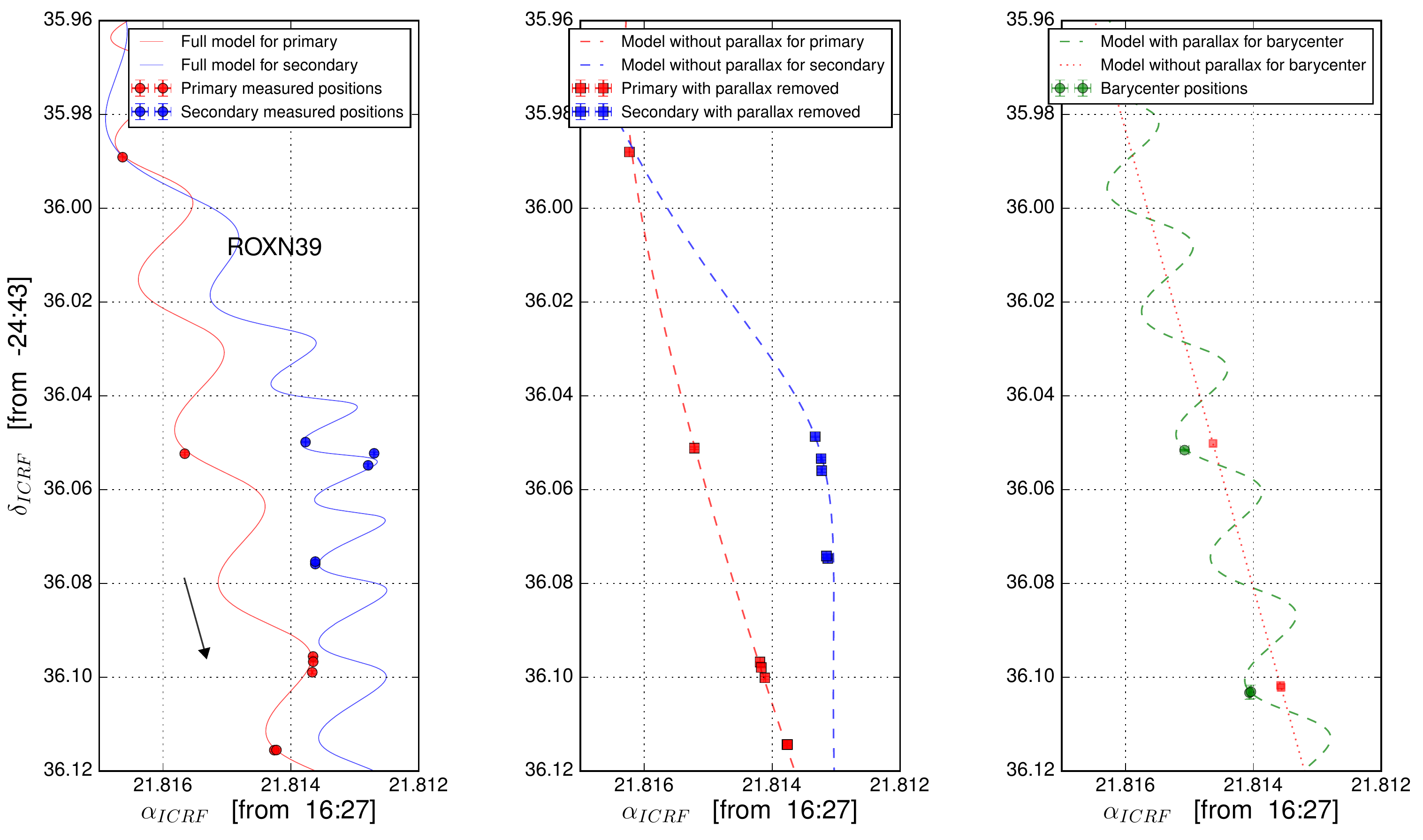}
\caption{Same as Figure \ref{fig:lfam15},  but for ROXN 39.}
\label{fig:roxn39}
\end{center}
\end{figure*}

The components of the system ROXN 39 have been detected  separately in 7 and 5 epochs, respectively, and simultaneously in only 3 epochs. The ``Full model'' fit and measured positions of individual components are shown in Figure \ref{fig:roxn39}, while Figure \ref{fig:orbits} shows the same model and relative positions. 
We did not fit the ``Relative model'' to this system because of the small number of simultaneous detections of both components. Consequently, the orbital parameters are less constrained, compared to the other systems described above.

\subsubsection{S1}\label{sec:s1}


This source has been detected in a total of 8 epochs, and was also observed at 14 epochs as part of projects BL128 and BT093 (Table \ref{tab:doar21}). 
The distance estimate by \cite{Loinard_2008} of $d= 116.9^{+7.2}_{-6.4}$~pc, based on the data from the first 6 of these old epochs, is significantly different from the distance obtained here for others sources in Lynds 1688 (which range from 130 to 140 pc), and hence requieres  a careful inspection. We re-calibrated the data used in \cite{Loinard_2008}, as well as the additional 8 unpublished observations obtained as part of BL128 and BT093. Similarly to DoAr~21, source positions measured at these 14 older epochs were corrected, before fitting the data,
by the positional offset of the calibrator J1625$-$2527 relative to its old position.

\begin{figure}[!ht]
\begin{center}
 \includegraphics[width=0.5\textwidth,angle=0]{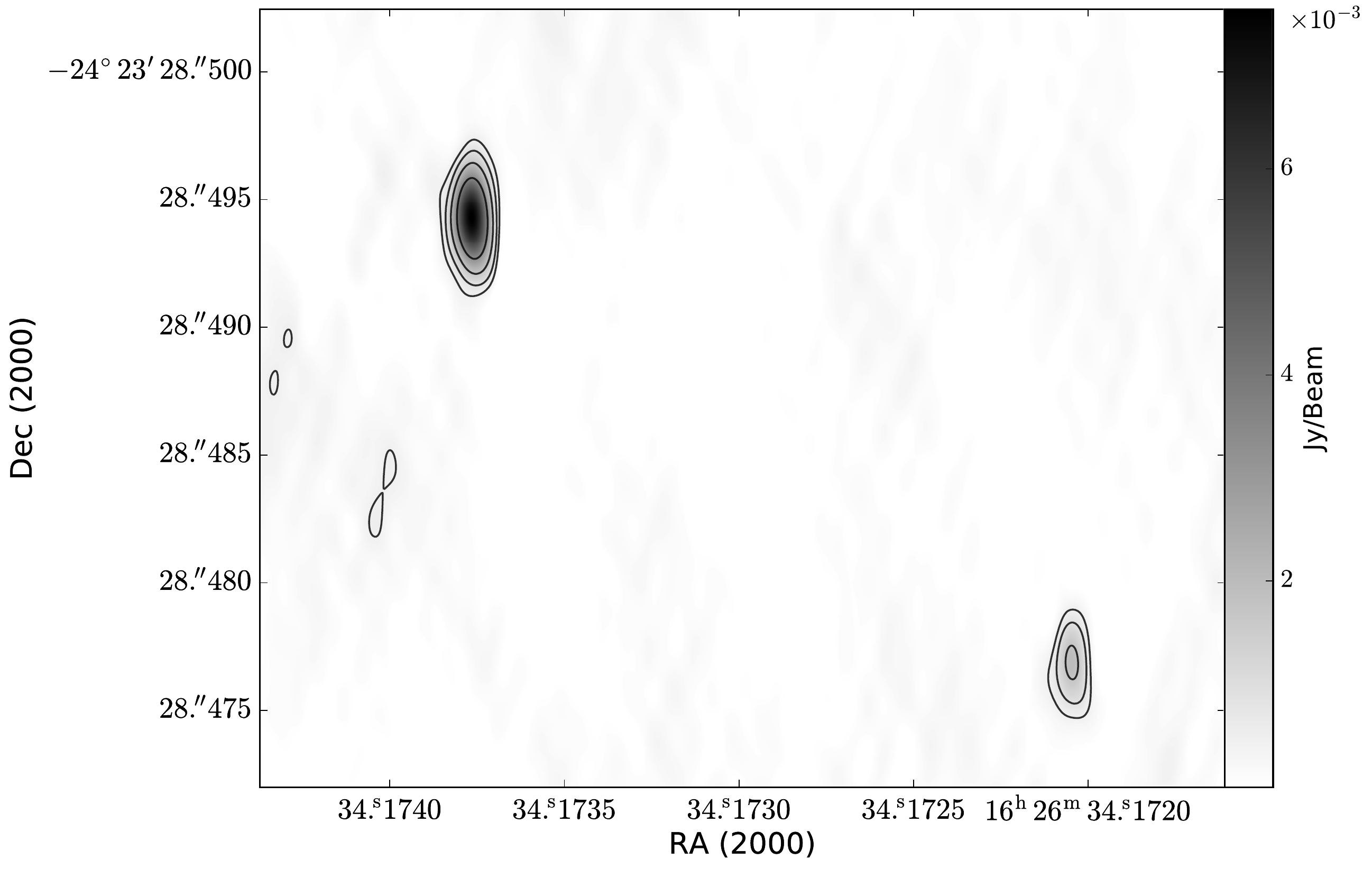}
\caption{The two components of S1 detected on June 21, 2007. The contours are 4, 8, 16, and 32$\sigma$, where $\sigma=1.08\times10^{-4}~\mu$Jy is the rms noise of the image.  }
\label{fig:s1bin}
\end{center}
\end{figure}

\begin{figure*}[!tb]
\begin{center}
 \includegraphics[width=0.9\textwidth,angle=0]{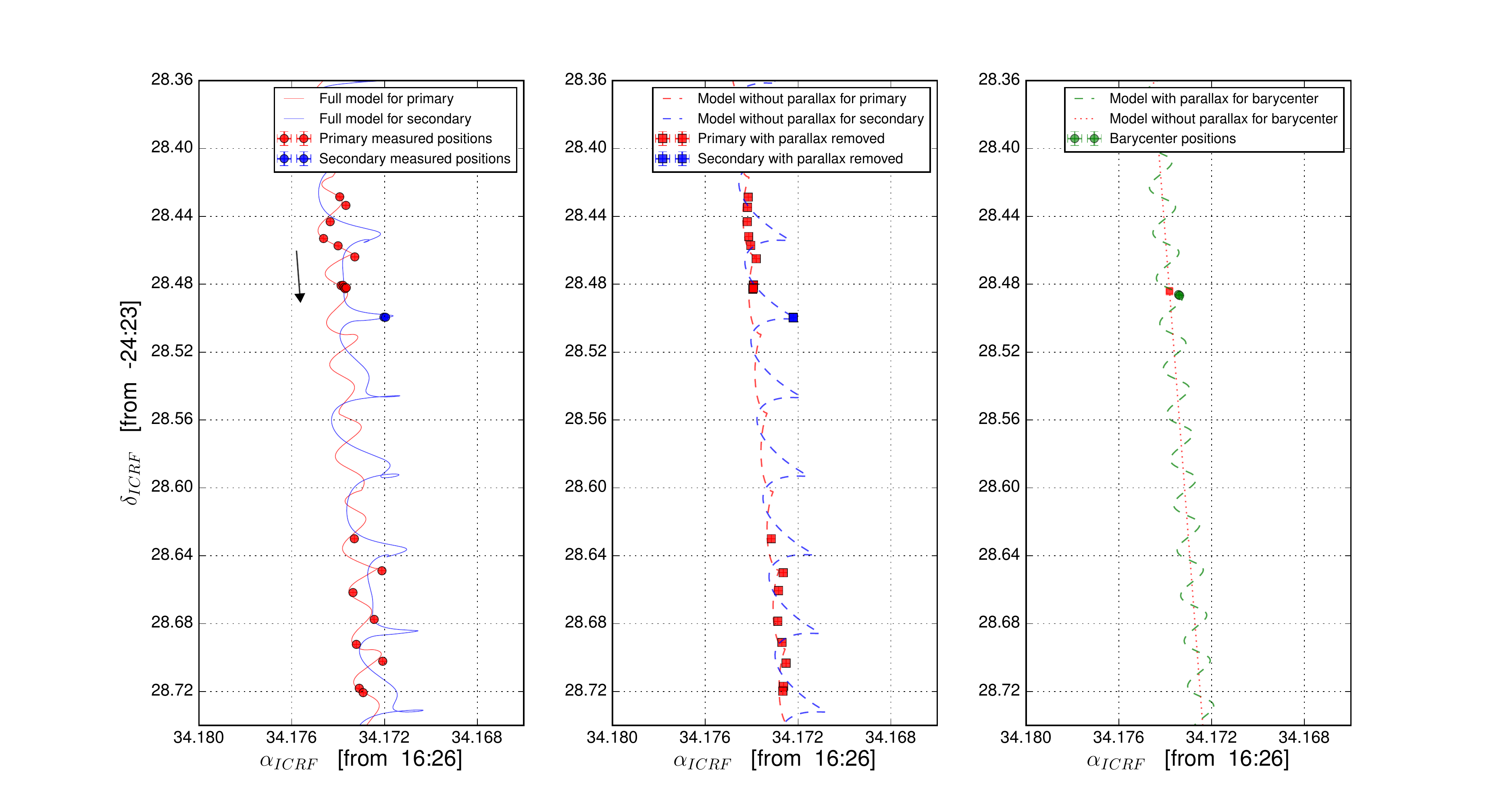}
\caption{Same as Figure \ref{fig:lfam15}, but for S1.}
\label{fig:s1}
\end{center}
\end{figure*}

A second source was detected in four epochs of the archival data, at an angular separation of about 20-30 mas from S1 (Figure \ref{fig:s1bin}). The detections are, however, only evident by self-calibrating the images, and the source is not present in the most recent epochs,  so they should be taken somewhat cautiously. It is interesting, however, that \cite{Richichi_1994} did report on the detection, using the lunar occultation technique in the NIR, of a companion to S1, at about 20 mas. Our detections of a second source in the system would be consistent with that earlier result. If we assume that the source is double, we can fit the orbital motion of the system jointly with the astrometric parameters, and use the positions of the putative second component to estimate individual masses. To that end, we discard the BL128 and 2nd BT093 secondary epochs, as well as the 2nd, 4th, and 6th BT093 primary epochs, because their corresponding images are of poor quality due to observing issues (cf.\ Table \ref{tab:doar21}), and source positions do not match up to the best fit solution.  The ``Full model'' fit, shown in Figure \ref{fig:s1}, yields $\varpi = 7.249\pm0.091$~mas and a distance of $138.0\pm1.7$~pc. The derived mass for the primary component is $5.8~M_\odot$, which is consistent with its B4 spectral type, while for the secondary we find a mass of $1.2~M_\odot$.  When ignoring the secondary component, the purely astrometric fit to all available data of the primary source yields $\varpi=8.335\pm0.522$~mas, and hence the distance derived by \cite{Loinard_2008}.  It is clear that this discrepancy in the results from the astrometry alone and the ``Full model'' fits is  due to the fact that the former does not take into account the multiplicity of the source. In the rest of the paper, we will use the results based on the astrometric plus orbital model, which are consistent with the distances obtained for other sources in  Lynds~1688. Since the two components are detected simultaneously in only two epochs, we are not able to fit the ``Relative model'' to relative positions, and only provide the resulting orbital parameters from the ``Full model'' in Table \ref{tab:orbits}.

\subsubsection{VSSG 11}

\begin{figure*}[!ht]
\begin{center}
{\includegraphics[width=0.32\textwidth,angle=0]{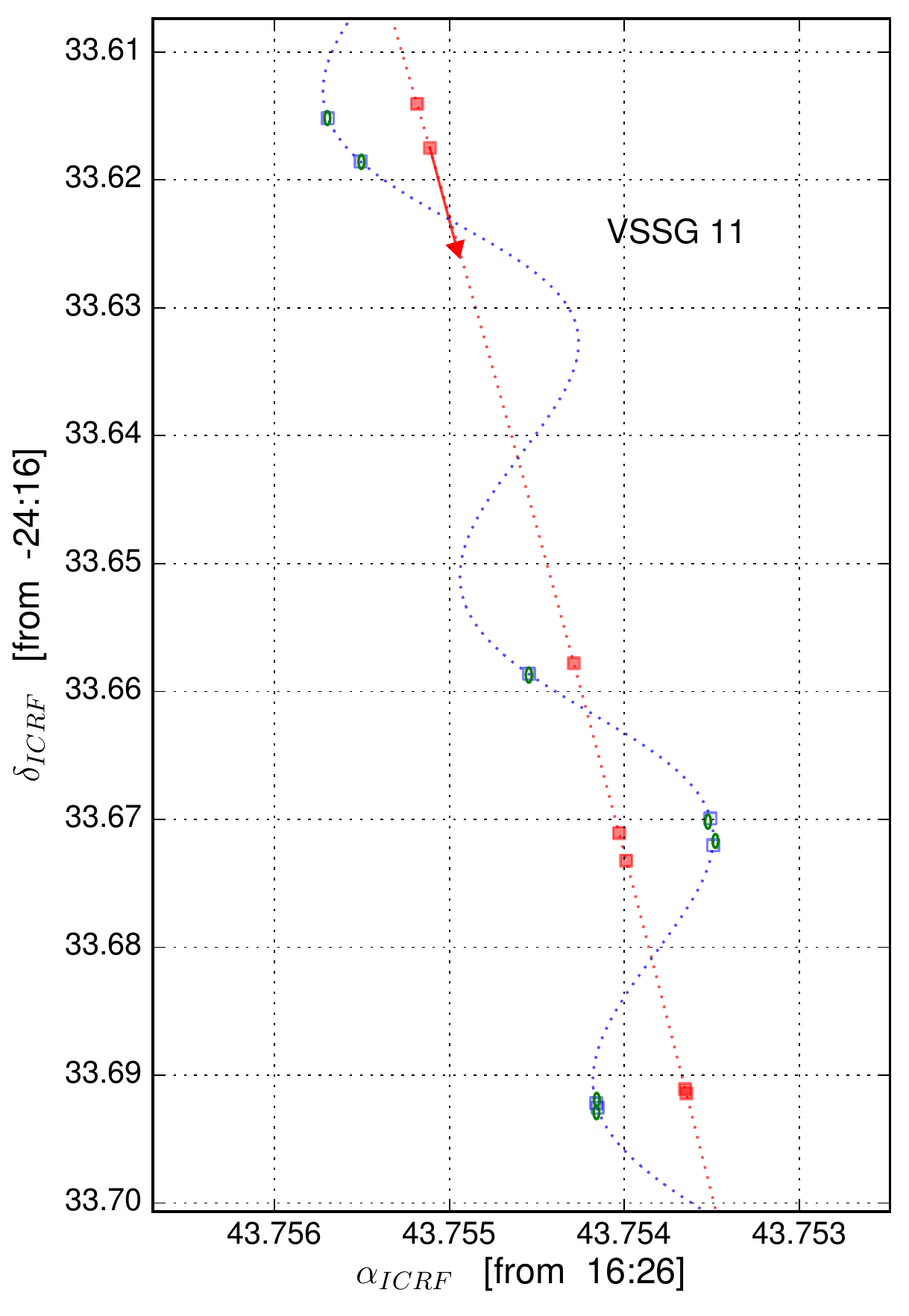}} 
{\includegraphics[width=0.275\textwidth,angle=0]{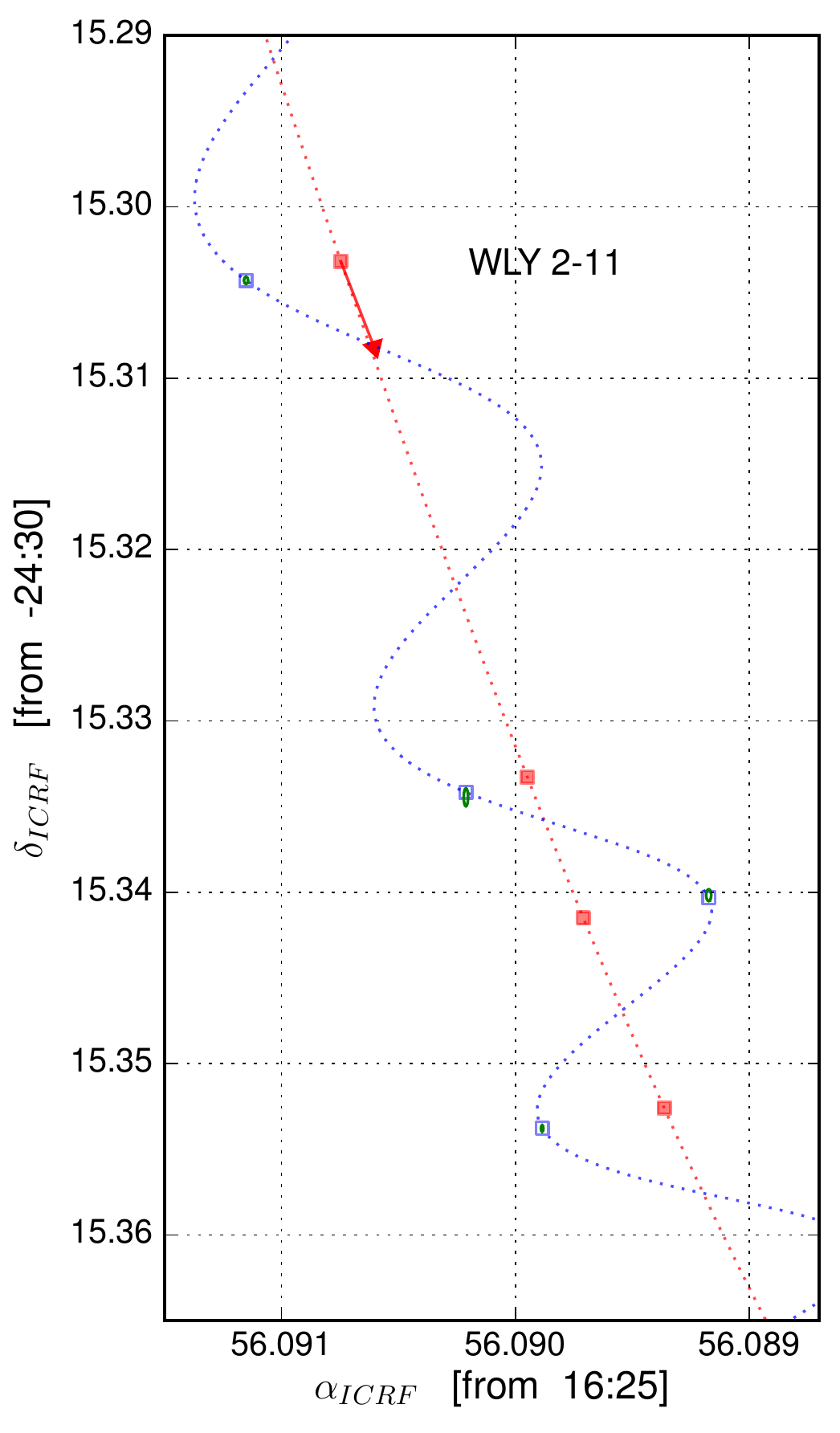}}
\caption{Same as Figure \ref{fig:all},  but for VSSG 11 and WLY 2-11.}
\label{fig:vssg11}
\end{center}
\end{figure*}

VSSG 11 has been detected in 7 observed epochs, and was found to be double in the last 3.  The second component is separated about 9 mas from the primary source. We perform the fits similarly to the other multiple systems discussed above. However, because our observations only cover a small fraction of the 
orbit,  the ``Full model'' fit to the secondary source does not converge. This produces considerably larger errors in the astrometric and orbital parameters than those derived for the other multiple systems. Also, we were not able to reproduce the observed separations between the components of the system. We then fit solely  parallax and accelerated proper motion. The resulting best fit is shown in Figure \ref{fig:vssg11}.

\subsubsection{WLY 2-11}

WLY2-11 has been detected in a total of 5 epochs. We resolve it into a double source in the last epoch, where the companion is detected at $> 10 \sigma$.  Since there are insufficient detections of primary and  secondary 
for attempting to model the orbit, we can only fit one of the components of the system.  We find that a good fit to the data is produced only when we discard the first epoch and consider for the fit  the fainter component detected in the last epoch. If, instead, we take the positions of the single component as measured at the first 4 epochs and the position of the brightest component detected in the last observation, the fit does not match all the measured points, producing large errors in the astrometric parameters. Thus, the detection at the first epoch probably corresponds to the companion source, which is detected as the brightest source in the last epoch. The fit, including acceleration terms, is shown in Figure \ref{fig:vssg11}.  Because of the few detections, we are not able to estimate systematic errors on the source positions.  Nevertheless, the derived parameter uncertainties
are comparable to those from the fits that do incorporate systematic errors on source positions. 

\subsection{YLW 15}\label{sec:ylw15}

\begin{figure}[!t]
\begin{center}
 \includegraphics[width=0.5\textwidth,angle=0]{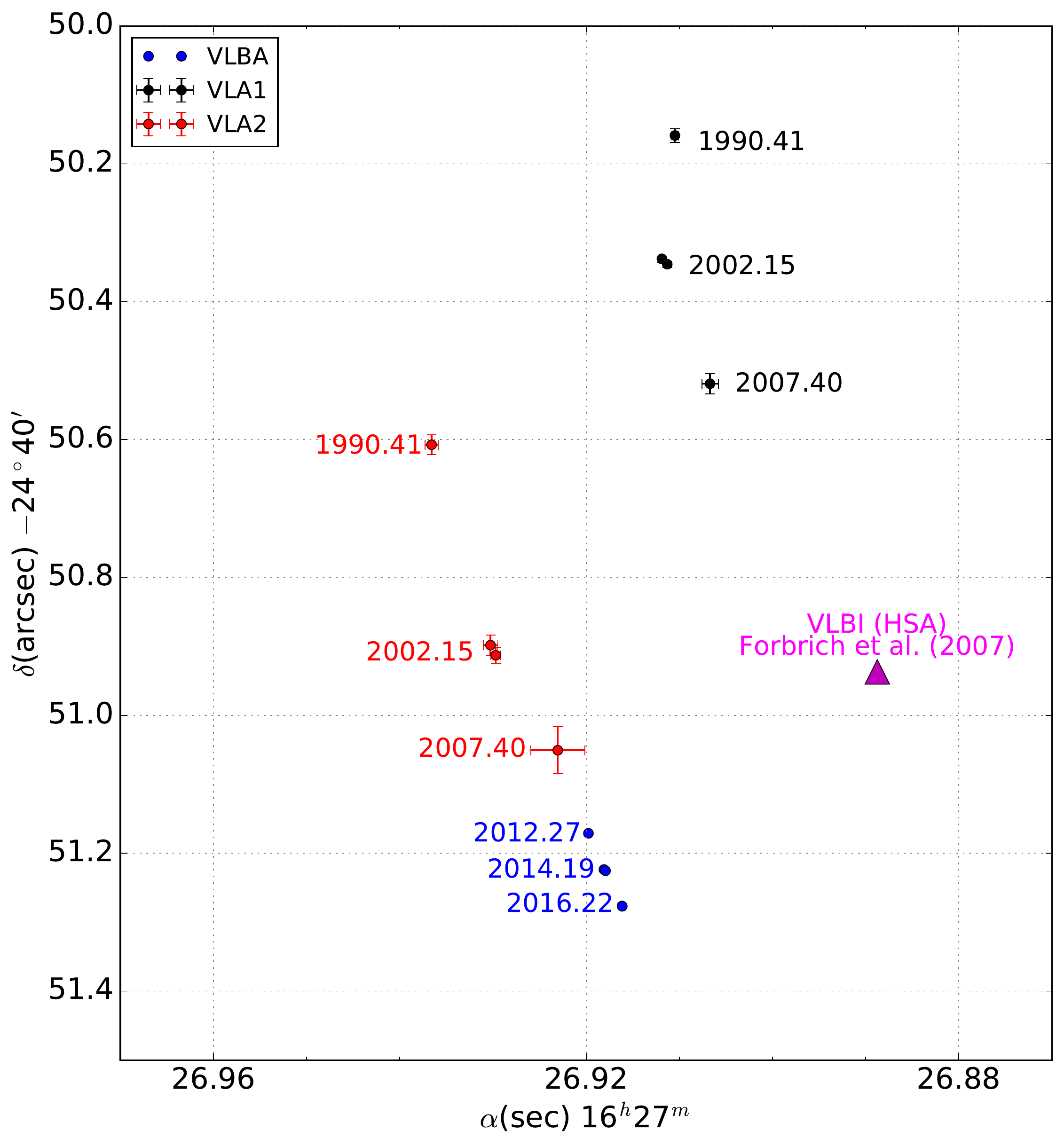}
\caption{Observed positions of the components of YLW 15. Red and black circles mark positions measured with the VLA at the indicated epochs. The blue circles show the positions obtained from our VLBA observations. The triangle marks the position reported by \cite{Forbrich_2007} from HSA observations.}
\label{fig:ylw15}
\end{center}
\end{figure}

We have observed the Class I protostar YLW~15 at eleven epochs, and clear detections  were obtained at five. Only one source is detected in the VLBA images, and since YLW 15 is known to be a binary system \citep{Curiel_2003}, it is important to determine which of the two stars is detected in our VLBA data. This can be achieved by comparing the astrometry of the present VLBA observations with that of the VLA data published by \cite{Curiel_2003}, which were taken between 1990 and 2002. Such a comparison is shown graphically in Figure \ref{fig:ylw15}. The positions of the two sources in the system (VLA 1 and VLA 2) are shown as a function of time as black and red symbols, respectively. We also calibrated and imaged the data from a VLA observation obtained in 2007 as part of project AF455. The VLBA positions are shown as blue symbols. It is clear that the source detected with the VLBA is located very near the expected position of VLA2 at the epochs of the VLBA observations, and more than half an arcsecond away from the expected position of VLA1. We conclude that the source detected with the VLBA is VLA2.

A marginal (7--9 $\sigma$) VLBI detection of YLW~15 has been reported in the past \citep{Forbrich_2007}. 
Indeed, the authors themselves mentioned that their detection was difficult to interpret, as the position did not coincide with the expected location of any of the two protostars in the system. This can also be seen in our Figure \ref{fig:ylw15}, where the position of the VLBI source reported by \cite{Forbrich_2007} is shown as a cyan triangle. 
Their observations were conducted, under project code BF083, with the High Sensitivity Array (HSA), consisting, for that run, of the Green Bank (GBT) 100-m radio telescope, the phased Very Large Array (equivalent in collecting area to a 130 m dish) as well as the VLBA. We calibrated these data following the standard procedures for incorporating non-VLBA antennas. 
We achieve rms noise levels between 10 and 22 $\mu{\rm Jy~beam}^{-1}$ (depending on the AIPS ROBUST parameter used for imaging), which are consistent with the value reported by \cite{Forbrich_2007} of 15.4 $\mu{\rm Jy~beam}^{-1}$. We do not detect the source, and confirm our suspicion that their detection was likely spurious.

Even though five data points are, in principle, enough to perform an astrometric fit, most of our VLBA detections of YLW15 were obtained close in time (two are separated by $\sim 3$ weeks and other two by only $\sim 3$ days). Moreover, the five detections were acquired around successive spring equinoxes, with no detection close to the fall equinox. As a result of this, the astrometric fit produces unreliable results. We will wait until we have more detections for the derivation of the astrometric parameters of this star.  

\begin{figure}[!t]
\begin{center}
 \includegraphics[width=0.35\textwidth,angle=0]{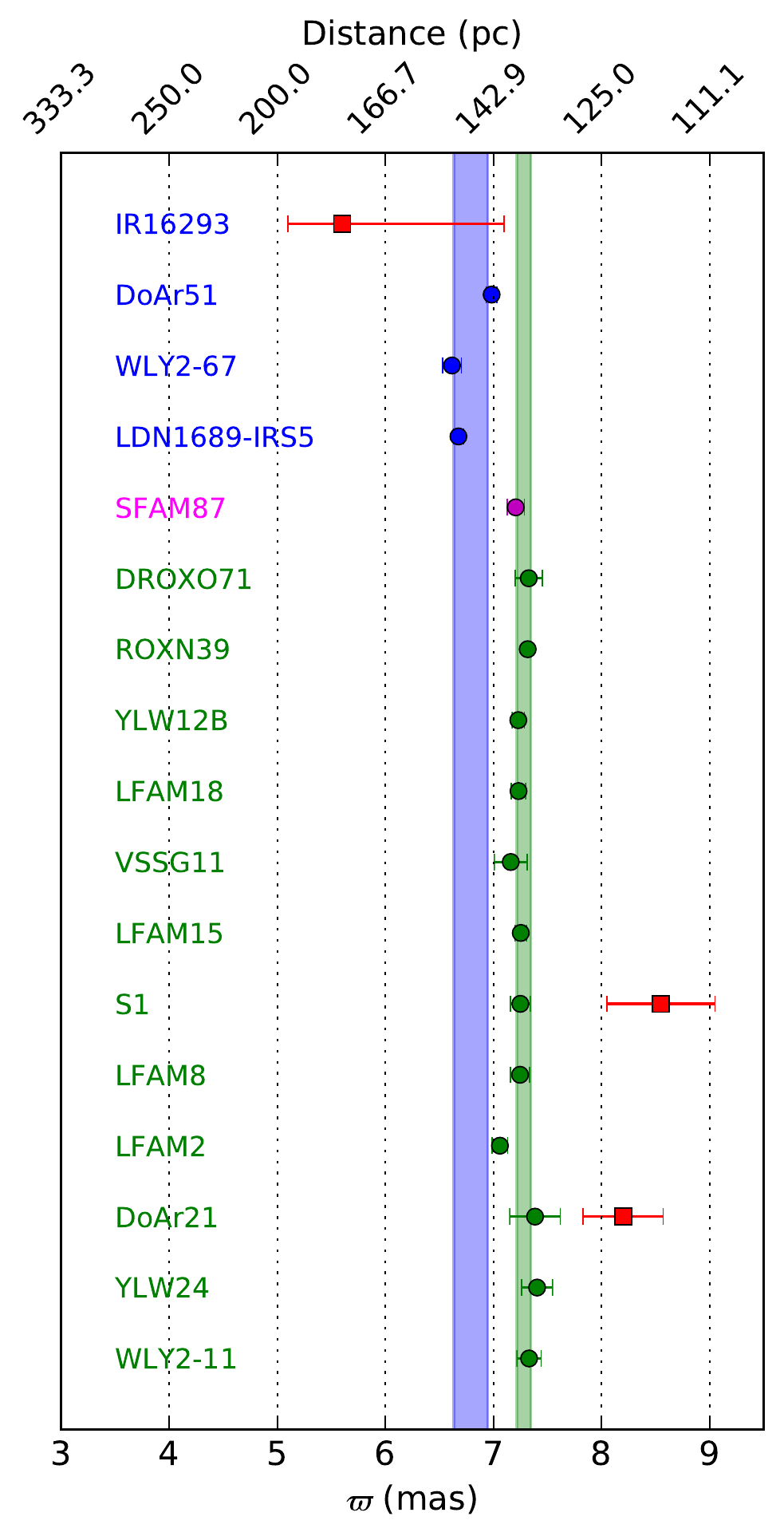}
\caption{Summary of the parallax measurements reported here. Green circles and characters are for sources in the core (Lynds 1688), while blue circles and characters are for sources in the eastern streamer (Lynds 1689). SFAM 87 is shown with a different color because it lies neither within the core nor within the boundaries of Lynds 1689. The red squares are for previously published parallaxes \citep{Loinard_2008, Imai_2007}. The green vertical bar shows the mean parallax value for sources in the core, and its standard deviation (see text). 
} 
\label{fig:prlx}
\end{center}
\end{figure}

\section{The distance to Ophiuchus}\label{sec:distance2oph}

In this paper, we report on 16 independent trigonometric parallax measurements. These results are listed in Table \ref{tab:parallaxes} and summarized graphically in Figure \ref{fig:prlx}. They largely surpass all previously published distance measurements for Ophiuchus.

Out of these 16 measurements, 12 are for YSOs in the Ophiuchus core (Lynds 1688). The parallaxes for these 12 sources are highly consistent (Figure \ref{fig:prlx}), and yield a weighted mean value $\varpi$ = 7.28 mas, with a weighted standard deviation $\sigma_\varpi$ = 0.06 mas. 
In terms of distance, this corresponds to $d$ = 137.3 pc, with a standard deviation $\sigma_d$ = 1.2~pc. In principle, the standard deviation around the mean value could reflect both the uncertainties in our distance measurements, and the true depth of the complex. Since our uncertainties on individual parallax measurements are typically larger than 0.06 mas, we argue that the measured weighted standard deviation is completely dominated by the uncertainties on individual parallaxes rather than by the true depth of the core. We will, therefore, adopt this value as our final uncertainty on the distance to the core. We note that Lynds 1688 is about 0.75 degree across (Figure \ref{fig:pm}), corresponding to 1.8 pc. Thus, our results indicate that it is not significantly more elongated along the line of sight than on the plane of the sky. 

We currently can only provide limited information on the location of the streamers relative to the core, since we only have measured parallaxes in the eastern streamer (Lynds 1689), and have just three independent measurements there (the parallaxes of LDN1689 IRS5, WLY 2-67, and DoAr51).
The weighted mean value of these three measurements is $\varpi$ = 6.79 $\pm$ 0.16 mas,
corresponding to $d$ = 147.3 $\pm$ 3.4 pc. This suggests that the eastern streamer is about 10 pc farther than the core, although more parallax measurements of sources in Lynds 1689 would be required to confirm this. Interestingly, \cite{Imai_2007} have measured the parallax of water masers in the protostar IRAS~16293--2422, located in the northern part of Lynds 1689, and found a value $\varpi$ = 6.5$_{-0.5}^{+1.5}$ mas, which is consistent within 1 sigma with our parallax estimate for Lynds 1689 (Figure \ref{fig:prlx}). Notice that we have not considered  SFAM 87 in the previous analysis.  It certainly lies somewhat outside of Lynds 1688 (Figure \ref{fig:pm}), and was considered as belonging to the Lynds 1689 ``fringe'' by \cite{McClure_2010}. These latter authors, however, do not explain how they arrived to such a conclusion, and we note that SFAM 87 does not formally lie within the boundaries of Lynds 1689. We find that its parallax is more consistent with the mean weighted parallax of the Ophiuchus core, suggesting that it is not part of Lynds 1689. This will be confirmed also when the distance to stars in the central parts of  this cloud  become available. 

The typical distance to Ophiuchus that is generally used in the pre-main sequence literature is 125--130~pc. Our new distance is about $12\%$ larger, which translates to a luminosity increase of $\sim25\%$, and makes young stellar objects slightly younger with respect to evolutionary tracks.

Finally, we should mention that the stellar population of Ophiuchus could be contaminated by stars from the Upper Sco association, which is located at a similar  distance ($\sim$140 pc; \citealt{de_Zeeuw_1999}) and overlaps the Ophiuchus region on the sky. The YSOs in Ophiuchus show extinctions in the range $3\leq A_V\leq 26$~mag (Table \ref{tab:yso}), whereas the associated members of Upper Sco typically have $A_V  \lesssim  2$~mag \citep{Walter_1994}.  The fact that our detected YSOs have larger extinctions than Upper Sco ensures that these objects are part of the Ophiuchus Complex. 



\subsection{Proper Motions}

\begin{figure*}[!t]
\begin{center}
 \includegraphics[width=0.7\textwidth,angle=0]{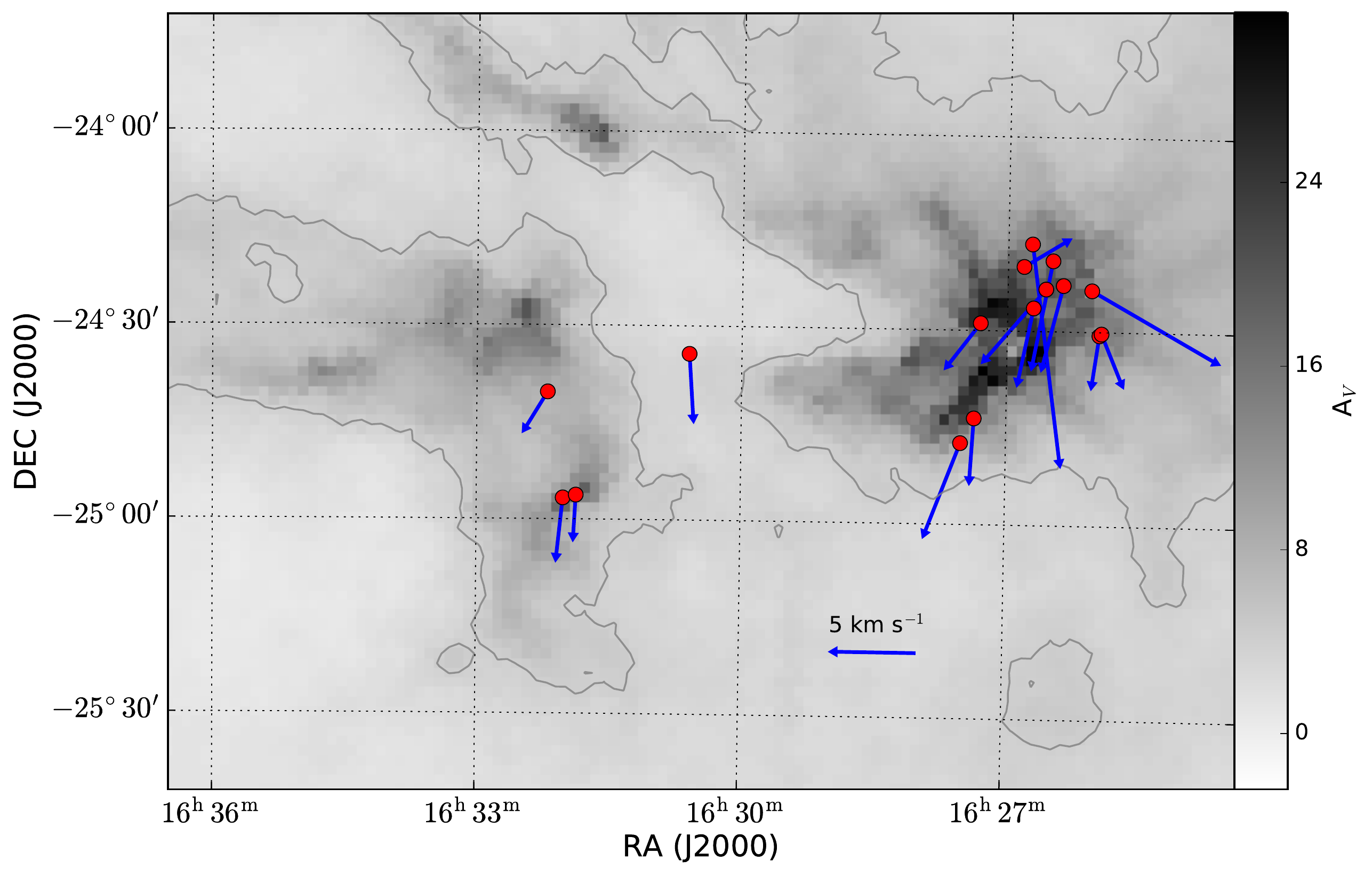}
\caption{Spatial distribution of YSOs in Ophiuchus with astrometric parameters derived in this work. The grey scale represents the extinction map obtained as part of the COMPLETE project \citep{Ridge_2006}, based on 2MASS data \citep{Skrutskie_2006}.  The grey contour indicates an $A_V$ of 4. The arrows correspond to the tangential velocity corrected by the solar motion. } 
\label{fig:pm}
\end{center}
\end{figure*}

Figure \ref{fig:pm} shows the distribution of the 16 individual sources with astrometric parameters measured in this paper. In order to calculate the motion of each source relative to its local environment, we need to remove the contribution of the solar peculiar motion. For this correction, we use the  formulation of \cite{Abad_2005}, and the solar motion relative to the LSR derived by \cite{Schonrich_2010}. They obtained rectangular components of the solar motion $(u_\odot,v_\odot,w_\odot)=(11.1\pm0.7,12.2\pm0.47,7.25\pm0.37)~{\rm km~s}^{-1}$, directed towards the Galactic center, the direction of Galactic rotation, and the Galactic north pole, respectively.  Corrected proper motions were transformed to tangential velocities (Table \ref{tab:pm}) and   overlaid on Figure \ref{fig:pm}.  It is clear that, with the exception of a few objects  that may belong to a different substructure of the complex, all sources share similar motions. 
The derived one-dimensional velocity dispersion in R.A. and DEC.\ of sources in Lynds 1688 are  2.8 and $3.0~{\rm km~s}^{-1}$, respectively. These are larger than the values found in other studies, which range from $\sim$1 to 2 ${\rm km~s}^{-1}$ \citep{Makarov_2007,Wilking_2015}.  However, our sample size is small and our results need to be confirmed with additional proper motions and parallaxes. In addition, we estimate that the associated errors of the velocity dispersions (assumed to be Gaussian distributed) are 1.8 and 2.0 ${\rm km~s}^{-1}$, in R.A.\ and DEC., respectively. Thus, within the errors, our results are still consistent with past measurements. 

\section{Summary}\label{sec:conclusions}

We have presented the first results from GOBELINS toward the region of Ophiuchus. We observed a total of 50 YSOs with the VLBA, and detected 26 of them. Most of our VLBA-detected YSOs are Class II-III, but three Class I sources have also been detected. 
The YSOs detected here are clearly non-thermal emitters (otherwise they would not be detected with the VLBA), so our observations have revealed the existence of a large population of non-thermal YSOs in Ophiuchus. 
About 30\% of our the VLBA-detected YSOs belong to tight multiple systems with angular separations from 0.6 to 44 AU. While this fraction appears to be consistent with  the binarity fraction in Ophiuchus measured in a recent infrared multiplicity survey, we note that most of the binaries detected with the VLBA are very tight systems with separations below 10 AU.  

The astrometry of 16 young stellar systems was presented. Absolute positions of single sources were modelled to derive parallaxes and proper motions, and source distances were then obtained with a few percent accuracy. For sources in multiple systems, we use individual positions and, in most cases, angular separations to model jointly orbital and astrometric parameters.  For these sources, the distance was measured with  $0.3-2\%$  accuracy. Because the VLBA  delivers absolute positions for each component, we were able to determine the individual masses in 6 of the 8 total multiple systems; masses range from $\sim1$ to 7~${\rm M}_\odot$. 

Twelve sources are associated with the Ophiuchus core (Lynds 1688). They yield a mean distance of 137.3$\pm$1.2 pc, and no indication of a detectable depth. Three sources for which  the astrometric elements could be measured are located in the eastern streamer (Lynds 1689); the measurements imply a distance of 147.3$\pm$3.4 pc for this cloud. This result suggests that the eastern streamer is 10 pc more distant than the core, but this needs to be confirmed when more parallaxes become available.

The measured proper motions of young stars in the core yield one-dimensional velocity dispersions in R.A. and DEC. of 2.8$\pm$1.8 and 3.0$\pm$2.0~${\rm km~s}^{-1}$, respectively. This result may indicate that our sources belong to different substructures of the complex. However, our result may suffer  from small-number statistics, and the associated errors are large. Indeed, our values are consistent within $1\sigma$ with velocity dispersions of $1-2~{\rm km~s}^{-1}$, which have been measured in the past.

Finally, we note that  6 YSOs have been detected only 2 or 3 times with the VLBA in the observations presented here. For these sources, no meaningful astrometric fit could yet be performed, but this will become possible once a few additional detections are obtained in the coming few years. Thus, we anticipate that we will soon be able to increase the number of individual trigonometric parallaxes in Ophiuchus. 

\acknowledgements{G.N.O.-L., L.L., L.F.R., R.A.G.-L., G.P., and J.L.R. acknowledge CONACyT, Mexico for financial support through grants 339802, 104497, 153522 and  I0017-151671. L.L. and R.A.G.-L.\ were supported by DGAPA, UNAM grant PAPIIT IG100913.  L.L. and G.N.O.-L. also acknowledge support from the von Humboldt Stiftung. N.J.E.\ was supported by NSF grant AST-1109116 to the University of Texas at Austin. P.A.B.G.\ acknowledges financial support from FAPESP. The National Radio Astronomy Observatory is operated by Associated Universities, Inc., under cooperative agreement with the National Science Foundation. This work made use of the Swinburne University of Technology software correlator, developed as part of the Australian Major National Research Facilities Programme and operated under licence.}

\clearpage

\clearpage
\begin{deluxetable*}{llllc}
\tabletypesize{\scriptsize}
\tablecaption{Observed epochs \label{tab:observations}}
\tablewidth{0pt}
\tablehead{ Project & Observation & \multicolumn{2}{c}{Observed fields centers} & Observed \\
code &  Date  & \colhead{R.A. ($\alpha_{2000}$)} &  \colhead{Decl. ($\delta_{2000}$)} & band \\
}\startdata
BL175B0 &13 mar 2012&16 27 55.92 & -24 47 24.82 & X\\
BL175B1 &25 mar 2012&16 27 30.82 & -24 47 27.21 & X\\
BL175B2 &09 apr 2012&16 27 24.36 & -24 42 13.39 & X\\
BL175B3 &21 apr 2012&16 25 49.10 & -24 38 31.00 & X\\
BL175B4 &24 apr 2012&16 27 15.70 & -24 38 45.68 & X\\
BL175B5 &29 apr 2012&16 28 04.65 & -24 34 56.66 & X\\
BL175B6 &01 may 2012&16 25 56.80 & -24 30 23.76 & X\\
BL175B7 &05 may 2012&16 26 07.63 & -24 27 41.73 & X\\
\enddata
\end{deluxetable*}
\clearpage

\floattable
\clearpage
\begin{deluxetable}{cccccccccc}
\tabletypesize{\scriptsize}
\tablecaption{Detected YSOs \label{tab:yso}}
\tablewidth{0pt}
\tablehead{ GBS-VLA&       Other & Minimum flux   & Maximum flux  & Minimum flux  & Maximum flux   & log [$T_b$ (K) ] & SED       & Num. of     & $A_V$   \\
        name$^{1}$ &  identifier & at 5 GHz (mJy) &at 5 GHz (mJy) & at 8 GHz (mJy)& at 8 GHz (mJy) &                  &   Class   & detc./obs.  &         \\
      (1)          &       (2)   &   (3)          &  (4)          &   (5)         &    (6)         &     (7)          &  (8)      &  (9)        &  (10)   \\
}\startdata
J162556.09-243015.3 & WLY2-11a & 0.13$\pm$0.05 & 0.27$\pm$0.06 & \multicolumn{2}{c}{--} & $>$7.0 & \multirow{2}{*}{Class III} & 4/5 & \multirow{2}{*}{13} \\ 
J162556.09-243015.3 & WLY2-11b & \multicolumn{2}{c}{0.86$\pm$0.05} & \multicolumn{2}{c}{0.60$\pm$0.09} & $$7.9 &  & 2/5 &  \\ 
J162557.51-243032.1 & YLW24 & 0.21$\pm$0.05 & 1.35$\pm$0.06 & \multicolumn{2}{c}{0.25$\pm$0.06} & $$8.0 & Class III & 4/5 & 13 \\ 
J162603.01-242336.4 & DOAR21 & 1.98$\pm$0.11 & 14.97$\pm$0.14 & 4.13$\pm$0.07 & 5.66$\pm$0.07 & $$9.2 & Class III & 7/7 & 12 \\ 
J162616.84-242223.5 & LFAMP1 & 0.15$\pm$0.06 & 0.47$\pm$0.04 & \multicolumn{2}{c}{--} & $$6.8 & Class II & 2/6 & 20 \\ 
J162622.38-242253.3 & LFAM2 & 0.30$\pm$0.05 & 0.38$\pm$0.07 & \multicolumn{2}{c}{$<$0.09} & $>$6.7 & Class II & 3/7 & 20 \\ 
J162625.62-242429.2 & LFAM4 & \multicolumn{2}{c}{0.66$\pm$0.12} & \multicolumn{2}{c}{$<$0.12} & $>$6.9 & Class I & 1/14 & 17 \\ 
J162629.67-241905.8 & LFAM8 & 0.37$\pm$0.06 & 1.18$\pm$0.13 & 0.26$\pm$0.07 & 0.30$\pm$0.05 & $$7.3 & Class III & 7/9 & 19 \\
\enddata
\tablecomments{Reported sources  have flux densities above $6\sigma$ and $5\sigma$ in the cases of one or several detections, respectively. Non-detections are indicated by giving an upper flux density limit of $3\sigma$.}
\tablenotetext{1}{GBS-VLA stands for the Gould's Belt Very Large Array Survey \citep{Dzib_2013}.}
\tablenotetext{2}{This star is resolved into a double source in past VLBA observations.
}
\end{deluxetable}
\clearpage
\clearpage
\begin{deluxetable*}{lcccc}
\tabletypesize{\scriptsize}
\tablewidth{0pt}
\tablecolumns{5}
\tablecaption{Measured source positions \label{tab:positions}}
\tablehead{ Julian Day & $\alpha$ (J2000.0) & $\sigma_\alpha$ & $\delta$ (J2000.0) & $\sigma_\delta$    \\
}
\startdata
\cutinhead{DROXO71} %
2456011.96538 & 16  27  30.83414340 & 0.00000470 & -24  47  27.142107 & 0.000161 \\ 
2456172.52613 & 16  27  30.83298666 & 0.00000234 & -24  47  27.153542 & 0.000071 \\ 
2456471.70936 & 16  27  30.83302924 & 0.00000177 & -24  47  27.178102 & 0.000063 \\ 
2456537.53271 & 16  27  30.83267868 & 0.00000191 & -24  47  27.182477 & 0.000065 \\ 
2456718.03765 & 16  27  30.83359624 & 0.00000260 & -24  47  27.198462 & 0.000085 \\ 
2456937.93345 & 16  27  30.83241321 & 0.00000472 & -24  47  27.212846 & 0.000216 \\ 
2457081.04343 & 16  27  30.83323569 & 0.00000437 & -24  47  27.227091 & 0.000170 \\ 
2457473.99603 & 16  27  30.83275199 & 0.00000469 & -24  47  27.256316 & 0.000188 \\
\enddata
\end{deluxetable*}
\clearpage

\clearpage
\begin{deluxetable*}{cccccccc}
\tabletypesize{\scriptsize}
\tablewidth{0pt}
\tablecolumns{6}
\tablecaption{Parallaxes and proper motions \label{tab:parallaxes} }
\tablehead{ GBS-VLA& Other identifier & Parallax & $\mu_\alpha\cos\delta$ & $\mu_\delta$      & a$_\alpha\cos\delta$ & a$_\delta$         &Distance \\
                   name      &                         & (mas)     & (mas yr$^{-1}$)               & (mas yr$^{-1}$)  & (mas yr$^{-2}$)          & (mas yr$^{-2}$) & (pc)          \\
                   (1)           &           (2)         &    (3)       &   (4)                                 &   (5)                    &   (6)                            &  (7)                    & (8)            \\
}
\startdata
J162556.09-243015.3 & WLY2-11        & 7.330  $\pm$ 0.112  & -9.78  $\pm$ 0.09 &  -25.11 $\pm$ 0.20  & 0.94 $\pm$ 0.38 & 6.02 $\pm$ 0.83 & 136.4 $\pm$ 2.1 \\ 
J162557.51-243032.1 & YLW24          & 7.404 $\pm$ 0.143  &  -7.26 $\pm$ 0.04 &  -25.29  $\pm$ 0.07  & -- & --  & 135.1 $\pm$ 2.6  \\ 
J162603.01-242336.4 & DoAr21          & 7.385 $\pm$ 0.234  & -19.63  $\pm$ 0.19 &  -26.92 $\pm$ 0.13 & -- & -- &135.4 $\pm$ 4.3  \\ 
J162622.38-242253.3 & LFAM2          & 7.060  $\pm$ 0.072  &  -5.64 $\pm$ 0.09 & -27.02  $\pm$ 0.30  & -- & --  & 141.6 $\pm$ 1.5   \\ 
J162629.67-241905.8 & LFAM8           & 7.246 $\pm$ 0.088  &  -5.89  $\pm$ 0.06 &  -29.54 $\pm$ 0.16 & -- & -- &138.0 $\pm$ 1.7  \\ 
J162634.17-242328.4 & S1                  & 7.249 $\pm$ 0.091  & -2.05  $\pm$ 0.02 &  -26.72  $\pm$ 0.04     & -- & -- & 138.0 $\pm$ 1.7\\  
J162642.44-242626.1 & LFAM15          & 7.253 $\pm$ 0.054 &  -6.31  $\pm$ 0.02  &  -26.95 $\pm$ 0.05  & -- & -- &137.9 $\pm$ 1.0   \\ 
J162643.76-241633.4 & VSSG11        & 7.160  $\pm$ 0.152 &  -10.48 $\pm$ 0.16 & -38.99 $\pm$ 0.35 & 0.31 $\pm$ 0.65 & -1.54 $\pm$ 1.09 & 139.7 $\pm$ 3.0 \\
\enddata
\tablenotetext{1}{Parallax is fixed at the value obtained for YLW 12Bab when solving for the other astrometric parameters. }
\end{deluxetable*}
\clearpage

\clearpage
\begin{deluxetable*}{llll}
\tabletypesize{\scriptsize}
\tablecaption{Archive VLBA data \label{tab:doar21}}
\tablewidth{0pt}
\tablehead{ Project code & Date of observation & Source & Relevant entries from the observing log   \\
}
\startdata
BL128 FA  & 2005 Jun 07 & LFAM 15\\
BL128 FB  & 2005 Sep 03 & LFAM 15\\
BL128 FC  & 2005 Dec 09 & LFAM 15\\ 
BL128 FD  & 2006 Mar 01 & LFAM 15\\
BL128 FG$^1$ & 2006 Dec 20 & DOAR 21 & PT may have some residual snow in the dish. \\                          
         &             &   & Cloudy at NL,LA, and PT.  \\
         &             &   & LA weather: overcast and snowing. 2 inches of snow. \\
         &             &   & NL precipitation.   \\
\enddata
\tablenotetext{1}{Epoch was discarded.}
\tablenotetext{2}{Data shows low amplitudes for IFs 3-4.}
\end{deluxetable*}
\clearpage

\clearpage
\begin{deluxetable*}{ccccccccccc}
\tabletypesize{\scriptsize}
\tablewidth{0pt}
\tablecolumns{6}
\tablecaption{Orbital elements \label{tab:orbits}}
\tablehead{  Name             & a     & P   & $T_0$& $e$   & $\Omega$& $i$    & $\omega$       & $M_1$       & $M_2$      & $M_T$       \\
                              & (mas) & (yr)&      &     & ($^o$)  & ($^o$) & ($^o$)         & (M$_\odot$) & (M$_\odot$)& (M$_\odot$) \\
                  (1)         & (2)   & (3) & (4)  & (5) & (6)     & (7)    & (8)            & (9)         & (10)       & (11)        \\
}
\startdata
\sidehead{LFAM 15}
Full   & 16.40$\pm$0.13 & 3.591$\pm$0.0183 & 2007.008$\pm$0.039 & 0.528$\pm$0.005 & 337.93$\pm$0.81 &110.30$\pm$0.49 & 235.54$\pm$1.02 & 0.4687 & 0.421 & 0.89 \\ 
& & &  & &  & & & $\pm$0.0146 & $\pm$0.010 & $\pm$0.01  \\
Rel. astr.     & 16.98$\pm$0.16 & 3.598$\pm$0.0099 & 2010.626$\pm$0.011 & 0.561$\pm$0.007 & 340.05$\pm$0.49 &109.77$\pm$0.27 &239.60$\pm$0.64 & --     & --  & 0.99    \\ 
& & &  & &  & & & & & $\pm$0.27  \\
\hline
\sidehead{YLW 12Bab}
Full & 12.70$\pm$0.09 & 1.425$\pm$0.0012 & 2005.174$\pm$0.009 & 0.444$\pm$0.003 & 135.15$\pm$1.03  &75.60$\pm$1.25 & 158.69$\pm$1.47 & 1.3969 & 1.258   & 2.66 \\ 
& & &  & &  & & & $\pm$0.0194 & $\pm$0.006 & $\pm$0.02 \\
Rel. astr.    & 12.54$\pm$0.06 & 1.424$\pm$0.0005 & 2013.720$\pm$0.003 & 0.442$\pm$0.003 & 135.38$\pm$0.32  &74.32$\pm$0.48 & 160.11$\pm$0.99 & --     & --  & 2.58     \\ 
& & &  & &  & & & & & $\pm$0.39 \\
\enddata
\end{deluxetable*}
\clearpage

\clearpage
\begin{deluxetable*}{ccccc}
\tabletypesize{\scriptsize}
\tablewidth{0pt}
\tablecolumns{5}
\tablecaption{Corrected proper motions and tangential velocities \label{tab:pm} }
\tablehead{ Name &   $\mu^{\rm corr}_{\alpha\cos\delta}$ & $\mu^{\rm corr}_{\delta}$   & $v^{\rm t}_\alpha$    & $v^{\rm t}_\delta$   \\
                 &       (mas yr$^{-1}$)             & (mas yr$^{-1}$)         &   (km s$^{-1}$)     & (km s$^{-1}$)     \\
       (1)       &           (2)                     &    (3)                  &   (4)               &   (5)             \\
}
\startdata
WLY2-11 & -1.96  $\pm$ 0.72 & -4.2   $\pm$ 0.75 & -1.26 $\pm$ 0.46 & -2.72  $\pm$ 0.49 \\ 
YLW24   & 0.63   $\pm$ 0.83 & -4.17  $\pm$ 0.84 & 0.41  $\pm$ 0.53 & -2.67  $\pm$ 0.54 \\ 
DoAr21  & -11.74 $\pm$ 0.89 & -5.91  $\pm$ 0.86 & -7.54 $\pm$ 0.57 & -3.79  $\pm$ 0.55 \\ 
LFAM2   & 1.87   $\pm$ 0.87 & -6.94  $\pm$ 0.73 & 1.26  $\pm$ 0.58 & -4.66  $\pm$ 0.49 \\ 
LFAM8   & 1.82   $\pm$ 0.83 & -8.96  $\pm$ 0.76 & 1.19  $\pm$ 0.54 & -5.86  $\pm$ 0.5 \\ 
S1      & 5.65   $\pm$ 0.88 & -6.1   $\pm$ 0.77 & 3.69  $\pm$ 0.57 & -3.99  $\pm$ 0.51 \\ 
LFAM15  & 1.37   $\pm$ 0.74 & -6.29  $\pm$ 0.71 & 0.9   $\pm$ 0.49 & -4.11  $\pm$ 0.46 \\ 
VSSG11  & -2.88  $\pm$ 0.92 & -18.67 $\pm$ 0.9  & -1.9  $\pm$ 0.61 & -12.36 $\pm$ 0.59 \\
\enddata
\end{deluxetable*}
\clearpage

\appendix 

\section{Non-thermal radio emission in young stars and GOBELINS detection statistics}\label{sec:mechanisms}

\subsection{Introduction}

YSOs are often detectable radio sources due to a variety of mechanisms. 
One common radio emission process in YSOs is thermal bremsstrahlung (free-free) continuum emission, either from photo-ionized gas around massive stars \citep[e.g.,][]{Churchwell_2002} or from shock-ionized gas in supersonic jets and outflows \citep[e.g.,][]{Rodriguez_1997}. The brightness temperature corresponding to thermal bremsstrahlung emission, however, is typically only 10$^4$ K, much too small to be detectable in VLBI observations. The situation with thermal dust continuum emission, which can be detected up to centimeter wavelengths from circumstellar disks \citep[e.g.,][]{Perez_2015}, is even worse, as such emission has brightness temperatures of at most several hundred K. Thermal line emission, from molecules or atoms, is similarly limited to brightness temperatures smaller than several hundred K. Thus, to find emission 
that could be detected in VLBI observations, one must turn to non-thermal mechanisms. For line emission, this implies focusing on strongly amplified maser lines, like those of water (H$_2$O), methanol (CH$_3$OH), formaldehyde (H$_2$CO), silicon monoxide (SiO), or hydroxyl (OH). Those lines (particularly those of water and methanol) are widespread in regions of high-mass star-formation, and can be detected and studied with VLBI observations up to distances of several kpc. Indeed, the BeSSeL project \citep{Brunthaler_2009} takes advantage of these lines to measure the parallax and proper motions of high-mass star-forming regions distributed across the entire Milky Way disk. In the Gould's Belt star-forming regions, however, only a handful of maser sources are known \citep[e.g.,][]{Moscadelli_2006}.  

Non-thermal {\em continuum} emission also exists, and broadly corresponds to the situation where electrons gyrate in a magnetic field. This type of radiation is called cyclotron, gyro-synchrotron, or synchrotron emission depending on whether the electrons are non-relativistic (with a $\gamma$ Lorentz factor about 1), mildly relativistic ($\gamma$ of a few), or ultra-relativistic ($\gamma$ $\gg$ 1), respectively \citep{Dulk_1985}. Such continuum non-thermal emission has been detected around a number of low-mass young stars \citep[e.g.,][and references therein]{Forbrich_2011}, and has been interpreted as coronal emission from active stellar magnetospheres. The most common mechanism appears to be gyro-synchrotron, although maser-amplified cyclotron emission \citep{Smith_2003} as well as synchrotron emission \citep{Massi_2006} have also been reported in rare instances. The magnetic field in which the electrons are gyrating is generated through the dynamo mechanism, which requires convection in the outer layers of the stellar interior \citep[e.g.,][]{Dormy_2013}. As a consequence, non-thermal coronal emission should occur only in low-mass stars, because intermediate and high-mass stars are fully radiative. This is true both for main sequence and pre-main sequence stars: while low-mass T Tauri stars approach the main sequence on fully convective Hayashi tracks, stars more massive than about 3 M$_\odot$ follow radiative Henyey tracks \citep[e.g.,][]{Palla_1993}. It is noteworthy, however, that a few intermediate-mass young stars have been found to exhibit non-thermal coronal emission \citep{Andre_1991, Dzib_2010}. The electrons gyrating in the magnetic field are thought to be accelerated to mildly relativistic speeds during energetic reconnection processes \citep{Parker_1957}, so the emission is often produced in flares \citep[e.g.,][]{Bower_2003} and is therefore highly variable. Once again, there are some exceptions to this general behavior \citep[e.g.,][]{Andre_1991}. Finally, it should be mentioned that most of the low-mass young stars where non-thermal emission has been reported are T Tauri stars (particularly Weak Line T Tauri stars), although a few younger Class I protostars have also been detected \citep{Forbrich_2006,Deller_2013}.

Coronal non-thermal radio continuum emission usually remains unresolved at the milli-arcsecond resolution of VLBI observations (e.g., \citealt{Loinard_2007}; but see \citealt{Andre_1991}). Theoretically, the emission is expected to be confined to the magnetosphere of the YSO, which extends to only a few stellar radii \citep{Bouvier_2007}. This is indeed very small: for instance, 10 R$_\odot$ corresponds to about 0.5 mas at 100 pc.

\subsection{Statistics}\label{sec:results}


As mentioned in the main text, we have firmly detected a total of 26 YSOs, corresponding to roughly half of the 50 YSOs targeted in our observations. This confirms the claim by \cite{Dzib_2013}, that about 50\% of the radio-bright YSOs in Ophiuchus are non-thermal emitters. We searched the literature and found that only 6 of these YSOs had previously been detected in VLBI experiments. These sources (DoAr 21, S1, LFAM 15,  VSSG 11, YLW 12 B, and SFAM 87) were known to be magnetically active stars, with centimetric non-thermal radio emission, before our observations \cite[e.g.][]{Andre_1992,Loinard_2008}. We note that \cite{Forbrich_2007} reported on the detection, with the HSA, of one of the components of the binary class 0/I source YLW 15 (a target that we also detect here). However, our analysis of the proper motions of the system components (see Section \ref{sec:ylw15}) suggests that the detection by \cite{Forbrich_2007} was likely spurious. The 26 detected YSOs correspond to a total of 34 individual young stars, because five detections are found to be tight binary systems, while one corresponds to a triple system (see Table \ref{tab:yso}). 

\cite{Dzib_2013} also published a list of YSOc detected in their VLA observations. They reported as YSOc those radio sources that are not associated with known young stars, but that show high VLA flux variations,  a negative spectral index, or circular polarization. Four of these sources were also detected in our VLBA observations, as well as another {27} (presumably background) sources.  We analyze the astrometry of these 31 objects in Section \ref{sec:bs} and find that,
as expected, most of them are extragalactic sources. 


In order to establish the nature of the emission, we have measured the brightness temperature ($T_b$) of the VLBA-detected sources, according to
\begin{equation}\label{eq:tb}
T_b = \frac{2c^2}{k_B\nu^2}\frac{S_{\rm total}}{\pi \theta_{\rm maj} \theta_{\rm min}},
\end{equation} 
where $\theta_{\rm maj}$ and $\theta_{\rm min}$ are the deconvolved sizes of the major and minor axes diameter, and $S_{\rm total}$ is the total flux density measured in the VLBA images. Since most of the sources have been detected at several epochs, we are reporting the highest measured brightness temperature. For unresolved sources, we give a lower limit to $T_b$ obtained using the corresponding beam size as an upper limit for the source size.  Brightness temperatures are given in columns (7) and (5) of Tables  \ref{tab:yso} and  \ref{tab:others}, respectively. All of the VLBA-detected  YSOs have $T_b > 10^6$ K, which is larger than the brightness temperature expected for thermal bremsstrahlung radiation ($T_b \apprle 10^4$ K), and consistent with the brightness temperature expected for non-thermal emission. Our study has, therefore, found a population of  non-thermal YSOs  
larger than previously reported. It is noteworthy that non-thermal emission is detected in sources in the Class I to Class III stages.

\subsection{Non-thermal emission as a function of evolutionary stage and multiplicity} 

\begin{figure}[!ht]
\begin{center}
 \includegraphics[width=0.45\textwidth,angle=0]{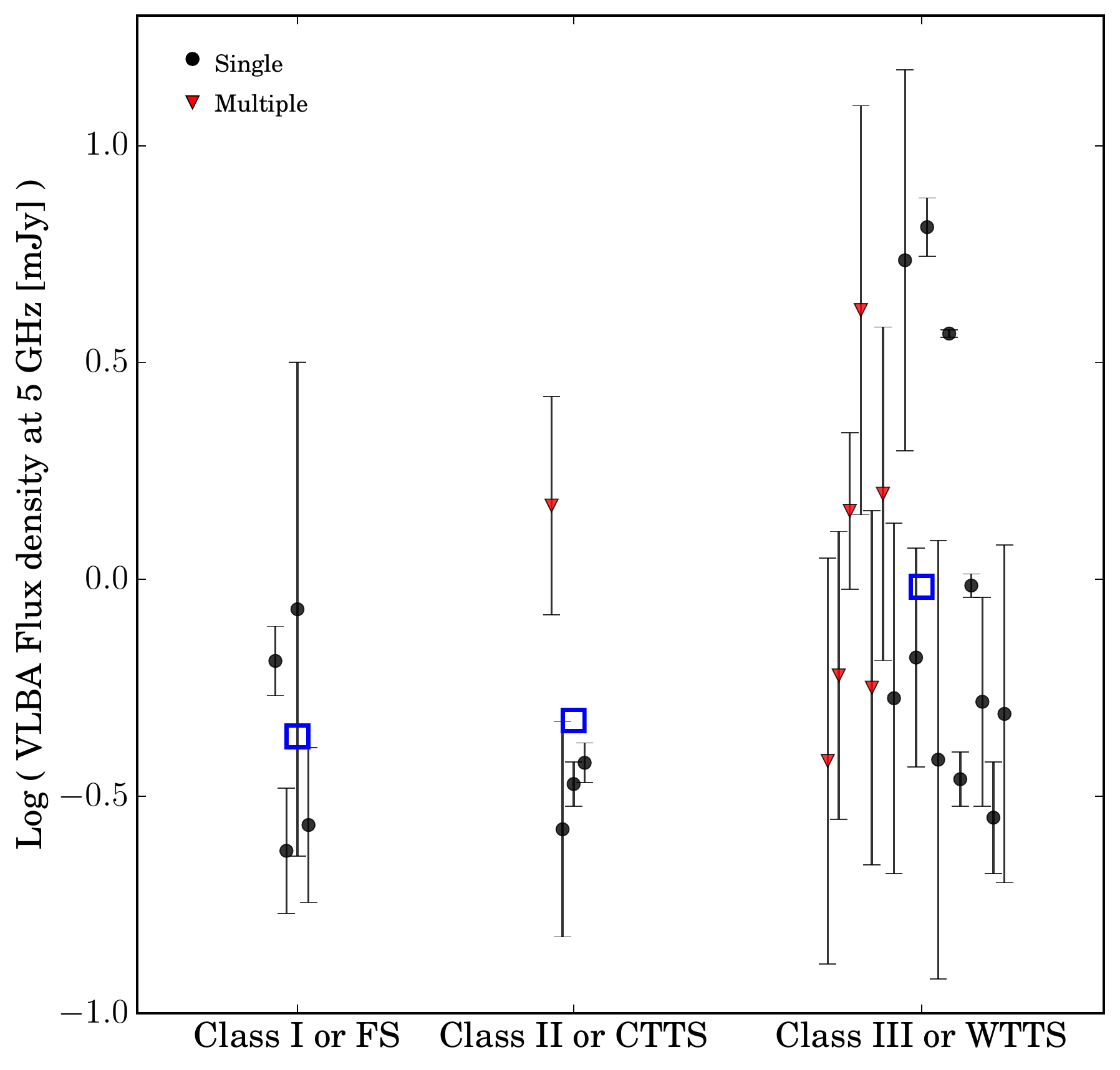}
\end{center}
 \caption{VLBA flux density at 5 GHz, as a function of evolutionary phase. The error bars represent the range in flux density from the maximum to the minimum measured values for variable sources, and measured flux uncertainties for sources with constant flux. Black circles correspond to single sources, while red triangles to multiple systems. We adopt for the VLBA flux density of multiple systems the sum of the flux densities of individual components. The black points are the mean VLBA flux density for each individual object, whereas the open blue squares are mean values for each object type. Here and in the rest of the paper, FS stands for Flat Spectrum, CTTS for classical T Tauri, and WTTS for weak-lined T Tauri.  }
\label{fig:vlba_sed}
\end{figure}

In Figure \ref{fig:vlba_sed}, we plot the VLBA flux density at 5 GHz of the detected YSOs as a function of their evolutionary phase, as measured from their infrared/millimeter spectral energy distribution (SED). It appears, at first sight, that older objects have, on average, stronger non-thermal radio emission. The significance of this correlation can be assessed by a Kolmogorov-Smirnov test on the three YSOs classes. The $p$ values derived from such a test are $>0.08$, so we cannot reject the null hypothesis that different classes are taken from the same distribution function. The relation is, therefore, not statistically significant. We also perform the Wilcoxon rank-sum statistic finding similar results. Thus, our conclusion about the significance of the correlation does not depend on the statistical test that we use.

From Table \ref{tab:yso}, it is clear that Class II and III sources are the most common types of YSOs with non-thermal radio emission.  Past studies had only found two cases of Class I protostars with non-thermal emission detectable in VLBI observations. These sources are CrA IRS 5 \citep{Deller_2013} and EC 95 \citep{Dzib_2010}. Here we have detected non-thermal emission from three other Class I objects, namely  LFAM 4, YLW 15,\footnote[2]{As we mentioned earlier, a VLBI detection was reported by \cite{Forbrich_2007} toward YLW 15, but that detection was most likely spurious (see  \ref{sec:ylw15}).}  and WLY 2-67. We have, therefore, more than doubled the number of protostars with confirmed non-thermal emission. In a recent study, \cite{Heiderman_2015} investigated whether the Class 0+I and Flat SED sources identified in the c2d \citep{Evans_2009} and Gould Belt \citep{Dunham_2015} surveys are in the embedded phase. These authors used the detection of the HCO$^+~J=3\rightarrow 2$ as a good indicator of the Stage 0+1, which corresponds to ``a star and a disk embedded in a dense, infalling envelope''  \cite{van_Kempen_2009}. Our 3 Class I objects with non-thermal radio emission meet the criterion to be in this embedded stage. Non-thermal emission has not been detected in Class 0 sources and at this point it is unclear if this results from a lack of such emission or if such young systems always contain thermal radio jets that systematically absorb underlying active coronas. 

Magnetic activity is known to occur throughout the early evolution of low-mass stars, so protostellar sources could easily produce non-thermal emission. On the other hand, accretion and outflow activity is also present in this protostellar stage. In order for the observer to see the radio emission from the stellar corona, the line of sight cannot cross regions where optically thick radio emission is present --i.e.\ the central portions of disks or wind/outflow systems. This might occur only for some privileged relative orientation of the system, and naturally explains the low fraction of detected protostars in our observations, in comparison with the larger number of detections of more evolved objects.

\medskip

There does seem to exist a correlation between non-thermal radio emission in young stellar objects and multiplicity. Perhaps the best documented case is that of the V773 Tau system \citep{Massi_2002,Torres_2012}, where the radio flux increases by more than one order of magnitude when the system passes periastron. We have found here that a significant fraction of the VLBA-detected young stars belong to very tight binary or multiple systems (with a separation of a few tens of mas -- a few AU; Table \ref{tab:yso}). 
Multiplicity may help to clear out the surrounding material, and result in non-thermal emission that is statistically less affected by free-free absorption,
but the exact mechanisms behind this remains unclear.
The flux from thermal jets becomes transparent at frequencies above a few GHz \citep{Reynolds_1986}. Thus,  coronal non-thermal emission from protostars with thermal emission should be detectable at higher frequencies.

In our VLBA observations, we found that 7 YSOs form multiple systems, and one more has evidence of multiplicity (See Table \ref{tab:yso}). This represents $\sim30\%$ of the total number of YSOs detected with the VLBA. The angular separations of the components in these systems range from 4 to 315 mas, corresponding to 0.6 to 44 AU at the distance of Ophiuchus. In order to investigate if this fraction is expected from the known population of multiple systems in Ophiuchus, or if it indicates that tight multiple systems are more likely to be non-thermal radio emitters than single stars or wider multiple systems, we compared it with the binary fraction reported in the literature. Recently, \cite{Cheetham_2015} compiled high-resolution multiplicity data and combined them with their results from an aperture masking survey; they obtained a binary fraction of  $35\pm6\%$ for spatial scales from $1.3-41.6$ AU. Taken at face value, our VLBA detection of 30\% of binaries with separations between 0.6 and 44 AU appears to be  consistent with the binary fraction derived from the multiplicity survey by \cite{Cheetham_2015}. However, we still favor the idea that very tight binaries are more often radio sources than single stars or more separated binary systems, because 7 of our 8 detected binaries have separations of a few AU, whereas only one has a separation larger than 10 AU. Thus, for separations below 10 AU, there does appear to be an excess of radio bright binaries. Our interpretation will be tested by considering the whole sample of multiple stars seen in the five regions observed by GOBELINS, and comparing with multiplicity studies in the infrared, where aperture masking observations can explore  angular separations similar to those attained with the VLBA.

\subsection{Radio luminosity function of non-thermal YSOs}

\begin{figure}[!t]
\begin{center}
 \includegraphics[width=0.5\textwidth,angle=0]{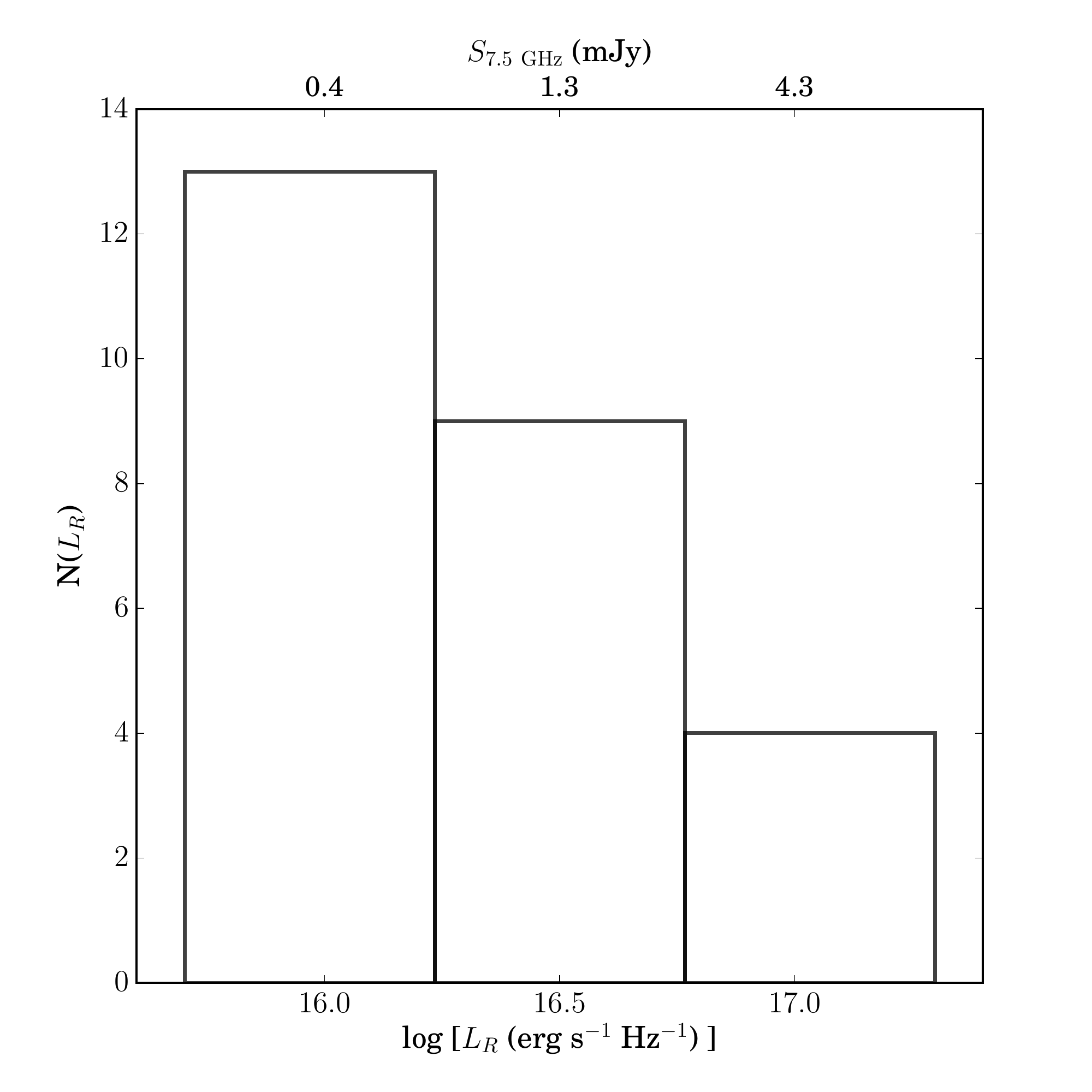}
\end{center}
 \caption{Luminosity function for the YSOs detected with the VLBA at 5 GHz.}
\label{fig:LR_function}
\end{figure}

We show in Figure \ref{fig:LR_function} the number of objects versus the radio luminosity at 5 GHz, i.e., the luminosity function, of the 26 YSOs detected with the VLBA. The number of objects appears to decrease  with increasing luminosity, following roughly a linear trend (in logarithmic luminosity). Assuming that the trend is valid for lower luminosities, we can infer that deeper observations with an improved sensitivity by an order of magnitude will detect $\sim 20$ more YSOs with non-thermal radio emission, i.e., will double the number of sources. It will  be interesting to construct the luminosity function for the other regions considered in GOBELINS,  and see if this is a general trend.  

\section{Other sources}\label{sec:bs}

In this section, we analyze the astrometry of 27 sources detected in our VLBA observations that are not classified as YSOs in the literature. 
These sources are listed in Table \ref{tab:others}.
Column 6 in that table provides a proposed classification based on the radio emission properties in the VLA observations of \cite{Dzib_2013}. Out of the 31 sources that are not known YSOs in that list, 18 are proposed to be extragalactic (E), 4 are YSO candidates (YSOc), and 9 have no proposed identifications. Many of these sources have been detected in several of our VLBA observations, so we can analyze their changes in position as a function of time to constrain their location along the line of sight. Extragalactic sources are expected to remain practically fixed on the celestial sphere, so any measured change in their positions should be of the order of or less than the rms astrometric errors. These errors  are given by the systematic errors found in the astrometric fits, and are $\apprle 1.2$ mas  for YSOs in the core, while for sources located in the L1689 streamer, we obtain systematic errors $\apprle 2.4$ mas. These larger errors in L1689 result from the larger separation between the main phase calibrator and the targets there. Galactic sources, on other hand, should show an appreciable parallax and a proper motion signature, particularly if they are within a few kpc. 

For each source in Table \ref{tab:others} that has been observed in at least two epochs (that is, 56 out of the 64 sources in Table \ref{tab:others}), we measured the shift in position ($\sqrt{\Delta \alpha\cos\delta^2  + \Delta\delta^2}$) between consecutive epochs, normalized it to one year, and averaged over all consecutive pairs of epochs. For instance, for a source with three detections, we averaged two displacements: that between epoch 1 and epoch 2, and that between epoch 2 and epoch 3. We will call this quantity the {\em position change rate} in the rest of this section. This position change rate should be zero (within the errors) for extragalactic sources and non-zero for (nearby) Galactic objects. However, since the position change rate contains information on both the parallax and the proper motions, and is based on a varying number of detections for different sources (from 2 to 5 detections), a non-zero value cannot be easily interpreted in terms of distance along the line of sight.  Sources for which a non-zero value is found should be analyzed in more detail. 

\begin{figure}[!t]
\begin{center}
 \includegraphics[width=0.5\textwidth,angle=0]{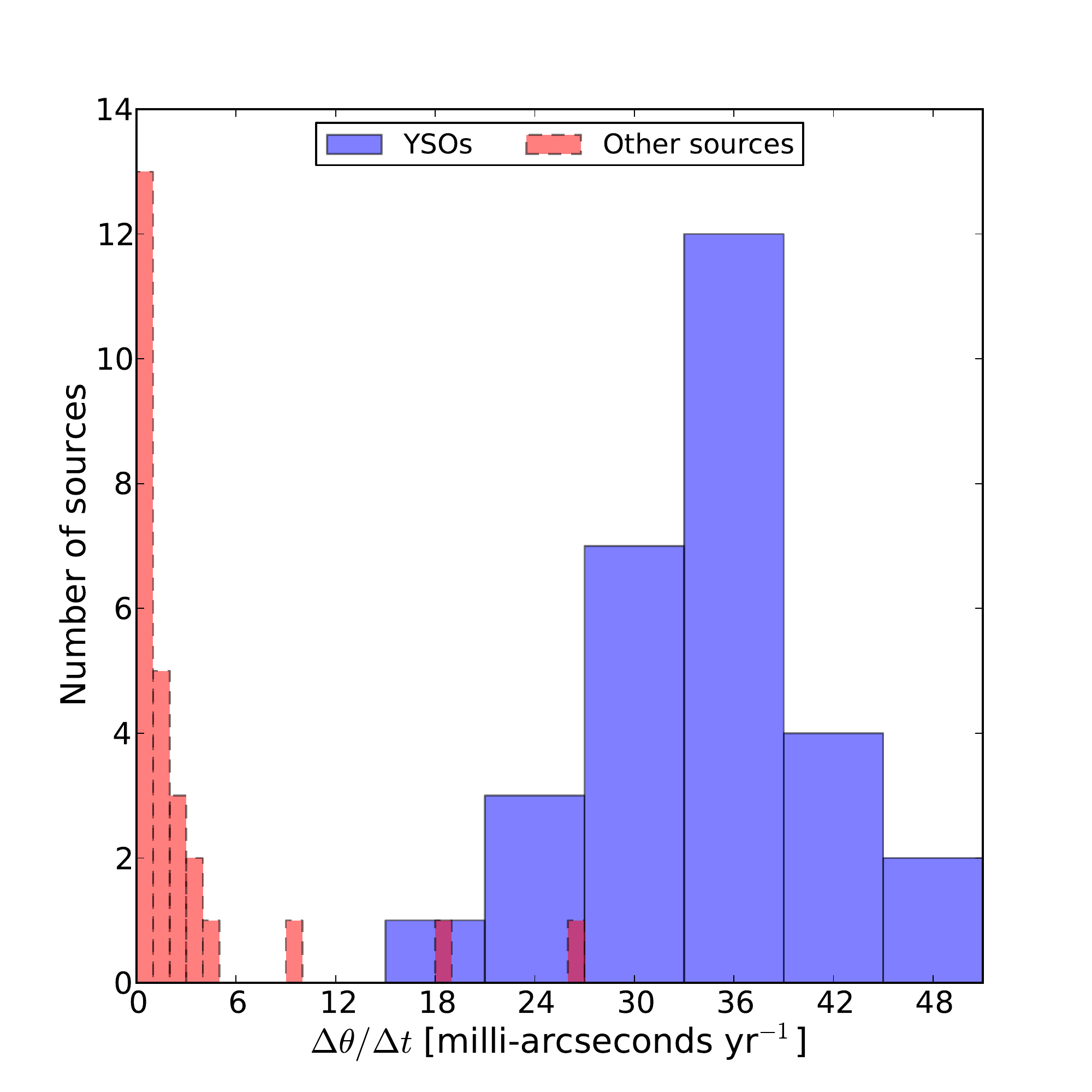}
\caption{Histogram of position change rate for all sources detected at least twice toward Ophiuchus. The sources previously identified as YSOs are shown as a blue histogram, whereas the other sources are shown as a red histogram.} 
\label{fig:hist}
\end{center}
\end{figure}

The results are shown in the form of a histogram in Figure \ref{fig:hist}, where we separate known YSOs from other sources. The two histograms are markedly different. For known YSOs, the histogram is roughly Gaussian and centered around a position change rate of $\sim$ 36 mas yr$^{-1}$. For the rest of the sources, on the other hand, the distribution is dominated by a peak around a position change rate of zero, to within a few mas yr$^{-1}$. Specifically, the first four bins in the histogram (23 sources) correspond to sources that do not show appreciable motion on the celestial sphere (i.e., they have position changes between consecutive epochs smaller on average than the astrometric noise), and we identify them as extragalactic. This new classification is shown in Column 6 of Table \ref{tab:yso}. Of course, for the four unclassified sources that have only been detected once in our VLBA observations, the position change rate cannot be measured, and we classify these sources as ``?'' in Column 6 of Table \ref{tab:others}.

There are three sources, however, that do show a non-zero position change rate, and we now discuss them in turn. The source with the largest position change rate ($\sim$ 27 mas yr$^{-1}$) is GBS-VLA~J163151.93-245617.4, and has been detected three times. However, as in the case of YLW15, the three detections occurred around the spring equinoxes. Thus, it is impossible to obtain any meaningful information on its parallax. We note, however, that its position change rate places it squarely within the range covered by the YSOs in Ophiuchus. In addition, because the detections occurred only on spring equinoxes (April 2013,  March 2014 and March 2016), the displacement is dominated by proper motions, with only a small contribution from parallax. The fact that the measured displacement ($\sim$ 27 mas yr$^{-1}$) is so similar to the proper motion of sources in Ophiuchus suggests that GBS-VLA~J163151.93-245617.4 is not only Galactic, but indeed an Ophiuchus member. Interestingly, this source had been classified as a YSO candidate by \cite{Dzib_2013} on the basis of the variability and spectral slope of the radio emission. Additional detections will be necessary to confirm this and obtain a trigonometric parallax but, for the time being, we classify GBS-VLA~J163151.93-245617.4  as an ``Oph" member in Column 6 of Table  \ref{tab:others}.

The source with the second largest position change rate ($\sim$ 19 mas yr$^{-1}$) is SFAM12, which was detected only twice. In this case, the two detections occurred around the spring equinox of 2015 and the fall equinox of 2015, respectively. Thus, the displacement results from the combination of both parallax and proper motion. This makes it unlikely that SFAM12 is an Ophiuchus member, since for sources in Ophiuchus, a larger position change rate would be expected (roughly 25 mas yr$^{-1}$ due to proper motion, and an additional 15 mas due to parallax, since the observations were obtained at opposite equinoxes). Thus, we classify SFAM12 as a Galactic source (``G" in Column 6 of Table \ref{tab:yso}), but unlikely to be associated with Ophiuchus itself. We note that SFAM12 was previously believed to be extragalactic. Finally, the third source with a definite non-zero position change rate ($\sim$ 10 mas yr$^{-1}$) is LFAM17. Since there were 5 detections of this target, an astrometric fit could be attempted. Formally, it suggests a parallax $\varpi$ = 0.87 $\pm$ 0.14 mas (corresponding to $d$ = 1.2$\pm{0.2}$~kpc), but the fit is poor. Certainly, however, this source is Galactic but much farther away than Ophiuchus. Thus, we classify it as ``G" in Column 6 of Table \ref{tab:yso}. We note that \cite{Dzib_2013} had classified it as a YSO on the basis of its radio properties. Both SFAM12 and LFAM17 are likely to be active stars (possibly, but not necessarily, young ones) located behind Ophiuchus.

To finish this section, we should mention source GBS-VLA~J163138.57-253220.0, which is the sole member of the fifth red histogram bin in Figure \ref{fig:hist}. Its position change rate is formally above the astrometric uncertainty. It was detected four times, so an astrometric fit could again be attempted. As in the case of LFAM17, the fit is poor; it results in a parallax $\varpi$ = 0.17 $\pm$ 0.05 mas (corresponding to $d$ = 5.9$_{-1.3}^{+2.4}$ kpc). This might suggest that GBS-VLA~J163138.57-253220.0 is a Galactic source at several kpc, but given the large errors and poor quality of the fit, we cannot completely discard that it be extragalactic. Thus, we classify it as ``G?" in Column 6 of Table \ref{tab:others}. In general, the identification proposed by \cite{Dzib_2013} for sources in Table \ref{tab:others} matches  well our new classification based on VLBA observations.


\bibliographystyle{aasjournal}
\bibliography{oph_paper_combined-aasv6.bib}

\clearpage
\begin{deluxetable*}{cccccccc}
\tabletypesize{\scriptsize}
\tablecaption{Other detected sources \label{tab:others}}
\tablewidth{0pt}
\tablehead{ GBS-VLA& Other identifier & Maximum flux   & Maximum flux   & log [$T_b$ (K) ]   & \multicolumn{2}{c}{Type$^{1}$} & Number of                     \\
              name &                  & at 5 GHz (mJy) & at 8 GHz (mJy) &                    & \multicolumn{2}{c}{of source}  &    detections/observations    \\
        (1)        &     (2)          &   (3)          &    (4)         &     (5)            & \multicolumn{2}{c}{(6)}        &  (7)                          \\
}
\startdata
%
J162540.94-244147.2 & SSTc2d J162540.9$-$244147 & 0.39 & $<$0.06 & 0.08 & E & ? & 1/1 \\
J162547.68-243735.7 & $-$ $$ & 0.45 & 0.37 & 0.05 & ? & E & 3/3 \\
J162633.48-241215.9 & SFAM12 $$ & 0.39 & $<$0.06 & 0.05 & E & G & 2/2 \\
J162635.33-242405.2 & LFAM13 $$ & 0.79 & $<$0.09 & 0.07 & E & E & 4/10 \\
J162646.36-242002.0 & LFAM17 $$ & 0.64 & $<$0.06 & 0.08 & YSOc & G & 5/6 \\
J162702.15-241927.8 & GDS J162702.1$-$241928 & 0.48 & $<$0.06 & 0.07 & E & E & 4/4 \\
J162713.06-241817.0 & $-$ $$ & 0.21 & $<$0.06 & 0.05 & ? & E & 2/9 \\
J162718.25-243334.8 & $-$ $$ & 0.52 & $<$0.06 & 0.16 & ? & ? & 1/6 \\
J162729.23-241755.3 & ROC25 $$ & 2.74 & $<$0.18 & 0.06 & E & E & 5/5 \\
J162734.55-242020.7 & ROC26 $$ & 1.25 & $<$0.18 & 0.06 & E & E & 5/5 \\
J163027.69-243300.2 & SSTc2d J163027.7$-$243300 & 0.25 & $<$0.06 & 0.05 & E & E & 2/4 \\
J163032.26-243127.9 & SSTc2d J163032.3$-$243128 & 0.90 & $<$0.06 & 0.06 & YSOc & E & 3/3 \\
J163033.26-243038.7 & SSTc2d J163033.2$-$243039 & 0.40 & $<$0.06 & 0.06 & YSOc & ? & 1/3 \\
J163036.26-243135.3 & $-$ $$ & 0.80 & $<$0.06 & 0.07 & ? & E & 3/3 \\
J163109.79-243008.4 & ROC49 $$ & 0.73 & $<$0.06 & 0.07 & E & E & 3/3 \\
J163115.25-243313.8 & $-$ $$ & 0.29 & $<$0.06 & 0.04 & ? & E & 2/4 \\
J163120.14-242928.5 & ROC52 $$ & 4.38 & $<$0.06 & 0.14 & E & E & 3/3 \\
J163130.62-243351.6 & SSTc2d J163130.6$-$243352 & 0.64 & $<$0.06 & 0.06 & E & E & 3/3 \\
J163138.57-253220.0 & $-$ $$ & 0.56 & 1.00 & 0.07 & ? & G? & 4/4 \\
J163151.93-245617.4 & $-$ $$ & 0.34 & 0.28 & 0.06 & YSOc & Oph & 3/4 \\
J163154.49-245217.1 & SSTc2d J163154.5$-$245217 & 0.35 & $<$0.06 & 0.05 & E & E & 2/3 \\
J163159.36-245639.7 & SFAM127 $$ & 1.72 & $<$0.06 & 0.09 & E & E & 2/2 \\
J163202.39-245710.0 & $-$ $$ & 0.18 & 0.34 & 0.06 & ? & E & 2/3 \\
J163210.77-243827.6 & SFAM130 $$ & 0.42 & $<$0.09 & 0.07 & E & E & 4/5 \\
J163211.08-243651.1 & SSTc2d J163211.1$-$243651 & 0.80 & $<$0.09 & 0.08 & E & E & 5/5 \\
J163212.25-243643.7 & $-$ $$ & 0.37 & $<$0.06 & 0.07 & ? & E & 2/5 \\
J163213.92-244407.8 & $-$ $$ & 0.36 & $<$0.05 & 0.06 & ? & ? & 1/6 \\
J163227.41-243951.4 & SSTc2d J163227.4$-$243951 & 0.51 & $<$0.06 & 0.06 & E & E & 3/5 \\
J163231.17-244014.6 & SSTc2d J163231.2$-$244014 & 0.51 & $<$0.06 & 0.05 & E & E & 2/5 \\
J163245.23-243647.4 & SFAM200 $$ & 0.47 & 0.26 & 0.07 & E & E & 3/3 \\
J163617.50-242555.4 & SFAM212 $$ & 1.82 & 2.50 & 0.09 & E & E & 5/5 \\
\enddata
\tablecomments{Reported sources  have flux densities above $6\sigma$ and $5\sigma$ in the cases of one or several detections, respectively. Non-detections are indicated by giving an upper flux density limit of $3\sigma$.}
\tablenotetext{1}{The first entry indicates the classification given by 
 \cite{Dzib_2013} from VLA observations. The second entry indicates the classification 
 given in this work by comparing the shift in source positions against the rms astrometric errors. Here {\it E} stands for extragalactic and {\it G}, for Galactic objects. 
 }
\end{deluxetable*}
\clearpage



\end{document}